\documentclass[prd,aps,preprint,longbibliography,
  showpacs,showkeys,lengthcheck,
  nofootinbib,tightenlines,onecolumn,notitlepage,
  preprintnumbers,superscriptaddress]{revtex4-2}

\usepackage{graphicx}
\usepackage[usenames,dvipsnames]{xcolor}
\graphicspath{{figures/}}

\usepackage{diagbox}
\usepackage{booktabs}

\usepackage{amsmath,amssymb,amsfonts}
\usepackage{mathrsfs}
\usepackage{slashed}
\usepackage{bm}
\usepackage{tensor}
\usepackage{multirow}
\usepackage{tikz}
\usetikzlibrary{arrows.meta, positioning, fit, shapes.geometric, calc}
\usepackage[bb=boondox]{mathalpha}
\usepackage{ytableau}

\usepackage{bbold}

\usepackage[draft,commentmarkup=uwave]{changes} % Use for draft to show changes
\definechangesauthor[color=orange!76!black]{Frank}
\definechangesauthor[color=green!50!black]{Wim}
\definechangesauthor[color=blue]{rev}  % used for resubmission

\usepackage{hyperref}
\hypersetup{colorlinks=true,citecolor=blue,linkcolor=blue,urlcolor=blue}

\usepackage{cancel}

\newcommand{\sh}[1]{#1\hskip -5pt  / }
\newcommand{\shl}[1]{#1\hskip -7pt  / }

\newcommand{\overbar}[1]{\mkern 1.5mu\overline{\mkern-1.5mu#1\mkern-1.5mu}\mkern 1.5mu}

\newcommand{\ProdQt}{\mathsf{T} }

\begin{document}

%\title{Covariant expressions for bilinear spinor amplitudes of any spin}
\title{Spinor Representations for Fields with any Spin:\texorpdfstring{\\}{} Lorentz Tensor Basis for Operators and Covariant Multipole Decomposition}

\author{Wim Cosyn}
\email{wcosyn@fiu.edu}
\affiliation{Department of Physics, Florida International University, Miami, FL 33199, USA}

\author{Frank Vera}
\email{fvera@jlab.org}
\affiliation{Theory Center, Jefferson Lab, Newport News, VA 23606, USA}
\affiliation{Department of Physics, Florida International University, Miami, FL 33199, USA}

\begin{abstract}
This paper discusses a framework to parametrize and decompose operator matrix elements for particles with higher spin $(j > 1/2)$ using chiral representations of the Lorentz group, \emph{i.e.} the $(j,0)$ and $(0,j)$ representations and their parity-invariant direct sum. Unlike traditional approaches that require imposing constraints to eliminate spurious degrees of freedom, these chiral representations contain exactly the $2j+1$ components needed to describe a spin-$j$ particle.  The central objects in the construction are the $t$-tensors, which are generalizations of the Pauli four-vector $\sigma^\mu$ for higher spin.  For the generalized spinors of these representations, we demonstrate how the algebra of the $t$-tensors allows to formulate a generalization of the Dirac matrix basis for any spin.  For on-shell bilinears, we show that a set consisting exclusively of covariant multipoles of order $0\leq m \leq 2j$ forms a complete basis.  We provide explicit expressions for all bilinears of the generalized Dirac matrix basis, which are valid for any spin value.  As a byproduct of our derivations we present an efficient algorithm to compute the $t$-tensor matrix elements.  The formalism presented here paves the way to use a more unified approach to analyze the non-perturbative QCD structure of hadrons and nuclei across different spin values, with clear physical interpretation of the resulting distributions as covariant multipoles.
\end{abstract}

\preprint{JLAB-THY-25-4259}

\maketitle

{\hypersetup{linkcolor=black}
% or \hypersetup{linkcolor=black}, if the colorlinks=true option of hyperref is used
\tableofcontents
}

\section{Introduction}
\label{sec:intro}

While the Standard Model only contains particles with spin 0,1/2 and 1, hadronic and nuclear bound states exhibit a much richer spin content, with spins larger than 1.  To quote two examples with large spin values, the most abundant isotope of Mendelevium has spin 8~\cite{ENSDF} and the $\Delta(2950)$ is a spin-15/2 resonance~\cite{ParticleDataGroup:2024cfk}.  A field-theoretical treatment of these higher-spin objects requires covariant fields, wave functions and wave equations with spin $>$ 1/2.  

The relativistic treatment of higher-spin particles has a long history~\cite{Dirac:1936tg,Fierz:1939ix,Rarita:1941mf}.  The traditional method to deal with spin $j>1/2$ one-particle states uses covariant objects which contain more than $2j+1$ degrees of freedom,  i.e. a four-vector $\epsilon^\mu(\lambda)$ for spin 1, a vector-spinor Rarita-Schwinger fermion for spin 3/2~\cite{Rarita:1941mf}, etc.  Spurious degrees of freedom are eliminated by imposing constraints on the wave functions in the form of ``equations of motion'', see e.g. Ref.~\cite{Pascalutsa:2006up} for a review of the spin-3/2 case.  General constructions for the wave functions are based on tensor products of four-vectors, belonging to Lorentz group representations of the form $(1/2,1/2)$,  with Dirac bispinors, belonging to Lorentz group representations of the form $(1/2,0) \oplus (0,1/2)$.  As spin values increase, more constraints must be imposed, such as:
\begin{enumerate}
\item $\left( \sh{p} -m \right) \psi(p,\lambda)=0$, for the case of fermion fields with spin 1/2 and above.
\item $\ p_\mu \epsilon^\mu(p,\lambda)=0$, for the case of fields with spin-1 and above.
\item $\ \gamma^\mu \psi_\mu(p,\lambda)=0$, for the case of fields with spin-3/2 and above. 
\item etc...  
\end{enumerate}
Maintaining such constraints consistently can be challenging in both analytic and numerical calculations (see e.g. Refs.~\cite{vanNieuwenhuizen:1981ae, Pascalutsa:1998pw}), and often leads to spurious singularities in amplitudes when using conventional constrained fields. A Lagrangian formulation for these Rarita-Schwinger fields was constructed by Singh and Hagen using auxiliary fields~\cite{Singh:1974qz,Singh:1974rc}.

Alternative approaches exist for treating higher-spin particles.  A $(j_1,j_2)$-representation of the Lorentz group contains spins $|j_1-j_2|,\cdots, j_1+j_2$.  Consequently, any $(j_1,j_2)$-representation for which $|j_1-j_2| \leq j \leq j_1+j_2$ can be used to represent spin-$j$ particles.  The simplest of these are the so-called maximally chiral representations $(j,0)$ and $(0,j)$, which contain the minimal $2j+1$ components to describe a spin-$j$ particle and all spin degrees of freedom reside in a single chiral sector. 
Therefore, these representations do not require additional constraints to eliminate spurious components.  These chiral fields (and their parity-invariant direct sum $(j,0)\oplus (0,j)$) were developed and studied a long time ago~\cite{Joos:1962qq,Barut:1963zzb,Weinberg:1964cn,Williams:1965rga}, but building consistent Lagrangians or field equations with them proved difficult~\cite{Tung:1967zz,Shay1968ALF,Hurley:1971ipy}.  
We note, however, that the interest in maximally chiral representations has persisted. Spinor-helicity methods based on the spin-1/2 chiral spinors have flourished in perturbative QCD and modern on-shell amplitude theory, and have been extended to {\it massive} higher-spin states \cite{Arkani-Hamed:2017jhn}. They have been successfully applied in the context of the Standard Model and beyond Standard Model \cite{Chalmers:1997ui,Chalmers:1998jb,Chalmers:2001cy}, in electroweak processes \cite{Beenakker:1991jk, Bohm:1993qx, Dittmaier:1997he}, as well as more recent work on massive higher-spin EFTs \cite{Ochirov:2022nqz, Cangemi:2023ysz}. By contrast, in the context of non-perturbative applications in hadronic and nuclear physics their use 
% of \((j,0)\oplus(0,j)\) representations 
has remained comparatively rare.

We want to stress that our interest here is not the construction of a(n effective) field theory based on the chiral representations, but their use in matrix elements of QCD operators for (massive) hadrons and nuclei.  These matrix elements are used to parametrize the non-perturbative partonic (spin) structure of QCD bound states.  Based on Lorentz covariance, they can be decomposed and parametrized with independent wave function bilinears, each multiplied by Lorentz invariant functions.  The invariant functions contain information about the non-perturbative structure of these strongly interacting bound systems (hadrons, nuclei). A classic example would be the parametrization of the hadron current in elastic electron-nucleon scattering:
\begin{align}
    \sum_f e_f \langle\, p',\lambda'\,|\, \bar{q}_f(0)\gamma^\mu q_f(0)\,|\,p,\lambda\,\rangle = \;  F_1(Q^2)\, \bar{u}(p',\lambda')\gamma^\mu u(p,\lambda)
     +  F_2(Q^2)\bar{u}(p',\lambda')\frac{\text{i}\sigma^{\mu\nu}q_\nu}{2m} u(p,\lambda)\,,
\end{align}
where $e_f$ is the fractional charge of quark flavor $f$, $m$ the hadron mass, $Q^2=-(p'-p)^2=-q^2$. $F_1$ and $F_2$ are the invariant Dirac and Pauli nucleon electromagnetic form factors and $\sigma^{\mu\nu} = \frac{i}{2}[\gamma^\mu,\gamma^\nu]$. A more contemporary pinnacle of this approach would be the off-forward spin-1/2 matrix element of a bilocal light-ray operator that parametrizes quark and gluon generalized transverse momentum distributions (GTMDs)~\cite{Meissner:2009ww}.

For hadrons of higher spin, similar efforts have focused on the spin 1 and spin 3/2 cases, given their phenomenological interest (vector mesons, deuteron nucleus, baryon decuplet).  The main approach to parametrize these matrix elements has been to parametrize the bilinears using the four-vectors for spin 1 and Rarita-Schwinger vector-spinor for spin 3/2.  The decomposition is essentialy carried out again for each spin case and operators separately, see Refs.~\cite{Berger:2001zb,Cosyn:2018rdm,Bacchetta:2000jk,Fu:2022bpf,Fu:2024kfx} for some examples.  For these higher spin cases, the physical interpretation of the different Lorentz invariant distributions becomes less clear. For instance, the multipole structure of different distributions is not immediately evident and requires additional case-by-case analysis.

We show here that the use of the chiral representations can pave the way to a more unified approach.  For our purposes, the hadron states in these matrix elements can be considered as asymptotic states and the question of a consistent Lagrangian for the chiral fields is not relevant at this point. As we show in this article, which lays out the foundations of the framework, one obtains essentially a generalization of the well-known Dirac algebra for the spin-1/2 case.  While this has been presented in the past~\cite{Williams:1965rga, Gomez-Avila:2013qaa}, our presentation here is a more intuitive and practical one which includes explicit expression for the spin-$j$ bilinears.  Moreover, we show that a natural basis for the on-shell spin-$j$ bilinears is built from two separate towers (even/odd) of covariant $\mathfrak{sl}(2,\mathbb{C})$ multipoles, this builds on work explored by Cotogno et al. in Ref.~\cite{Cotogno:2019vjb}.  

The generalized Dirac algebra structure  has the advantage that one can use the accrued intuition from the spin-1/2 case and apply it to any higher spin.  With the breakdown in covariant multipoles, it becomes immediately clear which structures appear in identical form in the lower spin cases, and which structures are new due to the increased spin degrees of freedom.  As such, a decomposition with a natural physical interpretation for the distributions or form factors can be obtained in a straightforward manner.  Additionally, the use of chiral spinors that do not require any additional constraints can improve the analyticity properties of these matrix elements, important for numerical implementations.  All this combined makes for an attractive, systematic and unified procedure to decompose the matrix elements for hadrons and nuclei with higher spin. 

This article provides all the details about the basic construction of the framework culminating in the expressions for all gamma matrix basis bilinears. Subsequent work will apply this formalism to specific operators.
As the chiral spinors are central to our framework but not commonly used, we start with a review of the chiral fields formalism in Section~\ref{sec:constructions}.  After a brief overview of the irreducible representations of the Lorentz group, we focus in the left and right chiral representations, denoted as $(j,0)$ and $(0,j)$, and the corresponding causal fields.  The $t$-tensors are discussed, which are the generalization of Pauli four-vector $\sigma^\mu$ to any spin.  These tensors are central to the framework and can be used to build the spinors and propagators for massive particles of any spin~\cite{Weinberg:1964cn}.  
We discuss the $\tilde p^\mu$ parameters that can be used to generate the standard helicity and light-front spinors; these parameters and their use in this context have, to our knowledge, not been presented before.  Here, the $\tilde p^\mu$ are a 4-tuple that package the effect of the standard boost, and contracting $\tilde p^\mu$ with the $t$-tensors result in the chiral spinors, see Eqs.~(\ref{eq:canonical_boosts_t}) and further.

\begin{figure}[ht]
    \centering
    \resizebox{\textwidth}{!}{%
    \begin{tikzpicture}[
        node distance=2.cm and 2.5cm,
        block/.style={rectangle, draw, minimum width=4cm, minimum height=1.5cm, align=center},
        smallblock/.style={rectangle, draw, minimum width=3.5cm, minimum height=1.5cm, align=center},
        specialblock/.style={ellipse, draw, minimum width=4.5cm, minimum height=2cm, align=center},
        arrow/.style={-{Stealth[length=3mm]}},
        % Section color styles
        secII/.style={draw=blue!70, fill=blue!10},
        secIII/.style={draw=blue!70, line width=2pt, fill=blue!10},
        secV/.style={draw=green!70!black, fill=green!10},
        secVI/.style={draw=orange!70!black, fill=orange!10},
        secVII/.style={draw=purple!70!black, fill=purple!10},
        ]
        
        % Define the top row nodes - Section II
        \node[block, secII] (propagator) {Propagator $\Pi(p)$ [Eq.~(\ref{eq:prop_ttensor})]};
        \node[block, secII, right=of propagator] (chiral) {SL(2,$\mathbb{C}$) chiral irreps,\\[1em]Spin $j: (j,0)$, $(0,j)$};
        \node[block, secII, right=of chiral] (fields) {Chiral fields $\psi(x),\chi(x)$ [Eq.~(\ref{eq:causal_fields})]\\[1em] Chiral spinors  $u_L(p,\lambda)$,$u_R(p,\lambda)$ [Eq.~(\ref{eq:uspinors_def})] }  ;
        
        % Define the second row with proper spacing - Section II+III with special shape for basis
        \node[specialblock, secIII, below=2cm of chiral] (basis) {$t^{(\mu)}, \bar{t}^{(\mu)} \rightarrow \mathfrak{u}(2j+1)$ basis  };
        \node[block, secV, below right = 0.2cm and -0.2cm of basis] (matrixelements) {Expand matrix elements $\langle \lambda_f |\hat{\mathcal{O}}|\lambda_i\rangle$};
        \node[block, secII, below= 7cm of fields] (bispinors) {Bispinors $u(p,\lambda)$ [Eq.~(\ref{eq:bispinor_def})]};
        
        % Define the left column with irreducible and reducible - Section IV
        \node[block, secV, below left = 2.5cm and -3cm of propagator] (irreducible) {Cubic $\rightarrow$ irreducible\\[1em]$t^{(\mu)}\bar{t}^{(\nu)}t^{(\rho)}\sim\prod_i \mathcal{C}^{\mu_i \nu_i \rho_i \sigma_i} t_{(\sigma)}$  [Eq.~(\ref{eq:t_reduction_cubic})]};
        \node[block, secV, below right=2cm and -3cm of irreducible] (reducible) {
            Quadratic $\rightarrow$ reducible \\[1em]
            $t^{(\mu)}\bar t^{(\nu)}\sim \prod_i \mathcal{Q}^{\mu_i \nu_i \sigma_i}t_{(\sigma)}$ [Eq.~(\ref{eq:def_t_reduction_quadratic})]\\[0.8em]
            $\qquad\sim \sum_{m=0}^{2j} \mathsf{T}_m^{(\mu\nu)}$ [Eq.~(\ref{eq:quadratic_products_tbart_with_T})] \\[1em]
            $\equiv \text{SL}(2,\mathbb{C})$ multipoles [Eq.~(\ref{eq:sl(2,C)_multipole_tTensors})]
        };
        
        % Define the middle column with gamma matrix - Section IV
        \node[block, secV, below right=2.5cm and -2.5cm of basis, text width=4.6cm] (dirac) {
                \centering $\gamma$-matrix basis $\Gamma_i$:\\[1em]
                \begin{flushleft}
                $\bullet \quad \gamma^{(\mu)}, \gamma^{(\mu)}\gamma_5$  [Eq.~(\ref{eq:gamma_mu})]\\[1em]
                $\bullet \quad \gamma_5,\; \mathbb{1}=\mathcal{G}_0,$\\[0.5em] $\quad\phantom{\bullet} \mathcal{G}_m^{(\mu\nu)},\; m=1,\ldots,2j$\\[0.5em]$\quad\phantom{\bullet}$  [Eq.~(\ref{eq:multipoles_gamma})]
                \end{flushleft} 
            };        
                % Define the right column nodes - Section V
        \node[block, secVI, below right=2.5cm and -2cm of dirac] (bilinears) {Bilinears for matrix elements\\[1em]
                        $\langle p_f,\lambda_f|\hat{\mathcal{O}}|p_i, \lambda_i\rangle = \underset{j}{\sum} F_j \,\bar{u}(p_f,\lambda_f) \Gamma_j u(p_i,\lambda_i)$ [Eq.~(\ref{eq:bilinear_gen_reduction})]};
        
        % Define the bottom row nodes - Sections V and VI
        \node[smallblock, secVI, below right=2.9cm and -3cm of reducible] (explicit) {Explicit expressions\\for bilinears of any spin $j$};
        \node[smallblock, secVII, below left= 2cm and -3cm of bilinears] (independent) {Multipole basis is sufficient\\[1em] $\bar u(p_f,\lambda_f) \mathcal{G}_m^{(\mu\nu)}u(p_i,\lambda_i)$ [Eq.~(\ref{eq:bi_indexed_bilinear})]};
        \node[smallblock, below=of explicit] (applications) {Applications: \\specific operators};
        
        % Draw the connections with arrows
        % Top row connections
        \draw[arrow] (chiral) -- node[above] {} (fields);
        \draw[arrow] (chiral) -- node[above] {} (propagator);
        
        % Second row connections
        \draw[arrow] (chiral) -- node[left] {} (basis);
        \draw[arrow] (basis) -- node[above] {\large $p_{\mu}$} (propagator);
        \draw[arrow] (fields.south)+(2,0) -- node[right] {\large $\mathsf{P}$} (bispinors.north);
        
        % Left column connections
        \draw[arrow] (basis) -- node[above] {\large Algebra} (irreducible);
        \draw[arrow] (irreducible) -- node[right] {\large $\;\rho_i=0\;$} (reducible);
        \draw[arrow] (reducible) -- (explicit);
        
        % Modified arrow from irreducible to explicit (going around reducible)
        \draw[arrow] (irreducible.south)+(-1.25,0) |- ($(explicit.west)+(-3,0)$) |- (explicit.west);
    
        % Middle column connections - Modified to hit specific points with restored P labels
        \draw[arrow] (basis.south)+(-1,+0.05) -- ++(-1,-2.5)  node[right, pos=0.7] {\large $\mathsf{P}$} |- ($(dirac.west)+(0,0.5)$);
        
        % Modified arrow from reducible to dirac with restored P label
        \draw[arrow] (reducible.east)+(0,-1) -- node[above, pos=0.5] {\large $\mathsf{P}$} ($(dirac.west)+(0,-0.7)$);
          
        % Thick arrow from basis to bispinors
        \draw[arrow] (basis) -- node[above] {\large $\tilde{p}_\mu$} (fields);
        
        % Tensor product symbol instead of cross
        \node[circle, draw, inner sep=0.1cm] (tensor) at ($(dirac.south) + (3.1cm,-1cm)$) {$\otimes$};
        
        % Connections to and from the tensor product
        \draw[arrow] (dirac) -- (tensor);
        \draw[arrow] (bispinors) -- (tensor);
        \draw[arrow] (tensor) -- (bilinears);
        
        % Bottom row connections
        \draw[arrow] (bilinears) -- node[right] {\large conditions} node[left] {\large on-shell\,} (independent);
        \draw[arrow] (bilinears) -- (explicit);
        \draw[arrow] (independent) -- (applications);
        
        % Additional connections
        \draw[arrow] (explicit) -- (applications);
        \draw[arrow] (bilinears) -- (applications);
        \draw[arrow] (basis.east)+(-0.3,-0.5) -- (matrixelements);
    
        % Add legend for sections - including special shape for basis
        \node[below left=3cm and -0.5cm of applications](colorlegend) {\Large Color legend:};
        \node[secII, rectangle, draw, minimum width=2cm, minimum height=0.7cm, align=center, 
               right=0.4cm of colorlegend] (legendII) {Section II};
        \node[secIII, ellipse, draw, line width=2pt, minimum width=2.2cm, minimum height=0.9cm, align=center, 
              right=0.4cm of legendII] (legendIII) {Sections II, III};
        \node[secV, rectangle, draw, minimum width=2cm, minimum height=0.7cm, align=center, 
              right=0.4cm of legendIII] (legendV) {Section V};
        \node[secVI, rectangle, draw, minimum width=2cm, minimum height=0.7cm, align=center, 
              right=0.3cm of legendV] (legendVI) {Section VI};
        \node[secVII, rectangle, draw, minimum width=2cm, minimum height=0.7cm, align=center, 
              right=0.3cm of legendVI] (legendVII) {Section VII};

    \end{tikzpicture}
        }
    \caption{Flowchart illustrating the structure of the paper and formalism.  Equations are meant to be read as schematic here.  Equation numbers link to the full expressions. Arrows labeled with $\mathsf{P}$ stand for parity-invariant extensions that combine a left- and right-chiral representation.  From the way the oval blob connects to many other parts of the flowchart, one can appreciate the central role the $t$-tensors play in the construction. }
    \label{fig:flowchart}
\end{figure}
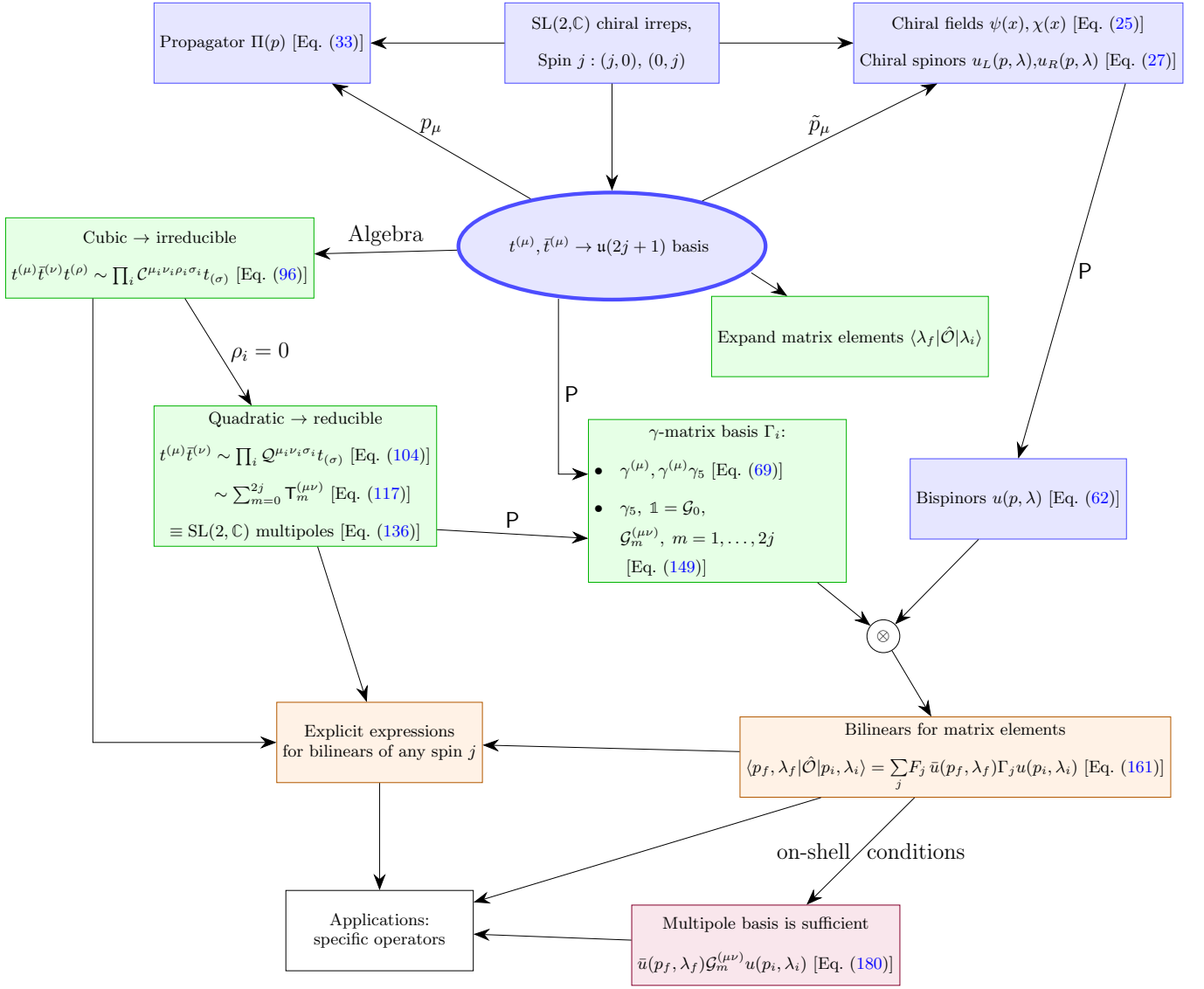

Background on the helicity and light-front spinors is included in Appendix~\ref{sec:hel_lf_spinors}.  We illustrate in Fig.~\ref{fig:flowchart} how the $t$-tensors relate to all the different objects used in the formalism and the corresponding sections of the paper.  We hope that this helps clarify the overall structure and interconnections in this article.  Section~\ref{sec:constructions} concludes by introducing the generalized Dirac gamma matrices which will enter as part of the independent bilinears later.

We then move on to present the new results developed in this work:
\begin{enumerate}
    \item  In Sec. ~\ref{sec:t_tensors_gen}, we show that the rank-$2j$ $t$-tensors contain a basis for Hermitian matrices of order $2j+1$ (the $\mathfrak{u}(2j+1)$ algebra), which can be used to parametrize any Hermitian operator matrix element for spin-$j$ particles, including all the QCD operators of interest.  We include in this section an efficient algorithm to construct the $t$-tensors for any spin, based on the fact that in a lightcone-spherical basis all the matrices contained within the $t$-tensors have exactly one non-zero matrix element.  A second algorithm to construct them is included in Appendix~\ref{sec:t_algorithm}, which is based on the anticommutators (Jordan algebra) of the rotation matrices.  This algorithm was one we arrived at in the early stages of this work and is less efficient than the one presented in Sec.~\ref{sec:t_tensors_gen}.  We choose to include it however, because it helps elucidate some aspects of the $t$-tensors, which can be hard to grasp at first.

    \item The massive spinor-helicity (MHS) formalism of Ref.~\cite{Arkani-Hamed:2017jhn} also uses objects that reside in the maximally chiral representations and is widely used in high-energy physics.  Consequently there should be a map between MSH and the chiral spinors considered here.  This is detailed in Section~\ref{sec:MSH}, in addition to illustrating that all the different types of chiral spinors used here also obey massive spinor-helicity-like equations. 

    \item Section~\ref{sec:t_gamma_tensors_algebra} contains the most important results in this work.  
    Products of $t$-tensors naturally appear in calculations with chiral (bi)spinors and the generalized gamma matrices. 
    Consequently their algebra is of interest.  In Sec.~\ref{sec:Cubic_products}, we focus first on the cubic product of $t$-tensors which is shown to be irreducible and can be written as an invariant linear combination of $t$-tensor elements for any spin:
    \begin{align}
        t^{\mu_{1} \cdots \mu_{2j} }  \bar{t}^{\rho_{1} \cdots \rho_{2j} } t^{\sigma_{1} \cdots \sigma_{2j} } &= \frac{1}{[(2j)!]^2}  
        \underset{\left\{(\rho),(\sigma) \right\}}{\mathcal{S}} 
        \left( \prod_{l=1}^{2j} {\mathcal{C}}^{\mu_{l} \rho_{l} \sigma_{l} \alpha_{l}} \right) t_{\alpha_{1} \cdots \alpha_{2j}} \,,\\
        {\mathcal{C}}^{\mu \rho \sigma \alpha } &= g^{\mu \rho } g^{\sigma \alpha } - g^{\mu \sigma } g^{\rho \alpha } + g^{\mu \alpha } g^{\rho \sigma } + {\rm i} \epsilon^{\mu \rho \sigma \alpha } \,.
    \end{align}
    The proof of this result is presented in Appendix \ref{sec:proof_cubic_reduction}.  Next, in Sec.~\ref{sec:quadratic_rpoducts_t}, we focus on the quadratic products of $t$-tensors, which are shown to be reducible.  The quadratic products contain a series of bi-indexed tensors (having symmetrized pairs of two antisymmetric indices) that transform covariantly within themselves and turn out to be in one-to-one correspondence to $\mathfrak{sl}(2,\mathbb{C})$ multipoles, see Sec.~\ref{sec:quadratic_t_vs_multipoles}.  These results for the $t$-tensor algebra are then used in Sec.~\ref{sec:dirac_basis} to identify a basis for the generalized Gamma matrices in a manner that exhibits this multipole structure.  Useful expressions and relations for the coefficients that appear in the algebra of the $t$-tensors are collected in Appendices~\ref{sec:coeff_relations} and ~\ref{sec:generalized_quadratic_reductions_and_CR}.

    \item As a practical application of the derived results, we present expressions for all bilinears using the gamma matrix basis in Sec.~\ref{sec:gen_bilinears}.  The expressions are valid for the different types of spinors (canonical, helicity, light-front) and can be used in further studies of QCD operator matrix elements for particles of any spin.  Additionally, we discuss on-shell identities in Sec.~\ref{sec:onshell_dentities}. These allow us to further reduce the number of independent bilinears that needs to be used in the decomposition of matrix elements to only those corresponding to the $\mathfrak{sl}(2,\mathbb{C})$ multipoles.  
        
\end{enumerate}

We wrap up the paper in Sec.~\ref{sec:discussion} with a discussion of our results and outlook for future extensions and applications.  As the work presented here is quite technical and the details of the results heavily depend on the conventions in notation and definitions, we summarize all our choices in the dedicated Appendix~\ref{sec:notation}. We encourage readers to consult it before delving into the detailed derivations.

%%%%%%%%%%%%%%%%%%%%%%%%%%%
%%%% Section II      %%%%%%
%%%%%%%%%%%%%%%%%%%%%%%%%%%

\section{Review of the construction for general spinors}
\label{sec:constructions}

The goal of this section is to provide a self-contained and pedagogical review and summary of Weinberg's construction for spinor representations of any spin for massive particles~\cite{Weinberg:1964cn}.  
Parts of this construction are anticipated in the work of Joos~\cite{Joos:1962qq} (generalized Dirac equation) and Barut-Muzinich-Williams~\cite{Barut:1963zzb} (intertwining maps between spinorial and Lorentz tensor representations for any spin).  This section serves as a reminder of the properties of the Lorentz group, its chiral and bispinor representations, causal fields and propagators.  For spin 0 and spin 1/2, the construction corresponds to the well-known formalism associated with Klein-Gordon, respectively Weyl and Dirac fields, but for spin $j > 1/2$ the use of these chiral and bispinor representations is not common.  
If the reader is familiar with the chiral spinor construction, they can skip ahead to the next section, although Sec.~\ref{sec:boosts_spinors} does include some results for the values of parameters $\tilde p$ for helicity and light-front spinors which to our knowledge have not appeared elsewhere.

\subsection{Irreducible representations of the Lorentz group}
\label{sec:irreps_Lorentz}

The starting point to discuss the finite-dimensional irreducible representations of the Lorentz group are the well known commutation relations for the $\mathfrak{so}(3,1) = \mathfrak{sl}(2,\mathbb{C})$ algebra of the homogeneous Lorentz group:
\begin{align} 
\label{eq:lorentz_JK}
\left[\mathbb{J}_l,\mathbb{J}_m \right]&= {\rm i} \epsilon_{lmn}\mathbb{J}_n \,, \nonumber\\
\left[\mathbb{J}_l,\mathbb{K}_m \right]&= {\rm i} \epsilon_{lmn}\mathbb{K}_n \,, \nonumber\\
\left[\mathbb{K}_l,\mathbb{K}_m \right]&= - {\rm i} \epsilon_{lmn}\mathbb{J}_n \,.
\end{align}
Here 
% Roman indices 
% % $ \{l,m,n\} = \{1,2,3\} \leftrightarrow  \{x,y,z\}$ 
% $l,m,n \in \{1,2,3\}$
% are used, 
$\mathbb{J}_i$ are the generators of rotations and $\mathbb{K}_j$ are the generators of pure (rotationless) boosts.  The Lorentz algebra $\mathfrak{sl}(2,\mathbb{C})$ can be complexified as $\mathfrak{sl}(2,\mathbb{C})_\mathbb{C} \cong \mathfrak{sl}(2,\mathbb{C}) \otimes \mathbb{C}$ by introducing
\begin{align} 
\label{eq:lorentz_AB}
&\mathbb{A}_m= \frac{1}{2} ( \mathbb{J}_m + {\rm i} \mathbb{K}_m )\,, 
&\mathbb{B}_m= \frac{1}{2}( \mathbb{J}_m - {\rm i} \mathbb{K}_m )\,,\nonumber\\
%%%%%%%%
&\mathbb{J}_m =  \mathbb{A}_m + \mathbb{B}_m \,, 
&\mathbb{K}_m = -{\rm i}( \mathbb{A}_m - \mathbb{B}_m )\,.
\end{align}
Using the commutation relations of Eq.~(\ref{eq:lorentz_JK}) leads to the following commutation relations for the generators $\mathbb{A}_m,\mathbb{B}_n$:
\begin{align} 
\left[\mathbb{A}_l,\mathbb{A}_m\right] &= {\rm i} \epsilon_{lmn} \mathbb{A}_n \,, \nonumber\\
\left[\mathbb{B}_l,\mathbb{B}_m\right] &= {\rm i} \epsilon_{lmn} \mathbb{B}_n \,, \nonumber\\
\left[\mathbb{A}_l,\mathbb{B}_m\right] &=0 \,.
\end{align}
This shows that $\mathbb{A}_m$ and $\mathbb{B}_m$ form two commuting subalgebras. Each subalgebra obeys the $\mathfrak{su}(2)$ commutation relations.   The Lie algebra direct sum decomposition $\mathfrak{sl}(2,\mathbb{C})_\mathbb{C} \cong \mathfrak{so}(3,1)_\mathbb{C} \cong \mathfrak{su}(2)_\mathbb{C} \, \oplus \, \mathfrak{su}(2)_\mathbb{C}$ is thus obtained. The irreducible representations (irreps) of SL($2,\mathbb{C})$, the double cover of the Lorentz group, are labeled by a pair of heighest weights $(j_A,j_B)$ that identify the corresponding $\mathfrak{su}(2)$ irreps of the two subalgebras.  While the finite-dimensional representations of the $\{\mathbb{A}_m,\mathbb{B}_m\}$ algebras are unitary, the factor i in the equation for $\mathbb{K}_m$ in Eq.~(\ref{eq:lorentz_AB}) shows that finite dimensional representations of the boosts will not be unitary.  However, the rotations remain unitary. This is also reflected in the non-compact nature of the homogeneous Lorentz group (boost parameters are not bound, while rotation parameters are).

Of particular focus in Weinberg's construction and the work presented here are the chiral representations of the Lorentz group.  These chiral representations correspond to the irreps labeled as $(j,0)$, denoted \textit{left-handed}, and $(0,j)$, denoted \textit{right-handed}.  This choice of handedness follows the standard convention.  They are particular as they transform as singlets under one of the two $\mathfrak{su}(2)$ subalgebras, which leads to the following relation between the chiral \emph{representations} $[\mathbb{J}_m],[\mathbb{K}_m]$ of the generators of pure boosts and rotations:
\begin{subequations}    \label{eq:JK_chiral}
\begin{align}
 \text{left-handed $(j,0)$:}&  \hspace{0.3cm}  [\mathbb{K}_m] \equiv  - {\rm i} [\mathbb{J}_m] \,, \\
%
%\vspace{0.2cm}   	
 \text{ right-handed $(0,j)$:}&    \hspace{0.3cm}  [\mathbb{K}_m] \equiv +  {\rm i} [\mathbb{J}_m]
 \,.
\end{align}
\end{subequations}

In Table~\ref{tab:lorentz_irreps}, we give an overview of the objects that correspond to the representations with the smallest $(j_A,j_B)$ values.  Although most of these objects are commonly used in covariant calculations, the spin-1 chiral spinors corresponding to $(1,0)$ and $(0,1)$ are not commonly used.  Objects like these will be the focus of this work.

\setlength{\tabcolsep}{20pt}
\renewcommand{\arraystretch}{1.5}
\begin{table}[ht]
    \centering
    \begin{tabular}{c|l}
        $(j_A,j_B)$ &  Representation\\
        \hline
        \hline
         $(0,0)$  & Scalar \\
         \hline
         $(\frac{1}{2},0)$  & Left-handed Weyl spinor, defining SL($2,\mathbb{C}$) representation \\
         $(0,\frac{1}{2})$  & Right-handed Weyl spinor \\
         $(\frac{1}{2},0) \oplus (0,\frac{1}{2})$  & Dirac spin-1/2 bispinor \\
         \hline
         $(\frac{1}{2},\frac{1}{2})$  & Four-vector, defining SO(3,1) representation \\
         \hline
         \multirow{2}{*}{$(1,0)$}  & Anti-self-dual 2-form \\
         &Left-handed spin-1 spinor \\
         \multirow{2}{*}{$(0,1)$}  & Self-dual 2-form \\
         &Right-handed spin-1 spinor \\
         \hline
         \multirow{2}{*}{$(1,0) \oplus (0,1)$}  & Parity-conserving 2-form (e.g. $F_{\mu\nu}$) \\
         & Spin-1 bispinor\\         
         \hline
    \end{tabular}
    \caption{Summary of the lowest dimensional representations of the Lorentz group.}
    \label{tab:lorentz_irreps}
\end{table}

\subsection{Chiral representations} 
\label{sec:chiral_reps} 

In Ref.~\cite{Weinberg:1964cn}, Weinberg used the chiral irreducible representations to construct causal fields for massive particles.   
These fields are unitary and infinite-dimensional representations of the Poincar\'{e} group.  He derived their Feynman rules and provided expressions for their propagators and spinors.  As the chiral spinors are objects that contain the minimal number of degrees of freedom ($2j+1$ for a spin-$j$ particle), these fields do not obey any free equation of motion besides the Klein-Gordon equation, which is satisfied for each field component.  This fixes the on-mass shell condition $p^2=m^2$ for the massive particle. 
Feynman rules and interacting Lagrangians for maximally chiral fields have been constructed in other contexts as well \cite{Dittmaier:1997he, Ochirov:2022nqz}, though these are not our focus here. 
In what follows, we summarize the most important results that Weinberg derived in Ref.~\cite{Weinberg:1964cn}, specific details can be found in that reference.\footnote{It is worth emphasizing that the construction for massless particles and fields proceeds differently because of their different little group~\cite{Weinberg:1964ev}.  For theories with long-range interactions, Weinberg in Ref.~\cite{Weinberg:1965rz} recovers gauge invariance and the Maxwell (massless spin-1) or Einstein (massless spin-2) equations. For a recent chiral formulation that includes massive and massless particles see Ref.~\cite{Arkani-Hamed:2017jhn}.}

For the chiral representations of Eq.~(\ref{eq:JK_chiral}), rotations $R$ take the form of the well-known Wigner $D$-matrices
\begin{equation} \label{eq:WignerD}
   D^{(j)}[R] = \overbar D^{(j)}[R] = e^{-{\rm i}\bm \theta \cdot \bm J^{(j)}} \,, 
\end{equation}
where in general $D^{(j)}{[\Lambda]}$ ($\overbar{D}^{(j)}{[\Lambda]}$) represent the left-handed (right-handed) irreducible matrix representation in dimension $2j+1$ of Lorentz transformation $\Lambda$, \emph{e.g.}, rotations in the case above.
The ${J}^{(j)}_i$ are the $(2j+1)$-dimensional representation of the generators of rotation in the basis of states with spin projections $\lambda$ and $\lambda'$
\begin{equation}
    \left({J}^{(j)}_i\right)_{\lambda'\lambda} = \langle j, \lambda' | \mathbb{J}_i | j, \lambda \rangle \,.
\end{equation}
If spin is quantized along the $z$-axis, these correspond to the familiar Pauli matrices for spin 1/2, the generators of rotations in the spherical basis for spin 1, etc.

Applying Eq.~(\ref{eq:exp_map}) to the left- and right-handed chiral representations of pure boost transformations, we obtain using Eqs.~(\ref{eq:JK_chiral})
\begin{subequations}
\label{eq:chiral_boost}
\begin{align} 
& D^{(j)}{[L({\bm p})]} = e^{-\rho \,\hat{\bm p} \cdot \bm J^{(j)}} \,,  \\ 
& \overbar{D}^{(j)}{[L({\bm p})]} = e^{+\rho \,\hat{\bm p} \cdot \bm J^{(j)}} \,.
\end{align}
\end{subequations}
These boosts are obtained from the Wigner $D$-matrices of Eq.~(\ref{eq:WignerD}) by analytic continuation~\cite{Polyzou:2012ut}. $L(\bm p)$ is the canonical standard boost\footnote{As they are connected to the identity, $L(\bm p)$ are elements of SO(3,1) and denote proper orthochronous Lorentz transformations} that transforms the four-momentum of a massive particle with mass $m$ from its rest frame momentum $\overset{\circ}{p}{\vphantom{p}}^\mu$ to the on-shell four-momentum $p^\mu$:
\begin{align}
   & p^\mu = \tensor{L(\bm p)}{^\mu_\nu}\,\overset{\circ}{p}{\vphantom{p}}^\nu \,,\\
    %%%
    &\overset{\circ}{p}{\vphantom{p}}^\mu\equiv(m,\bm 0) \,, \nonumber\\
    &p^\mu=(E_p,\bm p) \equiv (\sqrt{m^2+\bm p^2},\bm p) \,.
\end{align}

The boost parameters $\rho$ (rapidity) and $\hat{\bm p}$ (boost direction) are defined by
\begin{subequations}
\label{eq:rho_p}
\begin{align}
\sinh \rho&=\frac{|\bm p|}{m} \,, \\
\hat{\bm p}&=\frac{\bm p}{|\bm p|} \,,
\end{align}
\end{subequations}
and we have $\cosh \rho = \frac{E_p}{m}$.

For the chiral representations of Eq.~(\ref{eq:JK_chiral}), the generators of rotations are Hermitian and those of pure boosts are skew-Hermitian.  As a consequence, rotation matrices are unitary,
and the boost matrices in Eqs.~(\ref{eq:chiral_boost}) are Hermitian.
Equation~(\ref{eq:chiral_boost}) shows that the two chiral representations for pure boosts are related to each other by
\begin{align} \label{eq:D_Dbar}
&D^{(j)}[L^{-1}({\bm p})] = D^{(j)}  [L(-\bm p)] = \overbar{D}^{(j)}[L(\bm p)] \,.
\end{align}
%where $\bar{p}=(E,-\bm p)$.
For general Lorentz transformations $\Lambda$ (combinations of boosts and rotations), the two chiral representations are related as
\begin{align}\label{eq:chiral_relation}
 \overbar{D}^{(j)}[\Lambda]=\left(D^{(j)}[\Lambda^{-1}] \right)^{\dagger}\,.
\end{align}

For the general linear group one can consider four independent representations on $d$-dimensional (complex) vector spaces. Starting with the defining or fundamental representation on a vector space $V$, we also have those on the complex conjugate $V^*$, the dual $\widetilde{V}$, and the dual complex conjugate $\widetilde{V}^*$ vector spaces. When considering a subgroup of the general linear group, constraints that apply to the elements of the subgroup can result in 
%relations/identifications/
isomorphisms between two (or more) of the four representations above, which reduces the number of independent representations.  For SL(2,$\mathbb{C}$),  the fundamental representation has 2 by 2 complex matrices with unit determinant.  The unit determinant requirement implies the presence of an invariant rank-2 tensor,\footnote{We always use the word tensor in relation to Lorentz transformations. It can refer either to vector or chiral representations, which can always be inferred by context.}
the two-dimensional Levi-Civita tensor $\epsilon^{ij}$.  For all chiral representations of higher spin, this invariant tensor can also be introduced and is denoted as $C$ here. Its properties are detailed in the next paragraph.  The existence of $C$ leads to the equivalence of the representations $V$ and its dual $\widetilde{V}$. Similarly, $V^*$ and its dual $\widetilde{V}^*$ are also equivalent.  In analogy to the spin-1/2 case~\cite{Srednicki:2004hg}, $C$ can be used to raise and lower indices on spinors. We follow the standard spinor index notation,\footnote{ 
% We use lowercase letters from the beginning of the Roman alphabet, $c,\, d,\, e,\, f$, to denote spinor indices for generic spin in the left-chiral representation. The indices $a,\, b$ are reserved for spin 1/2. For the right-chiral representation we use the same convention with dotted indices. For spin-$j$ representation indices take integer values from $-j$ to $j$.
See App.~\ref{sec:notation} for conventions regarding index notations
}
\begin{subequations}
\label{eq:V_reps}
    \begin{align}
        &\psi_c \in V \,,\\
        &\psi^c \equiv \tensor{C}{^{cd}}\psi_d \in \widetilde{V} \,,\\
        &\chi^{* \dot c} \in \widetilde V^* \,,\\
        &\chi^*_{\dot{c}} \equiv \tensor{(C^{-1})}{_{\dot c \dot d}}\chi^{*\dot{d}} \in V^*
        \,,
    \end{align}
\end{subequations}
where $C^{cd} = C^{\dot{c}\dot{d}}$.
The relation between the left and right chiral representations presented in Eq.~(\ref{eq:chiral_relation}) shows we can identify the left chiral spinors with $\psi_c$ and the right chiral spinors with $\chi^{*\dot{c}}$.  In the remainder of this section we keep the spinorial indices explicit in many expressions as it helps to show what representations certain objects belong to, as well as identifying invariants.  In later sections we will often drop these indices for readability. From the context, the indices can always be restored if desired.

The defining equation for the invariant tensor $C$ is
\begin{equation}\label{eq:def_C}
    \quad C \bm{J}^{(j)} C^{-1} = - \bm{J}^{(j) *} = -\bm {J}^{(j)T} \,,
\end{equation}
where we used the hermiticity of the $\bm J^{(j)}$ in the last step.
$C$ is unitary and satisfies
\begin{align}
& C C^{\dagger }= C^{\dagger } C= \mathbb{1}^{(j)}, \nonumber\\
& C C^*=(-)^{2 j} \mathbb{1}^{(j)} \,.
\end{align}
The matrix $C$ can be stated up to a phase factor which is conventionally chosen such that
\begin{align}\label{eq:def_charge_conj_C}
& C^{cd}= (-)^{j-c}\delta_{-c,d} \,,\\
& C^{\dagger}={C}^{-1}=C^T=(-)^{2j}C \,,
\end{align}
where indices run from $-j,\ldots,j$.
Note that in this convention the matrix $C$ is real. The only non-zero elements are $\pm 1$ that alternate sign along the anti-diagonal, starting with $+1$ for the top right element.
Using Eq.~(\ref{eq:def_C}), the equivalence of representations can now be shown as
\begin{subequations}
\label{eq:charge_conj_J}
\begin{align}
    &\tensor{\left( \overbar D^{(j)*}[\Lambda]\right)}{^c_d} = \tensor{C}{^{ce}} \tensor{\left(D^{(j)}[\Lambda]\right)}{_e^f} \tensor{C}{^{-1}_{fd}} \,, &\widetilde V \sim V \,,\\ 
    &\tensor{\left(D^{(j)*}[\Lambda]\right)}{_{\dot c}^{\dot d}} = \tensor{C}{^{-1}_{\dot c \dot e}} \tensor{\left(\overbar D^{(j)}[\Lambda]\right)}{^{\dot e}_{\dot f}} \tensor{C}{^{\dot f \dot d}} \,, &V^* \sim \widetilde V^*
    \,.
\end{align}
\end{subequations}
from where, using Eq.~(\ref{eq:chiral_relation}), it follows that $C$ is invariant under Lorentz transformations.

The chiral boosts of Eq.~(\ref{eq:D_Dbar}) play a central role in the construction of the causal fields by providing an intertwining map between the fields and the creation and annihilation operators~\cite{Weinberg:1995mt}.   The creation and annihilation operators form an infinite-dimensional unitary representation of the Lorentz group. 
For the annihilation operator $a(\bm p,j\lambda)$, labeled with momentum $\bm p$ and SU(2) little group spin quantum number $\lambda$:
\begin{equation}\label{eq:transform_annihilation}
    U(\Lambda)\, a(\bm p,j\lambda) \,U^{-1}(\Lambda) = D_{\lambda\lambda'}^{(j)}[R^{-1}_\text{w}(\bm p,\Lambda)]\, a(\Lambda \bm p,j\lambda')
    \,,
\end{equation}
where the \emph{momentum-dependent} Wigner rotation (little group transformation) takes the form
\begin{equation}\label{eq:def_Wignerrot}
    R_\text{w}(\bm p,\Lambda) \equiv L^{-1}(\Lambda \bm p)\,\Lambda\, L(\bm p)
    \,.
\end{equation}
The creation/annihilation operators and single particle states have the following normalizations
\begin{subequations}
\begin{align}
\label{eq:a_commutator}
    [a(\bm p,j\lambda),a^\dagger(\bm p',j'\lambda')]_\pm = (2\pi)^3\delta_{jj'}\delta_{\lambda\lambda'}\delta(\bm p-\bm p')\,,\\
    %%%
    |p,j\lambda \rangle = \sqrt{2E_p}\,a^\dagger(\bm p,j\lambda)|0\rangle  \,,\\
    %%%%%
    \langle p',j'\lambda' | p, j \lambda \rangle = (2\pi)^3(2E_p)\delta_{jj'}\delta_{\lambda\lambda'}\delta(\bm p-\bm p') \,.
\end{align}    
\end{subequations}
By multiplying the creation/annihilation operators with the representation of the standard boost included in the Wigner rotation of Eq.~(\ref{eq:def_Wignerrot}), the $(2j+1)$-component causal fields $\varphi$ (left-chiral) and $\chi^*$ (right-chiral) transform 
in their spinor index with the finite dimensional (but non-unitary) representations $D[\Lambda]$ or $\overbar{D}[\Lambda]$: 
\begin{subequations}
\label{eq:causal_fields}
    \begin{align}
        \varphi_{c}(x) = \frac{m^j}{(2\pi)^{3}}\int \frac{\text{d}^3\bm p}{(2E_p)^{1/2}} &\sum_\lambda \left[ \tensor{D^{(j)}[L(\bm p)]}{_{c}^\lambda}\, a(\bm p, j\lambda) e^{-{\rm i}px}
        % \right. \nonumber \\
        % & \left. \hspace{1cm}
        +\tensor{\left(D^{(j)}[L(\bm p)]C^{-1}\right)}{_{c}^\lambda} b^\dagger(\bm p, j\lambda) e^{{\rm i}px} \right] \,,\\
        %%%%%%%%%%%%%%%%%%%%%%%%%%%%%%%%%
        \chi^{*\dot{c}}(x) =  \frac{m^j}{(2\pi)^{3}} \int \frac{\text{d}^3\bm p}{(2E_p)^{1/2}} 
        &\sum_\lambda \left[ \tensor{\overbar D^{(j)}[L(\bm p)]}{^{\dot{c}}_{\dot{\lambda}}}\, a(\bm p, j\lambda) e^{-{\rm i}px}
        % \right. \nonumber \\
        % & \left. \hspace{1cm}
        +(-1)^{2j} \tensor{\left(\overbar D^{(j)}[L(\bm p)]C^{-1}\right)}{^{\dot{c}}_{\dot{\lambda}}} b^\dagger(\bm p, j\lambda) e^{{\rm i}px} \right] \,,
        %%%%%%%%%%%%%%
    \end{align}
\end{subequations}
\begin{subequations}
    \begin{align}
        \label{eq:lhfield_transf}
        &U(\Lambda) \varphi_{c}(x) U(\Lambda)^{-1} = \tensor{D^{(j)}[\Lambda^{-1}]}{_{c}^{d}}\varphi_{d}(\Lambda x)\,,\\  
        %%%%%%%%%%%%%%%%%%%%%%%%%%%%%%
        &U(\Lambda) \chi^{*\dot c}(x) U(\Lambda)^{-1} = \tensor{\overbar D^{(j)}[\Lambda^{-1}]}{^{\dot c}_{\dot d}}\chi^{*\dot d}(\Lambda x)
        \,.
    \end{align}
\end{subequations}
Here, we leave the spin-$j$ index on the fields implicit, $a(p,\lambda)$ is the particle annihilation operator, and $b^\dagger(p,\lambda)$ the antiparticle creation operator.  Contrary to the transformations of the creation/annihilation operators, where the Wigner rotations depend on the argument $\bm p$ of the operators, the transformations on the spinor indices of the fields carry no dependence on the field coordinate $x$ and are completely determined by the transformation $\Lambda$.  

We can now identify the coefficients of the creation/annihilation operators as the chiral spinors $u_{L}, u_{R}$:
\begin{subequations}
\label{eq:uspinors_def}
\begin{align}
    &u_{L, c}(p,\lambda)  \equiv m^j \tensor{D^{(j)}[L(\bm p)]}{_{c}^{d}}\delta_{d}^\lambda = \tensor{D^{(j)}[L(\bm p)]}{_{c}^{d}} \, \overset{\circ}{u}_{{L},d}(\lambda) \,, &[\text{left chiral}]\\
    %%%%%
    &u_{R}^{\dot c}(p,\lambda)  \equiv m^j \tensor{\overbar D^{(j)}[L(\bm p)]}{^{\dot{c}}_{\dot{d}}}\delta^{\dot{d} \lambda} = \tensor{\overbar D^{(j)}[L(\bm p)]}{^{\dot{c}}_{\dot{d}}}\, \overset{\circ}{u}_{R}^{\dot d}(\lambda)\,, & [\text{right chiral}]
\end{align}    
\end{subequations}
where we introduce the rest frame spinor
\begin{subequations}\label{eq:chiralspinor_rf}
\begin{align}
& \overset{\circ}{u}_{{L},c}(\lambda) = \overset{\circ}{u}_{R}{\vphantom{u}}^{\dot c}(\lambda) = m^j \delta_{c}^\lambda = m^j \delta^{\dot{c}\lambda} \,, 
   \\ 
& \overset{\circ}{\phi}(\lambda) \equiv \overset{\circ}{u}_{L}(\lambda) = \overset{\circ}{u}{\vphantom{u}}_{R}(\lambda)
\,,
\end{align}
\end{subequations}
with the notation of the last equation explicitly stating that in the rest frame there is no distinction between the two chiral spinors.  The choice to introduce the Kronecker delta function with these indices is guided by the connection with the massive spinor-helicity formalism, which is detailed in Sec.~\ref{sec:MSH}.  There, the position of the SU(2) little group polarization index $\lambda$ plays a role and matches with the choice of upper indices in these expressions.
Ultimately, the correspondence between the three labels in Eq.~(\ref{eq:chiralspinor_rf}), namely the two Lorentz spinor indices ($c$, $\dot c$) and the little group index ($\lambda$), is feasible because they all belong to the same SU(2) representation.

Up to the normalization factor $m^j$, these chiral spinors correspond to the columns of the matrices for the chiral standard boosts of Eq.~(\ref{eq:chiral_boost}).  The chiral spinors have a $(2j+1)$-valued spinor index $c$, and $\lambda$ (also ($2j+1$)-valued) is its little group spin projection quantum number.  Similar to any wave function appearing in a field\footnote{This includes the well known Dirac spinor and polarization four-vectors for massive spin 1 particles.}, due to carrying both Lorentz and little group indices, the chiral spinors have the following transformation property
\begin{subequations}
\label{eq:spinor_transf}
\begin{align}
    &\tensor{D^{(j)}[\Lambda]}{_{c}^{d}} u_{{L},d}(p,\lambda) = \sum_{\lambda'} u_{{L},c}(\Lambda p,\lambda') D^{(j)}_{\lambda'\lambda}[R_\text{w}(\bm p,\Lambda)] \,,\\
    &\tensor{\overbar D^{(j)}[\Lambda]}{^{\dot c}_{\dot d}} u_{R}^{\dot d}(p,\lambda) = \sum_{\lambda'}  u_{R}^{\dot c}(\Lambda p,\lambda') \overbar D^{(j)}_{\lambda'\lambda}[R_\text{w}(\bm p,\Lambda)]
    \,.
\end{align}    
\end{subequations}
This can be inferred from writing the spinor as the overlap
\begin{align} \label{eq:spinor_notinvariant}
    & \langle 0 | \varphi_{c}(0) |\bm p, \lambda \rangle = u_{{L},c}(p,\lambda)
    \,,
\end{align}
and apply a Lorentz boost to the field, see Eqs.~(\ref{eq:lhfield_transf}) and (\ref{eq:transform_annihilation}).

Other essential objects in Weinberg's construction are the matrices $\Pi^{(j)}(p)$ and $\overbar \Pi^{(j)}(p)$, which encode the outer product of the chiral spinors, and correspond to the numerator of propagators in Weinberg's formalism \cite{Weinberg:1964cn}. They appear in the commutators of the chiral fields as
\begin{subequations}
\label{eq:commutators}
\begin{align}
    &[\varphi_{c}(x),\varphi^*_{\dot d}(y)]_\pm = \frac{1}{(2\pi)^3}\int \frac{\text{d}^3\bm p}{2E_p} \,\Pi_{c \dot d}(p)\left[e^{-{\rm i}p(x-y)}\pm e^{{\rm i}p(x-y)} \right]\,,\\
    %%%%%%
    &[\chi^{*\dot c}(x),\chi^{d}(y)]_\pm = \frac{1}{(2\pi)^3}\int \frac{\text{d}^3\bm p}{2E_p} \,\overbar\Pi^{\dot{c} d}(p)\left[e^{-{\rm i}p(x-y)}\pm e^{{\rm i}p(x-y)} \right]
    \,,
\end{align}    
where the $\pm$ in the left-hand side denotes the choice of sign for fermions and bosons and we left the spin index $j$ on the $\Pi$ implicit.
\end{subequations}
From Eqs.~(\ref{eq:causal_fields}) and (\ref{eq:a_commutator}), it can be inferred that these matrices take the form
\begin{subequations}
\label{eq:prop_def}
\begin{align}
&  \Pi_{c 
\dot d}(p) = m^{2 j} 
% \sum_{e, \dot e}
\tensor{D^{(j)}[L(\bm{p})]}{_{c} ^{e}} \delta_{e \dot e} \tensor{D^{(j)*}[L(\bm{p})]}{_{\dot d} ^{\dot e}} = m^{2 j} \left(e^{-2 \rho \, \hat{\bm p} \cdot \bm J^{(j)}}\right)_{c d} \,, \\
%%%%%%%%%%%%%%%%%%%
&  \overbar\Pi^{\dot c d}(p) = m^{2 j} \tensor{\overbar{D}^{(j)}[L(\bm{p})]}{^{\dot c}_{ \dot e}} \delta^{\dot{e} e} \tensor{\overbar{D}^{(j)*}[L(\bm{p})]}{^{d}_{e}} = m^{2 j} \left(e^{2 \rho \, \hat{\bm p} \cdot \bm {J}^{(j)}}\right)_{c d}
\,.
\end{align}
\end{subequations}
The position of the $e,\dot{e}$ indices warrants a comment, as we are summing over two upper indices, which does not seem to be an invariant operation.  However, both those indices refer to rest frame spinor indices which can be identified to the little group polarization indices (see Eq.~(\ref{eq:uspinors_def})), where the difference between the representations disappears (see Eq.~(\ref{eq:chiralspinor_rf})).  On the right-hand side, the $\bm J^{(j)}$ do not carry information about any of the representations considered in Eq.~(\ref{eq:V_reps}), so we opted to not distinguish them between the two equations.
The crux of Weinberg's construction is that the $\Pi^{(j)}$ and  $\overbar \Pi^{(j)}$ can equivalently be written as a symmetric and traceless rank-$2j$ tensor in space-time indices (denoted as $t$-tensors from here on) contracted with $2j$ copies of the four-momentum $p^\mu$.  We will not repeat the proof here (see Ref.~\cite{Weinberg:1964cn}) and only state the final result:
\begin{subequations}\label{eq:prop_ttensor}
\begin{align}
&  m^{2 j} \left(e^{-2 \rho \, \hat{\bm p} \cdot \bm J^{(j)}}\right)_{c d}= \Pi_{c \dot d}(p)  = \tensor{\left(t^{\mu_{1} \mu_{2} \cdots \mu_{2j}}\right) }{_{c \dot d}} p_{\mu_{1}} p_{\mu_{2}} \cdots p_{\mu_{2 j}} \,,  \\
&  m^{2 j} \left(e^{2 \rho \, \hat{\bm p} \cdot \bm {J}^{(j)}}\right)_{c d}= \tensor{\overbar\Pi}{^{\dot c d}}(p) = \tensor{\left(\bar{t}^{\mu_{1} \mu_{2} \cdots \mu_{2j}}\right)}{^{\dot c d}} p_{\mu_{1}} p_{\mu_{2}} \cdots p_{\mu_{2 j}}
\,.
\end{align}    
\end{subequations}
We want to stress that these $t$-tensors have both Lorentz and spinor indices, meaning each element of the tensor with specific Lorentz indices is a square matrix of order $2j+1$. 
The $t^{\mu_{1} \mu_{2} \cdots \mu_{2j} }$ and $\bar{t}^{\mu_{1} \mu_{2} \cdots \mu_{2j} }$ are 
the generalization to arbitrary spin  of the so-called Pauli four-vector or Infeld-Van der Waerden symbols\footnote{We thank C\'{e}dric Lorc\'{e} for making us aware of the latter terminology.} $\sigma^\mu \equiv t^\mu = (\mathbb 1^{(2)}, \bm \sigma)$ and $\bar\sigma^\mu \equiv \bar t^\mu =(\mathbb 1^{(2)},-\bm \sigma)$ formed with the Pauli matrices $\bm \sigma$~\cite{Infeld:1933zz}. 

The  $t$-tensors have the following properties:
\begin{enumerate}
\item The matrices in the $t$-tensors are Hermitian. This can be seen by taking the complex conjugated transpose of Eq.~(\ref{eq:prop_ttensor}).
Since the generators of rotations $\bm {J}^{(j)}$ are Hermitian, and the parameters in the exponential [Eq.~(\ref{eq:rho_p})] are real, then the left-hand side of the equation remains unchanged. On the right-hand side the momenta are real, thus we have
\begin{subequations}
 \label{eq:prop_ttensor_Hermitian}
\begin{align}
\left(t^{\mu_{1} \mu_{2} \cdots \mu_{2j}} \right)^\dagger p_{\mu_{1}} p_{\mu_{2}} \cdots p_{\mu_{2 j}} = \; & m^{2 j} \left(e^{-2 \rho \, \hat{\bm p} \cdot \bm J^{(j)}}\right)^\dagger = m^{2 j} e^{-2 \rho \, \hat{\bm p} \cdot \bm J^{(j)}} 
\nonumber \\
= \; & t^{\mu_{1} \mu_{2} \cdots \mu_{2j}} p_{\mu_{1}} p_{\mu_{2}} \cdots p_{\mu_{2 j}} \,,  \\
\left(\bar{t}^{\mu_{1} \mu_{2} \cdots \mu_{2j}}\right)^\dagger p_{\mu_{1}} p_{\mu_{2}} \cdots p_{\mu_{2 j}} = \; & m^{2 j} \left(e^{2 \rho \, \hat{\bm p} \cdot \bm {J}^{(j)}}\right)^\dagger = m^{2 j} e^{2 \rho \, \hat{\bm p} \cdot \bm {J}^{(j)}} 
\nonumber \\
= \; & \bar{t}^{\mu_{1} \mu_{2} \cdots \mu_{2j}} p_{\mu_{1}} p_{\mu_{2}} \cdots p_{\mu_{2 j}}
\,.
\end{align}    
\end{subequations}

\item  They are symmetric under any permutation $\pi$ of the space-time indices, and are Lorentz traceless in any pair of indices:  
\begin{align}\label{eq:t_symmetric}
t_{c \dot d}^{\mu_{1} \mu_{2} \cdots \mu_{2 j}} = t_{c \dot d}^{\mu_{\pi(1)} \mu_{\pi(2)}\cdots \mu_{\pi(2 j)}}
\,,
\end{align}
\begin{align}\label{eq:t_traceless}
g_{\mu_k\mu_l}t_{c \dot d}^{\mu_{1} \cdots \mu_{k} \cdots \mu_{l} \cdots \mu_{2 j}} = 0
\,.
\end{align}

\item Evaluating Eqs.~(\ref{eq:prop_ttensor}) with $\rho=0$ [see Eq.~(\ref{eq:rho_p})], one obtains that the $t$-tensors always contain the identity as the element with all zero indices:
\begin{equation} \label{eq:tzero_identity}
    t^{0\cdots 0} = \bar t^{0 \cdots 0} = \mathbb{1}^{(j)}
    \,.
\end{equation}

\item  They act as intertwiners~\cite{Jeevanjee:1494077} between the SL(2,$\mathbb{C}$) spinor outer product representation 
\begin{equation}
    (0,j)^*\otimes (0,j) \sim (j,0)\otimes (0,j) \sim (j,j)
    \,,
\end{equation}
and symmetric traceless Lorentz tensors of rank $2j$.  Consequently, they exhibit the following transformation properties under Lorentz transformations:  
\begin{subequations}
\label{eq:t_cov}
\begin{align}
& \tensor{\left( D^{(j)}[\Lambda]\right)}{_c ^e} \tensor{\left(t^{\mu_{1} \cdots \mu_{2 j}}\right)}{_{e \dot f}} \tensor{\left( {D^{(j)\dagger}}[\Lambda]\right)}{^{\dot f}_{\dot d}} =  \tensor{\left(t^{\nu_1 \cdots \nu_{2j}}\right)}{_{c \dot d}} \, \Lambda_{\nu_1}{}^{\mu_1} \cdots \Lambda_{\nu_{2j}}{}^{\mu_{2j}} \,,\\
& \tensor{\left( \overbar D^{(j)}[\Lambda]\right)}{^{\dot c}_{\dot e}} \tensor{\left( \bar{t}^{\mu_{1} \cdots \mu_{2 j}}\right)}{^{\dot e f}} \tensor{\left( \overbar D^{(j)\dagger}[\Lambda]\right)}{_{f}^{d}} =  \tensor{\left(\bar{t}^{\nu_1 \cdots \nu_{2j}}\right)}{^{\dot{c} d}} \, \Lambda_{\nu_1}{}^{\mu_1} \cdots \Lambda_{\nu_{2j}}{}^{\mu_{2j}}
\,.
\end{align}    
\end{subequations}
The covariant transformation property of the $t$-tensors can be contrasted with that of the individual spinor of Eq.~(\ref{eq:spinor_transf}), where SU(2) little group transformations appear through the momentum-dependent Wigner rotation.
In the outer product of Eq.~(\ref{eq:commutators}), little group indices are contracted and these Wigner rotations cancel out, which leads to the invariance of the $\Pi(p)$ and $\overbar\Pi(p)$, leading in turn to Eq.~(\ref{eq:t_cov}).

\item  If we introduce the parity conjugated four momentum $\bar p^\mu \equiv (E_p, -\bm p)$, from Eq.~(\ref{eq:prop_def}) $\overbar \Pi^{(j)}(p) = \Pi^{(j)}(\bar p) $ follows. Using Eq. (\ref{eq:prop_ttensor}), one finds by identifying equal powers of $\bm J^{(j)}$ that elements of $t$ and $\bar{t}$ are related by  
\begin{align}\label{eq:t_tbar_sign_relations}
\bar{t}^{\mu_{1} \mu_{2} \ldots \mu_{2 j}} = t_{\mu_{1} \mu_{2} \ldots \mu_{2 j}} 
\,.
\end{align}
Using the matrix $C$ to relate left and right chiral boosts, this can be formulated covariantly as
\begin{align}\label{eq:t_chargeconj}
&\bar{t}^{\mu_1 \cdots \mu_{2j} } = C \left(t^{\mu_1 \cdots \mu_{2j} }\right)^* C^{-1} 
\,.
\end{align}    
We will refer to Eq.~(\ref{eq:t_chargeconj}) as the barring operation. Extending the barring operation to include general expressions is straightforward. In general, we have 
\begin{align}\label{eq:barring_op}
\overbar{\left(\cdots\right)_{1} \, {t}^{\mu_1 \cdots \mu_{2j} } \, \left(\cdots\right)_{2}} & = C \left[\left(\cdots\right)_{1} \, t^{\mu_1 \cdots \mu_{2j} } \, \left(\cdots\right)_{2}\right]^* C^{-1} 
\nonumber \\
& = C(\cdots)_{1}^*C^{-1} \, C \left(t^{\mu_1 \cdots \mu_{2j} }\right)^* C^{-1} \, C(\cdots)_{2}^*C^{-1}
\nonumber \\
& = \overbar{(\cdots)}_{1} \, \bar{t}^{\mu_1 \cdots \mu_{2j} } \, \overbar{(\cdots)}_{2} 
\,.
\end{align}    
If the expression inside parenthesis does not contains $t$-tensors, the barring operation reduces to complex conjugation. Explicitly
\begin{equation}
\overbar{(\cdots)}_{\text{no $t$-tensor}}=C(\cdots)_{\text{no matrices}}^* C^{-1}=(\cdots)^* 
\,.
\end{equation}

\item Chiral fields for larger spin values can be built out of tensor products of lower spin chiral fields applying SU(2) tensor product reductions using  Clebsch-Gordan coefficients.  This leads to the following recursion relations that the $t$-tensors  obey~\cite{Barut:1963zzb,Williams:1965rga}
\begin{subequations}
\label{eq:t_recursion}
\begin{align}
    &\tensor{\left(t^{\mu_1 \cdots \mu_{2j}}\right)}{_{c \dot d}} = 
    \tensor{[j,j-\tfrac{1}{2},\tfrac{1}{2}]}{_c ^{e a}}\,
    \tensor{[j,j-\tfrac{1}{2},\tfrac{1}{2}]}{_{\dot d} ^{\dot f \dot b}}\;
    \tensor{\left(t^{\mu_1 \cdots \mu_{2j-1}}\right)}{_{e \dot f}}\,
    \tensor{\left(t^{\mu_{2j}}\right)}{_{a \dot b}} \,,\\
    %%%%%%%%%%%%%%%%%%%%%%%%
    &\tensor{\left(\bar{t}^{\mu_1 \cdots \mu_{2j}}\right)}{^{\dot c d}} = 
    \tensor{[j,j-\tfrac{1}{2},\tfrac{1}{2}]}{^{\dot c} _{\dot e \dot a}}\,
    \tensor{[j,j-\tfrac{1}{2},\tfrac{1}{2}]}{^{d} _{f b}}\;
    \tensor{\left(\bar{t}^{\mu_1 \cdots \mu_{2j-1}}\right)}{^{\dot{e} f}}\,
    \tensor{\left(\bar{t}^{\mu_{2j}}\right)}{^{\dot{a} b}}
    \,.
\end{align}
\end{subequations}
Here the Clebsch-Gordan coefficients were introduced with explicit spinor indices
\begin{equation}\label{eq:CG_def}
    \langle j_1 m_1, j_2 m_2 | j m \rangle \equiv \tensor{[j,j_1,j_2]}{_m ^{m_1 m_2}}
    =(-1)^{2j}\tensor{[j,j_1,j_2]}{^{\dot m} _{\dot m_1  \dot m_2}}
    \,.
\end{equation}
This reflects their transformation properties
\begin{subequations}\label{eq:CG_transf}
\begin{align}
    \tensor{\left(D^{(j)}[\Lambda]\right)}{_c^{d}}\tensor{[j,j_1,j_2]}{_{d}^{ef}} = \tensor{[j,j_1,j_2]}{_c^{e'f'}}\tensor{\left(D^{(j_1)}[\Lambda]\right)}{_{e'}^{e}}\tensor{\left(D^{(j_2)}[\Lambda]\right)}{_{f'}^{f}} \,,\\
    %%%%%%%%%%%%%%%%%%%%
    \tensor{\left(\overbar{D}^{(j)}[\Lambda]\right)}{^{\dot c}_{\dot d}}\tensor{[j,j_1,j_2]}{^{\dot c}_{\dot e \dot f}} = \tensor{[j,j_1,j_2]}{^{\dot c}_{\dot e'\dot f'}}\tensor{\left(\overbar D^{(j_1)}[\Lambda]\right)}{^{\dot e'}_{\dot e}}\tensor{\left(\overbar D^{(j_2)}[\Lambda]\right)}{^{\dot f'}_{\dot f}}
    \,.
\end{align}    
\end{subequations}
A rank-$2j$ symmetric and traceless Lorentz tensor is the tensor with the minimal number of Lorentz indices that contains the same $(j,0)\otimes (0,j)$ representation as the outer product of spin-$j$ chiral spinors.  This means the recursion of Eq.~(\ref{eq:t_recursion}) is unique if carried out all the way down to all spin-1/2 building blocks.  At each step in the reduction the intermediate spins are coupled to their maximal $j=j_1+j_2$ value.  It is possible to write down other towers of reduction, splitting off spins $j_2>1/2$ in intermediate steps that are then reduced themselves in later steps. But once reaching the final step when everything is reduced to spin-1/2 factors, the coefficients in the reduction will be the same. 

Eq.~(\ref{eq:t_recursion}) shows another perspective to think about the $t$-tensors.  They collect the Clebsch-Gordan coefficients used to construct the higher-spin representations in such a way that the overall object transforms covariantly in its indices.  
%We will use this recursion in Sec.~\ref{sec:t_tensors_gen} to describe an algorithm to efficiently construct the $t$-tensors.

\item The recursion formula Eq.~(\ref{eq:t_recursion}) provides us with an independent definition of the $t$-tensors, which we will use in Sec.~\ref{sec:t_tensors_gen} to obtain a simple formula to explicitly calculate the $t$-tensors.
Using this definition, it is straightforward to prove by induction that all Lorentz tensor components of the $t$-tensors are Hermitian matrices. Contrary to Eq.~(\ref{eq:prop_ttensor_Hermitian}), this proof only concerns the $t$-tensors themselves, in particular it is independent of any parametrization, as well as independent of $\Pi^{(j)}$ and $\overbar \Pi^{(j)}$. 
% From the recursion formula Eq.~(\ref{eq:t_recursion}), it is straightforward to prove by induction that all Lorentz tensor components of the $t$-tensors are Hermitian matrices.  
\end{enumerate}

\subsection{Representation of boosts and spinors}
\label{sec:boosts_spinors}

Note that Eqs.~(\ref{eq:prop_ttensor}) are convenient expressions to calculate the matrices $\Pi^{(j)}(p)$ using the $t^{\mu_1 \cdots \mu_{2j}}$ tensors. Because the Eqs.~(\ref{eq:prop_def}) define a relation between the $\Pi^{(j)}(p)$ (evaluated with an on-shell momentum) and the boosts of Eqs.~(\ref{eq:chiral_boost}), it is possible to find four parameters  that contracted with each of the $2j$ space-time indices in $t^{\mu_1 \cdots \mu_{2j} }$ and $\bar{t}^{\mu_1 \cdots \mu_{2j} }$ provide us with the representation of the boosts. 

Explicitly, from Eqs.~(\ref{eq:D_Dbar}), (\ref{eq:charge_conj_J}) and (\ref{eq:t_chargeconj}), we have 
\begin{subequations}\label{eq:canonical_boosts}
\begin{align}
\label{eq:canonical_boosts_t}
& D^{(j)}{[L(\bm{p})]} = \exp\left[ - \rho \, \hat{{\bm p}}\cdot \bm{J}^{(j)} \right] \equiv {t}^{\mu_1 \cdots \mu_{2j} }  \tilde{p}_{\mu_1} \cdots \tilde{p}_{\mu_{2j}} 
\,,
\\
%%%%
& \overbar{D}^{(j)}{[L(\bm{p})]} = \exp\left[ + \rho \, \hat{{\bm p}}\cdot \bm{J}^{(j)} \right] \equiv \bar{t}^{\mu_1 \cdots \mu_{2j}} \tilde{p}_{\mu_1} \cdots \tilde{p}_{\mu_{2j}} 
\,.
\label{eq:canonical_boosts_onlytbar}
\end{align}    
\end{subequations}

The values for $\tilde p$ follow from comparing 
\begin{align}
\Pi(p) = \; & m^{2j} \left(D^{(j)}[L(\bm{p})]\right)^2 =  m^{2j} \exp\left[ - 2\rho \, \hat{\bm p}\cdot \bm{J}^{(j)} \right] 
% \nonumber \\
% = \; & 
=t^{\mu_{1} \mu_{2} \cdots \mu_{2j} } p_{\mu_{1}} p_{\mu_{2}} \cdots p_{\mu_{2 j}}\,, 
&\{\rho,\hat{\bm p}\} \leftrightarrow \frac{p^\mu}{m} \,, \\
%%%%%%%%
D^{(j)}{[L(\bm{p})]} = \; & \exp\left[ - \rho \, \hat{\bm p}\cdot \bm{J}^{(j)} \right] = {t}^{\mu_1 \cdots \mu_{2j} }  \tilde{p}_{\mu_1} \cdots \tilde{p}_{\mu_{2j}}, 
&\{\frac{\rho}{2},\hat{\bm p}\} \leftrightarrow \tilde p^\mu 
\,,
\end{align}
and the use of the half angle formulas for hyperbolic functions; see Eq.~(\ref{eq:rho_p}).  On the right of these equations, we highlight the map between the parameters in the exponential map and the parameters multiplying the $t$-tensors.  Alternatively, one can do the calculation in the $j=1/2$ Weyl spinor representation using Eq.~(\ref{eq:canonical_boosts}) and compare to the textbook expressions for Weyl spinors.  

Although the $t$-tensors appear in the same form in both the expressions of the outer product $(\Pi(p))$ and the boosts, we want to stress that mathematically they play slightly different roles.  The boosts are automorphisms on $V$ or $\widetilde{V}^*$ while the $\Pi \in V\otimes V^*$.  This dual role is familiar from the spin-1/2 case where $\sigma^\mu$ and $\bar{\sigma}^\mu$ appear both in the generators of Lorentz boosts and the bilinear calculus.  For this reason we omit explicit spinor indices on the $t$-tensors in most equations from here on.  The context should always make clear what role they fulfill and indices can be reinstated if desired.

We want to stress that, although Eqs.~(\ref{eq:canonical_boosts}) are written such that we use the Minkowski metric of Eq.~(\ref{eq:metric}) in the 4-index contractions, the $\tilde{p}^\mu$ is \emph{not} a four-vector but rather a four-parametric set used to parameterize the boosts. 
In Eq.~(\ref{eq:ptilde canonical}), this is clear from the first component and the overall factor multiplying the expression which is not a Lorentz-scalar.  
As highlighted earlier, this is a general consequence of the fact that, contrary to the $\Pi(p)$, the boosts/spinors carry both Lorentz and little group indices, see Eqs.~(\ref{eq:spinor_transf}). 

As such, these parameters $\tilde p^\mu$ depend on the convention used to define the standard boosts. 
For the canonical boosts (instant form), the $\tilde p_c$ are real and take the form\footnote{Note that the label $c$ makes reference to the canonical parameterization of the boosts and should not be read as an index.} 
\begin{equation}\label{eq:ptilde canonical}
\tilde{p}_c^\mu = \frac{1}{[2m(m+E_p)]^{1/2}} (E_p+m,\bm p) = (\cosh \frac{\rho}{2},\sinh \frac{\rho}{2} \hat{\bm p})
\,.
\end{equation} 

So far, the above formulas for $\tilde p$ are valid for canonical boosts. Cases which are not discussed by Weinberg in Ref.~\cite{Weinberg:1964cn}, are standard boosts such as helicity and light-front helicity which also include rotations in their definitions.  In Appendix~\ref{sec:hel_lf_spinors} we give an overview of the definitions and properties of these spinors. 
In summary, we have to extend the boosts of Eq.~(\ref{eq:canonical_boosts}) to the more general
\begin{align}\label{eq:general_boosts_t}
{D}^{(j)}[\mathcal{B}(\bm{\xi},\bm{\theta})] = \exp\left[ - \left({\bm{\xi} + {\rm i}\bm{\theta}}\right) \cdot \bm{J}^{(j)} \right] = {t}^{\mu_1 \cdots \mu_{2j} }  \tilde{p}_{\mu_1} \cdots \tilde{p}_{\mu_{2j}}
\,,
\end{align}
where $\mathcal{B}(\bm{\xi},\bm{\theta})$ defines the standard boost under consideration, parametrized by $\bm \xi$ (pure boost), $\bm \theta$ (rotation).
Since now we deal with a general Lorentz transformation, the expression for the right-chiral boost follows from Eq.~(\ref{eq:chiral_relation}),
\begin{align}\label{eq:general_boosts_tbar}
\overbar{D}^{(j)}[\mathcal{B}(\bm{\xi},\bm{\theta})] = \exp\left[ + \left({\bm{\xi} - {\rm i}\bm{\theta}}\right) \cdot \bm{J}^{(j)} \right] = \bar{t}^{\mu_1 \cdots \mu_{2j} }  \left(\tilde{p}_{\mu_1} \cdots \tilde{p}_{\mu_{2j}}\right)^* 
\,,
\end{align}
which shows that the generalized boost parameters of the left- and right-chiral representations are related by complex conjugation.

  Similar to the canonical case, the determination of these parameters is easiest to carry out for the chiral spin-1/2 case.  We obtain the following results:
\begin{itemize}
    \item For the original Jacob \& Wick helicity spinors~\cite{Jacob:1959at}
    \begin{align}\label{eq:ptilde_h}
        &\tilde{p}_h^{\,0} = \frac{\cos\frac{\theta}{2}}{\left[2m(m+E_p)\right]^{1/2}}\, (E_p+m) \,,\nonumber\\
        &\tilde{p}_h^{\,x} = \frac{\sin\frac{\theta}{2}}{\left[2m(m+E_p)\right]^{1/2}}\left[-{\rm i}(E_p+m)\sin\phi+|\bm p|\cos\phi\right] \,,\nonumber\\
        &\tilde{p}_h^{\,y} = \frac{\sin\frac{\theta}{2}}{\left[2m(m+E_p)\right]^{1/2}}\left[{\rm i}(E_p+m)\cos\phi+|\bm p|\sin\phi\right] \,,\nonumber\\
        &\tilde{p}_h^{\,z} = \frac{\cos\frac{\theta}{2}}{\left[2m(m+E_p)\right]^{1/2}}|\, \bm p| 
        \,,
    \end{align}
    or
    \begin{equation}
        \tilde p_h^\mu = \left(\cosh\frac{\rho}{2}\cos\frac{\theta}{2},\sinh\left(\frac{\rho}{2}-i\phi\right) \sin\frac{\theta}{2},\sinh\left(\frac{\rho}{2}+i\phi\right) \sin\frac{\theta}{2},\sinh\frac{\rho}{2}\cos\frac{\theta}{2}\right).
    \end{equation}
    \item For the revised helicity ($h'$) spinors
\begin{align}\label{eq:ptilde_h'}
    &\tilde p^{\,0}_{h'} = \frac{\cos\frac{\theta}{2}}{\left[2m(E_p+m)\right]^{1/2}}\left[(E_p+m)\cos\frac{\phi}{2}+{\rm i}|\bm p|\sin\frac{\phi}{2} \right]\nonumber \,,\\
    &\tilde p^{\,x}_{h'} = \frac{\sin\frac{\theta}{2}}{\left[2m(E_p+m)\right]^{1/2}}\left[-{\rm i}(E_p+m)\sin\frac{\phi}{2}+|\bm p|\cos\frac{\phi}{2} \right]\nonumber \,,\\
    &\tilde p^{\,y}_{h'} = \frac{\sin\frac{\theta}{2}}{\left[2m(E_p+m)\right]^{1/2}}\left[{\rm i}(E_p+m)\cos\frac{\phi}{2}+|\bm p|\sin\frac{\phi}{2} \right]\nonumber \,,\\
    &\tilde p^{\,z}_{h'} = \frac{\cos\frac{\theta}{2}}{\left[2m(E_p+m)\right]^{1/2}}\left[{\rm i}(E_p+m)\sin\frac{\phi}{2}+|\bm p|\cos\frac{\phi}{2} \right]
    \,,
\end{align}
    or
    \begin{align}
        \tilde p_{h'}^\mu = \left(\cosh\left(\frac{\rho+i\phi}{2}\right)\cos\frac{\theta}{2},
        \sinh\left(\frac{\rho-i\phi}{2}\right) \sin\frac{\theta}{2},\right.
        \left.i\cosh\left(\frac{\rho-i\phi}{2}\right) \sin\frac{\theta}{2},\sinh\left(\frac{\rho+i\phi}{2}\right)\cos\frac{\theta}{2}\right).
    \end{align}
For helicity spinors the left and right chiral spinors are related through the substitution $|\bm p| \leftrightarrow -|\bm p|$ in the spinor expressions.  We can see that $\bar t^{\mu_1\cdots \mu_{2j}} \tilde p^*_{\mu_1} \cdots \tilde p^*_{\mu_{2j}}$ indeed achieves this substitution for both $\tilde p_h$ and $\tilde p_{h'}$.  
    \item For the light-front helicity spinors
    \begin{equation}\label{eq:ptilde_LF}
        \tilde p_{\text{LF}}^\mu = \frac{1}{2(mp^+)^{1/2}}\left((p^++m),p^L,{\rm i}p^L,(p^+-m) \right)
        \,.
    \end{equation}
\end{itemize}
It is worth noting that the boost parameters for the original helicity spinors and the light-front spinors have a similar structure: the time and longitudinal component are real whereas only the transverse components are complex.  

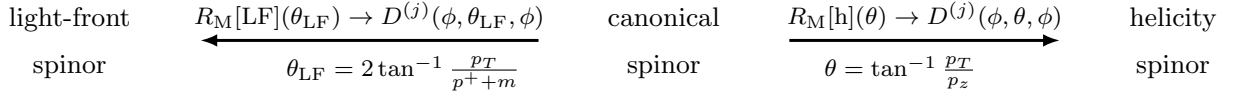
\begin{figure}[ht]
    \centering
    \resizebox{\textwidth}{!}{%
\begin{tikzpicture}
  % Define nodes for the spinors
  \node (lf) at (0,0) {\begin{tabular}{c}
    light-front\\
    spinor
  \end{tabular}};
  
  \node (cs) at (7,0) {\begin{tabular}{c}
    canonical\\
    spinor
  \end{tabular}};
  
  \node (hs) at (13,0) {\begin{tabular}{c}
    helicity\\
    spinor
  \end{tabular}};
  
  % Draw arrows between spinors
  \draw[<->, line width=1pt, {Latex[length=2mm]}-] 
    (lf.east) -- node[above] {\footnotesize $R_\text{M}[\text{LF}](\theta_{\rm LF}) \rightarrow D^{(j)}(\phi,\theta_{\rm LF},\phi)$} (cs.west);
    
  % Arrow from Canonical to Helicity Spinor
  \draw[<->, line width=1pt, -{Latex[length=2mm]}] 
    (cs.east) 
    -- node[above] {\footnotesize $R_\text{M}[\text{h}](\theta)\rightarrow D^{(j)}(\phi,\theta,\phi)$}  
    (hs.west) ;
      
  % Add equations below

 \node[below left=-0.6cm and .1cm of cs] {\footnotesize $\theta_{\text{LF}} = 2\tan^{-1}{\frac{p_T}{p^+ + m}}$};
 
  \node[below right=-0.6cm and .3cm of cs]  {\footnotesize $\theta = \tan^{-1}{\frac{p_T}{p_z}}$};

\end{tikzpicture}
        }
    \caption{The figure shows schematically how to obtain the light-front or helicity spinors from canonical ones using rest frame rotations before applying the canonical boost. Here, the Melosh rotations are represented through their Euler angles, where $\phi$ is the azimuthal angle of the momentum, see Eq.~(\ref{eq:spinor_RF_meloshrotated}).}
    \label{fig:Melosh_angles}
\end{figure}

A second way of dealing with the other choices of standard boosts is provided by Melosh rotations, see Eq.~(\ref{eq:melosh}).  The helicity and light-front spinors can be obtained by a canonical boost applied to rotated rest-frame spinors, where the rotation is called the Melosh one, see Fig.~\ref{fig:Melosh_angles}.  This allows to use the real-valued canonical $\tilde p$ for all choices of spinors, but one has to use the appropriate Melosh-rotated rest-frame spinors in the calculations, see Eq.~(\ref{eq:spinor_RF_meloshrotated}).

To allow for formulas that work for any choice of standard boost, we keep the complex conjugation explicit and write\footnote{We should technically replace the $L(\bm{p})$ on the left side of these equations by the more general $\mathcal{B}(\bm{\xi},\bm{\theta})$, but we choose to keep the notation intuitive.}
\begin{subequations}\label{eq:boosts_t}
\begin{align}
& D^{(j)}{[L(\bm{p})]} \equiv {t}^{\mu_1 \cdots \mu_{2j} }  \tilde{p}_{\mu_1} \cdots \tilde{p}_{\mu_{2j}} = \Pi(\tilde p)
\,,
\\
%%%%
& \overbar{D}^{(j)}{[L(\bm{p})]} \equiv \bar{t}^{\mu_1 \cdots \mu_{2j}} \left(\tilde{p}_{\mu_1} \cdots \tilde{p}_{\mu_{2j}}\right)^* = \overbar \Pi(\tilde p^*)
\,.
\label{eq:boosts_onlytbar}
\end{align}    
\end{subequations}

It is worth emphasizing that the four parameters that appear in $\tilde{p}^\mu$ are \emph{identical} for any spin representation. The difference is in the number of copies of $\tilde{p}^\mu$ that needs to be contracted with the $t$-tensors, which increases for higher spin values (as $2j$).  
As a result, being able to construct all objects of interest (spinors, propagators) from these $t$-tensors allows to construct a general formalism that applies to any spin representation, and will only differ in the Lorentz-rank of the tensors involved.  
Eventually, expressions will be contracted with copies of $\tilde p$, but the exact form of $\tilde p$ only depends on the choice of spinors, and thus the standard boost that is being considered.  
This implies that the constructions that are presented in this paper can be applied to any spinor form, and as illustrated here in particular the most common ones used in calculations (canonical, helicity, light-front helicity).  

\subsection{Bispinors and generalized Dirac gamma matrices}
\label{sec:bispinors}

Parity $\mathsf{P}$ transforms $\mathbb J \overset{\mathsf{P}}{\to} \mathbb J, \mathbb K \overset{\mathsf{P}}{\to} -\mathbb K$ and consequently transforms a left chiral representation into a right one (and vice versa).   Direct sum bispinor representations $(j,0)\oplus(0,j)$ are more practical if one wants to  consider applications to theories where parity is an explicit symmetry.  In order to simplify the presentation we choose to work in the Weyl representation for the bispinors, where we follow the convention where the left chiral representation is on top and the right chiral at the bottom. The direct sum representation expressions for the generators of rotations $\mathcal{J}^{(j)}$ and canonical pure boosts $\mathcal{B}^{(j)}$ have Weyl representation
\begin{equation}
\bm{\mathcal{J}}^{(j)} = \left(\begin{array}{cc}
{\bm J}^{(j)} &  0 \\
0 & {\bm J}^{(j)}
\end{array} \right) 
\,,
\end{equation}
\begin{equation}
\bm {\mathcal{B}}^{(j)} = {\rm i}\bm{\mathcal{J}}^{(j)} \gamma_5 = \left(\begin{array}{cc}
-{\rm i}{\bm J}^{(j)} &  0 \\
0 & +{\rm i}{\bm J}^{(j)}
\end{array} \right)
\,,
\end{equation}
where 
\begin{equation}\label{eq:gamma5}
 \gamma_5 \equiv \left(\begin{array}{cc}
-\mathbb{1}^{(j)} & 0 \\
0  & \mathbb{1}^{(j)} 
\end{array} \right)
\,.
\end{equation}

By considering the action of the canonical boosts on the direct sum representation, we can write 
\begin{align}\label{eq:bispinor_def}
u(p,\lambda)  = \; & 
\begin{pmatrix}
u_{L,c}(p,\lambda)\\
u_{R}^{\dot c}(p,\lambda)
\end{pmatrix}
=
\mathcal{D}^{(j)}[L(\bm p)] \overset{\circ}{u}(\lambda) 
% \nonumber \\[0.5em]
% = \; &
= \left(\begin{array}{cc}
{D}^{(j)}[L(\bm p)] & 0 \\
0 & \overbar{D}^{(j)}[L(\bm p)]
\end{array}\right) \overset{\circ}{u}(\lambda)
= \left(\begin{array}{cc}
 {\Pi}(\tilde{p}) & 0 \\
0 & \overbar{\Pi}(\tilde{p}^*)
\end{array} \right) \overset{\circ}{u}(\lambda)
\,,
\end{align}
where we introduced the rest-frame bispinor\footnote{We want to remind that in the rest frame there is no distinction between the two chiral representations [Eqs.~(\ref{eq:chiralspinor_rf})], though we prefer to still make the distinction in the indices.}
\begin{equation}\label{eq:spinors_RF}
\overset{\circ}{u}(\lambda)=\left(\begin{array}{c}
\overset{\circ}{u}_{{L},c}(\lambda) \\
\overset{\circ}{u}\vphantom{u}_{R}{}^{\dot c}(\lambda)
\end{array}\right)
\,,
\end{equation}
and Eqs.~(\ref{eq:boosts_t}) were used.  Note that other conventions exist in the literature that put the right chiral on top, and/or introduce an additional phase between the two chiral components.

In addition to the pure boosts in Eq.~(\ref{eq:bispinor_def}), we can consider any Lorentz transformation in the bispinor representation
\begin{equation}
    \mathcal{D}^{(j)}[\Lambda]  = \left(\begin{array}{cc}
{D}^{(j)}[\Lambda] & 0 \\
0 & \overbar{D}^{(j)}[\Lambda]
\end{array}\right)
\,,
\end{equation}
% For this representation, 
it follows from Eq.~(\ref{eq:chiral_relation}) that $\left( \mathcal{D}^{(j)}[ \Lambda ]\right)^\dagger$ is here equivalent to $\left(\mathcal{D}^{(j)}[ \Lambda]\right)^{-1}$:  
\begin{align}\label{eq:inv_dagger_equiv}
\left(\mathcal{D}^{(j)}[ \Lambda ]\right)^\dagger = \beta \left(\mathcal{D}^{(j)}[ \Lambda ]\right)^{-1} \beta^{-1}
\,,
\end{align}
where\footnote{As matrices, we have from Eq.~(\ref{eq:tzero_identity}) that $\beta = \gamma^{0\cdots0}$, where the $\gamma^{(\mu)}$ are introduced below in Eq.~(\ref{eq:gamma_mu}). Their spinor indices, however, are different (see Eqs.~(\ref{eq:beta_def}) and (\ref{eq:gamma_mu})). Following Ref.~\cite{Srednicki:2004hg} for the spin-1/2 case, we prefer to distinguish the two roles here for clarity.}
\begin{equation}\label{eq:beta_def}
 \beta \equiv \begin{pmatrix}
0 & \tensor{\mathbb{1}}{^{(j)\dot c}_{\dot d}} \\
\tensor{\mathbb{1}}{^{(j)}_c^d} & 0
\end{pmatrix}; \qquad  
\beta^{-1} = \beta^T 
\,.
\end{equation}
The equivalence of Eq.~(\ref{eq:inv_dagger_equiv}) does not hold for each individual chiral representation, see Eq.~(\ref{eq:chiral_relation}).

As an application of Eq.~(\ref{eq:inv_dagger_equiv}), we can construct the conjugated/adjoint bispinor 
\begin{equation}\label{eq:adjoint_bispinor_def}
\bar{u}(p,\lambda) \equiv u^{\dagger}(p,\lambda) \beta = \overset{\circ}{u}^{\dagger}(\lambda) \left(\mathcal{D}[L(\bm p)]^{(j)}\right)^{\dagger} \beta
= 
\overset{\circ}{u}^{\dagger}(\lambda) \left(\begin{array}{cc}
0 &  {\Pi}\left(\tilde{p}\right) \\
\overbar{\Pi}\left(\tilde{p}^*\right) & 0
\end{array} \right) 
\,.
\end{equation}
As is well known in the spin-1/2 case, bilinears of bispinors can be constructed that transform as Lorentz tensors (up to little group transformations).  The generalization of this construction to arbitrary spin using the objects introduced here will be discussed in Sec.~\ref{sec:gen_bilinears}.

As the bispinors are objects that contain twice the $2j+1$ number of physical degrees of freedom, they obey wave function constraints, which take the form of equations of motion, {\emph{i.e.}} generalized Dirac equations
\begin{subequations}
\label{eq:gen_dirac_eq}
\begin{align}
\left( \gamma^{\mu_{1} \cdots \mu_{2j}} p_{\mu_{1}} \cdots p_{\mu_{2j}} - m^{2j} \right) u(p,\lambda) & = 0 \,,
\\
\bar{u}(p,\lambda) \left( \gamma^{\mu_{1} \cdots \mu_{2j}} p_{\mu_{1}} \cdots p_{\mu_{2j}} - m^{2j} \right)  & = 0 
\,.
\end{align}    
\end{subequations}
The associated equations of motion for the bispinor fields are commonly referred to as the Joos-Weinberg equations~\cite{Joos:1962qq,Weinberg:1964cn}.
In Eq.~(\ref{eq:gen_dirac_eq}), the generalized Dirac matrices were introduced. In the Weyl representation, they are
\begin{equation}\label{eq:gamma_mu}
 \gamma^{\mu_1\ldots \mu_{2j}} \equiv \left(\begin{array}{cc}
0 & \tensor{(t^{\mu_1\ldots \mu_{2j}})}{_{c \dot d}} \\
\tensor{(\bar{t}^{\mu_1\ldots \mu_{2j}})}{^{\dot c d}}  &  0 
\end{array} \right)
\,.
\end{equation}
The $\gamma^{(\mu)}$ obey
\begin{equation}\label{eq:parity_transf_gamma}
    \beta \gamma^{\mu_1\ldots \mu_{2j}} \beta^{-1} = \left( \gamma^{\mu_1\ldots \mu_{2j}}\right)^\dagger = \left(\begin{array}{cc}
0 & \bar{t}^{\mu_1\ldots \mu_{2j}} \\
{t}^{\mu_1\ldots \mu_{2j}}  &  0 
\end{array} \right) = \gamma_{\mu_1\ldots \mu_{2j}}
\,,
\end{equation}
which is the transformation rule of a proper tensor under parity, see Eq.~(\ref{eq:t_tbar_sign_relations}). 
This is to say that in this construction $\beta$ is a representation of the parity transformation for any spin. There are other well-known spin-1/2 equations that translate to any spin without modification. As expected, the generalized $\gamma_5$ of Eq.~(\ref{eq:gamma5}) transforms as a pseudoscalar
\begin{equation}\label{eq:parity_transf_gamma5}
    \beta \gamma_{5} \beta^{-1} = - \gamma_{5}
\,,
\end{equation}
and $\gamma^{(\mu)}\gamma_{5}$ transforms as a pseudo-tensor
\begin{equation}\label{eq:parity_transf_gamma_gamma5}
    \beta \gamma^{\mu_1\ldots \mu_{2j}} \gamma_{5} \beta^{-1} = - \gamma_{\mu_1\ldots \mu_{2j}} \gamma_{5}
\,.
\end{equation}
In addition, we have that $\gamma_5$ anticommutes with all gamma matrices:
\begin{equation}\label{eq:anti_CR_gamma_gamma5}
    \left\{ \gamma^{\mu_1\ldots \mu_{2j}}, \gamma_{5} \right\} = 0
\,.
\end{equation}

The causal bispinor field can be written as
\begin{equation}\label{eq:causal_bispinor_field}
        \psi(x) = \frac{1}{(2\pi)^{3}}\int \frac{\text{d}^3\bm p}{(2E_p)^{1/2}} \sum_\lambda \left[ u (p,\lambda) \, a(\bm p, j\lambda) e^{-{\rm i}px}+
        v(p,\lambda) b^\dagger (\bm p,j\lambda) e^{{\rm i}px} \right]
        \,,
\end{equation}
    \begin{align}\label{eq:bispinor_u_v}
        &u(p,\lambda) = 
        \left(\begin{array}{c}
        D^{(j)}(p) \, \overset{\circ}{u}_{L,c}(\lambda)\\
        \overbar{D}^{(j)}(p) \,\overset{\circ}{u}_{R}^{\dot c}(\lambda)\\
        \end{array}\right) \,,
        %%%%%%
        &v(p,\lambda) = 
        \left(\begin{array}{c}
        D^{(j)}(p)C^{-1} \, \overset{\circ}{u}_{L,c}(\lambda)\\
        (-1)^{2j}\overbar{D}^{(j)}(p)C^{-1} \, \overset{\circ}{u}_{R}^{\dot c}(\lambda)\\
        \end{array}\right)
        \,.
    \end{align}
An associated Lagrangian can be written through an auxiliary field~\cite{Shay1968ALF}.
The spinor $u$ and antispinor $v$ are related through charge conjugation
\begin{align}
\label{eq:bispinor_chargeconj}
    &\mathsf{C}\left(\bar{u}(p,\lambda)\right)^T = v(p,\lambda) 
    \,.
\end{align}    
Here, the bispinor charge conjugation matrix $\mathsf{C}$ is defined as
\begin{equation}
   \mathsf{C} =  \left(\begin{array}{cc}
(-1)^{2j}C & 0 \\
0 & C
\end{array}\right)
\,,
\end{equation}
see Eq.~(\ref{eq:def_charge_conj_C}) for the definition of $C$.  For the spin 1/2 case this corresponds to the charge conjugation convention
\begin{equation}
    \mathsf{C}^{(\frac{1}{2})} = -{\rm i}\gamma^2\gamma^0\,.
\end{equation}

To close this section, we want to emphasize that the construction discussed so far only requires two inputs.  The first is the generators of rotation in the spin representation of interest, the second is the expression of the boost parameter $\tilde p$ of choice.  Everything else follows from these and involves only elementary algebraic operations.

%%%%%%%%%%%%%%%%%%%%%%%%%%%
%%%% Section III      %%%%%%
%%%%%%%%%%%%%%%%%%%%%%%%%%%

\section{The \texorpdfstring{$t$}{t}-tensors contain the independent generators of \texorpdfstring{$\mathfrak{u}(2j+1)$}{u(2j+1)}}
\label{sec:t_tensors_gen}

Clearly the $t$-tensors play a central role in performing general calculations with chiral fields.  As shown in Sec.~\ref{sec:constructions}, the $t$-tensor appears in the expressions for the field commutators (and hence propagators) and spinors.  
In this section, we show that the $(2j+1)^2$ independent matrices that are elements of the $t$-tensors\footnote{The $t$-tensors were introduced in Eqs.~(\ref{eq:prop_ttensor}), and an independent definition is given in Eq.~(\ref{eq:t_recursion}). We provide below a simple formula for explicit construction in Eq.~(\ref{eq:t_matrix_value}).} represent a minimal set corresponding to a basis for the Lie algebra $\mathfrak{u}(2j+1)$.
This means they can be used as a basis to decompose matrix elements of operators between spin-$j$ states.  This type of matrix elements is ubiquitous in hadron physics.  The Lorentz invariant decomposition of local and bilocal QCD operators serves as the definition of Lorentz scalar non-perturbative objects, characterizing the quark and gluon content of hadrons through (electromagnetic, gravitational, generalized) form factors, (transverse momentum) parton distribution functions, generalized parton distributions, etc.  The coefficients multiplying these non-perturbative objects can be written using covariant expressions made out of $t$-tensors on the one-hand 
(as these contain a basis for $(2j+1) \times (2j+1)$ Hermitian matrices),
% The coefficients multiplying these non-perturbative objects can be written using covariant expressions with the $t$-tensor basis on the one-hand (as these contain all the Hermitian matrices of order $2j+1$), 
or using bispinor bilinears that have well-defined Lorentz transformation properties on the other hand.  This obviously implies there is a map between these covariant bilinears and Lorentz contractions of the $t$-tensors and kinematical four-vectors.  This correspondence is well known from the spin-1/2 case using the Pauli matrices. In Sec.~\ref{sec:gen_bilinears}, we extend that formalism to the general spin case, drawing many parallels and discussing physical interpretations in the process. To do so, we elucidate in this section details regarding the content of the $t$-tensors, by discussing their number of independent elements and by discussing algorithms to construct them.  

The number of independent Hermitian matrices contained in the $t$-tensors can be determined by counting the number of independent components of the $t$-tensors. The maximum number of independent components that a totally symmetric rank-$2j$ tensor in four dimensions can have is 
\begin{equation*}
\frac{(2j+1)(2j+2)(2j+3)}{6}  \,.    
\end{equation*}
The traceless property of the $t$-tensors implies\begin{align}\label{traceless condition - bis}
g_{\mu_1\mu_2}t^{\mu_{1} \mu_{2} \mu_{3} \cdots \mu_{2 j}} = 0\,.
\end{align}
As the $t$-tensor is completely symmetric, any contraction between two other indices reduces to this equation.  
The number of non-trivial constraints that Eq.~(\ref{traceless condition - bis}) provides corresponds to the (maximum) number of independent components in a symmetric rank-$(2j-2)$ tensor, being
\begin{equation*}
\frac{(2j-1)(2j)(2j+1)}{6}\,.    
\end{equation*}
It follows that the amount of independent components in the symmetric and traceless rank-$2j$ $t$-tensor is
\begin{align}
\frac{(2j+1)(2j+2)(2j+3)}{6} - \frac{(2j-1)(2j)(2j+1)}{6} = (2j+1)^2 \,.
\end{align}
We repeat that each of these tensor components is a Hermitian matrix (including the identity).  Below, we show explicitly that each independent component in the $t$-tensor contains an independent matrix.  Consequently, these $(2j+1)^2$ independent matrices form a complete basis for the dimension-($2j+1$) representation of the $\mathfrak{u}$(N) algebra, with N$\,=2j+1$.

Through the exponential map of Eq.~(\ref{eq:prop_ttensor}), it is clear that elements of the $t$-tensors with $k$ spatial indices are homogeneous polynomials of degree $k$ in the generators of rotations $\bm J^{(j)}$.  Due to the Cayley-Hamilton theorem applied to $\bm J$, the exponential map evaluates to a polynomial of degree $2j$.  
This implies the $t$-tensors will allow for a natural  $\mathfrak{su}(2)$ multipole expansion of expressions containing them.
In Appendix~\ref{sec:t_algorithm}, we present an algorithm that looks at the content of the $t$-tensors from the context of the Jordan algebra of the generators of rotations (i.e. their anticommutators), which contrary to the Lie algebra of the commutators is representation dependent.  In Sec.~\ref{sec:quadratic_t_vs_multipoles}, we show that a similar decomposition can be achieved for the covariant $\mathfrak{sl}(2,\mathbb{C})$ multipoles, see Ref.~\cite{Cotogno:2019vjb} for related work. 

In Ref.~\cite{Weinberg:1964cn}, Weinberg included closed expressions for $\Pi(q)$ and $\overbar \Pi(q)$. This allows to extract explicit expressions for the elements of the $t$-tensors through Eq.~(\ref{eq:prop_ttensor}).  He did not give expressions for spin $j>1$, however, stating \emph{``We won't bother extracting the $t^{\mu\nu\ldots}$ for $j>1$, because it is $\Pi(q)$ that we really need to know''}.  That was true for the Feynman rules of course. But if we want to use these tensors as a basis for polarized matrix elements, the individual elements do become objects of interest. It is useful to detail their properties, specifically with the interest of using them to perform a multipole decomposition of these matrix elements.

Using the recursion of Eq.~(\ref{eq:t_recursion}), we can carry out the complete reduction of the $t$-tensor to direct products of spin 1/2 representations. The full tensor product of Pauli matrices ($\sigma^\mu$) is obtained as~\cite{Barut:1963zzb}
\begin{align}\label{eq:singlet_fullreduction}
\tensor{\left(t^{\mu_{1} \cdots \mu_{2j} }\right)}{_{c_{2j}\dot d_{2j}}}  = \prod_{i=1}^{2j} \tensor{\left[j-\tfrac{i-1}{2},j-\tfrac{i}{2},\tfrac{1}{2}\right]}{_{c_{2j-i+1}}^{c_{2j-i} a_i}} 
\tensor{\left[j-\tfrac{i-1}{2},j-\tfrac{i}{2},\tfrac{1}{2}\right]}{_{\dot d_{2j-i+1}}^{\dot d_{2j-i}\dot b_i}} \tensor{(\sigma^{\mu_{i}})}{_{a_i \dot{b}_i}}
\end{align} 
where in the last ($i=2j;\, j-\frac{i}{2}=0$) factor, the uncontracted (spin-0) indices $d_0,c_0$ in the CG-coefficients are evaluated as 0, and the CG-coefficients themselves are taken as 
\begin{subequations}
\begin{align}\label{eq:singlet_fullreduction_convention}
\tensor{\left[\tfrac{1}{2},0,\tfrac{1}{2}\right]}{_{c_{1}}^{c_{0} a_{2j}}} & 
= \tensor{\left[\tfrac{1}{2},0,\tfrac{1}{2}\right]}{_{c_{1}}^{0 a_{2j}}} 
\equiv \delta^{a_{2j}}_{c_{1}},
\\
\tensor{\left[\tfrac{1}{2},0,\tfrac{1}{2}\right]}{_{\dot d_{1}}^{\dot d_{0}\dot b_{2j}}} & 
= \tensor{\left[\tfrac{1}{2},0,\tfrac{1}{2}\right]}{_{\dot d_{1}}^{0 \dot b_{2j}}} 
\equiv  \delta^{\dot b_{2j}}_{\dot d_{1}} \,.
\end{align} 
\end{subequations}

To obtain compact expressions for the matrices contained in the $t$-tensors, it is extremely useful to consider the Lorentz components of the $t$-tensors in the $\{+-R\,L\}$-basis\footnote{While taking complex combinations of $t$-tensor elements yields non-Hermitian matrices in general, what matters for the final results is that they are Hermitian in the Cartesian coordinates.}, see Eq.~(\ref{eq:coords_LCRL}).  To give an example, the $t$-tensor traceless condition in these coordinates, in combination with the symmetric nature of the tensors, yields the simple constraint
\begin{equation} \label{eq:traceless_spherical}
 t^{RL\mu_3\cdots \mu_{2j}}=t^{+-\mu_3\cdots \mu_{2j}} 
 \,.
\end{equation}
 The simplifications occur in the  $\{+-R\,L\}$-basis because each of the $\sigma^+, \sigma^-,\sigma^R, \sigma^L$ matrices in Eq.~(\ref{eq:singlet_fullreduction}) has only one non-zero element, with value 2. From  Eq.~(\ref{eq:singlet_fullreduction}) and the properties of the CG-coefficients, it then follows that any general $t$-tensor element in this basis will be a matrix with exactly one non-zero matrix element.  If we consider a $t$-tensor of Lorentz-rank $2j$, with $\mathfrak{p}$ indices +, $\mathfrak{m}$ indices -, $\mathfrak{r}$ indices $R$ and $\mathfrak{l}$ indices $L$ (with constraint $\mathfrak{p}+\mathfrak{m}+\mathfrak{r}+\mathfrak{l}=2j$), and if we label both spinor (matrix) indices with values $\{+j,\cdots,-j\}$, that single non-zero element is in position
\begin{align} \label{eq:t_index}
    &\frac{(\mathfrak{p}-\mathfrak{m})+(\mathfrak{r}-\mathfrak{l})}{2} \qquad \text{[row]} \,,\nonumber\\
    &\frac{(\mathfrak{p}-\mathfrak{m})-(\mathfrak{r}-\mathfrak{l})}{2} \qquad \text{[column]} 
    \,.
\end{align}
For each choice of row and column index, at least one set of valid indices $\{\mathfrak{p},\mathfrak{m},\mathfrak{r},\mathfrak{l}\}$ can be found that solves the above constraints. This demonstrates that the independent tensor elements of the $t$-tensor are also independent matrices, which in Cartesian components become the independent Hermitian matrices. This justifies the statement that the $t$-tensors contain a complete $\mathfrak{u}$(N) basis.  The fact that this position only depends on the differences $\mathfrak{p}-\mathfrak{m}$ and $\mathfrak{r}-\mathfrak{l}$ reflects the Lorentz traceless condition of Eq.~(\ref{eq:traceless_spherical}).

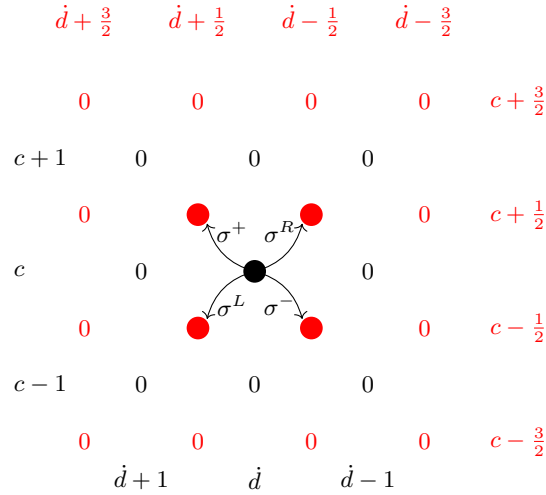
\begin{figure}[htb]
    \centering
\begin{tikzpicture}
    % Parameters
    \def\a{1.5}  % grid spacing
    
    % First Grid (Grid 1)
    \foreach \x in {-1,0,1} {
        \foreach \y in {-1,0,1} {
            \node[black] at (\x*\a,\y*\a) {0};
        }
    }
    
    % Second Grid (Grid 2, offset by a/2 in x and y)
    \foreach \x in {-1.5,-0.5,0.5,1.5} {
        \foreach \y in {-1.5,-0.5,0.5,1.5} {
            \node[red] at (\x*\a,\y*\a) {0};
        }
    }
    
    % Central point in Grid 1 (1,1)
    \fill[black] (0*\a,0*\a) circle (0.15);
    
    % Diagonal arrows from central point to nearest neighbors in Grid 2
    \fill[red] (-0.5*\a,-0.5*\a) circle (0.15);
    \fill[red] (-0.5*\a,0.5*\a) circle (0.15);
    \fill[red] (0.5*\a,-0.5*\a) circle (0.15);
    \fill[red] (0.5*\a,0.5*\a) circle (0.15);
    
    \draw[->] (0,0) to[bend left=30] (-0.42*\a,+0.42*\a);
    \draw[->] (0,0) to[bend left=30] (+0.42*\a,-0.42*\a);
    \draw[->] (0,0) to[bend right=30] (-0.42*\a,-0.42*\a);
    \draw[->] (0,0) to[bend right=30] (0.42*\a,0.42*\a);

    \node[black, above] at (0.22*\a,0.18*\a) {$\sigma^R$};
    \node[black, above] at (-0.2*\a,0.18*\a) {$\sigma^+$};
    \node[black, below] at (-0.2*\a,-0.13*\a) {$\sigma^L$};
    \node[black, below] at (0.22*\a,-0.13*\a) {$\sigma^-$};

    % Labels for first grid (black)
    % Rows: a+1, a, a-1 (a is the row with the arrow)
    \node[black, right] at (-2.2*\a,1*\a) {${c}+1$};
    \node[black, right] at (-2.2*\a,0*\a) {${c}$};
    \node[black, right] at (-2.2*\a,-1*\a) {${c}-1$};
    
    % Columns: b+1, b, b-1 (b is the column with the arrow) - REVERSED
    \node[black, above] at (1*\a,-2*\a) {$\dot{d}-1$};
    \node[black, above] at (0*\a,-2*\a) {$\dot{d}$};
    \node[black, above] at (-1*\a,-2*\a) {$\dot{d}+1$};
    
    % Labels for second grid (red)
    % Rows: multiple points to choose from now
    \node[red, right] at (2*\a,1.5*\a) {${c}+\frac{3}{2}$};
    \node[red, right] at (2*\a,0.5*\a) {${c}+\frac{1}{2}$};
    \node[red, right] at (2*\a,-0.5*\a) {${c}-\frac{1}{2}$};
    \node[red, right] at (2*\a,-1.5*\a) {${c}-\frac{3}{2}$};
    
    % Columns: multiple points to choose from - REVERSED
    \node[red, above] at (1.5*\a,2*\a) {$\dot{d}-\frac{3}{2}$};
    \node[red, above] at (0.5*\a,2*\a) {$\dot{d}-\frac{1}{2}$};
    \node[red, above] at (-0.5*\a,2*\a) {$\dot{d}+\frac{1}{2}$};
    \node[red, above] at (-1.5*\a,2*\a) {$\dot{d}+\frac{3}{2}$};
\end{tikzpicture}
    \caption{Illustration of how the recursion in the relation between $t$-tensors for different spins works in the $\{+-R\,L\}$-basis.  The black grid positions illustrate part of a spin-$j$ $t$-tensor element in the $\{+-R\,L\}$-basis which has its single non-zero element on row $c$ and column $\dot d$ (black dot).  Using the recursion formula to build a spin-($j+\frac{1}{2}$) $t$-tensor element (part of which is shown in the red grid) then corresponds to adding a Lorentz index: $t^{\mu_1\cdots \mu_{2j}} \to t^{\mu_1\cdots \mu_{2j},\mu_{2j+1}}$.  The position of the non-zero element in that rank-($2j+1$) $t$-tensor (nearest neighbor red dot) then depends on the choice $\{+-R\,L\}$ for the extra Lorentz index and is illustrated with the action of the arrows and the corresponding choice of the spin-1/2 matrix which results in the 4 red dots shown.  The action of the arrow corresponds to multiplication by the CG-coefficients in Eq.~(\ref{eq:singlet_fullreduction}) and the non-zero element of the Pauli matrix (which is 2 in the basis we use). }
    \label{fig:t_recursion}
\end{figure}

Fig.~\ref{fig:t_recursion} illustrates how the position Eq.~(\ref{eq:t_index}) can be inferred from the action of the tensor product in the recursion of Eq.~(\ref{eq:t_recursion}). Matrices $\sigma^+$ and $\sigma^-$ create diagonal shifts in row and column value: $(+\frac{1}{2},+\frac{1}{2})$ respectively $(-\frac{1}{2},-\frac{1}{2})$. $\sigma^R$ and $\sigma^L$ create antidiagonal shifts in row and column value: $(+\frac{1}{2},-\frac{1}{2})$ respectively $(-\frac{1}{2},+\frac{1}{2})$.   It is clear from Fig.~\ref{fig:t_recursion} that there are multiple ways of ``hopping'' to a certain position through this recursion algorithm, this reflects the redundancy in the elements of the symmetric and traceless Lorentz tensor.

The value of the non-zero matrix element can be deduced from the CG coefficients in Eq.~(\ref{eq:singlet_fullreduction}).  The fact that at each step in the recursion $j$ is coupled to the maximal $j=j_1+j_2$ means that the recursion is unique, i.e. we can also couple to intermediate $j_1,j_2\geq 1/2$ and then couple those tensors (eventually) to the total spin $j$, the final result will not change.  This simplifies both the recursion and the values of the CG coefficients in the recursions we use.  
First, we can consider 4 separate intermediate tensors: the first rank $\mathfrak{p}$ with all indices +, the second rank $\mathfrak{m}$ with indices -, the third rank $\mathfrak{r}$ with all indices $R$ and the fourth rank $\mathfrak{l}$ with all indices $L$.  As can be seen using the illustration of Fig.~\ref{fig:t_recursion}, these four tensors have their non-zero element in a corner of the matrix (top left for the one with only + indices, etc.).  The CG coefficients appearing in the recursion for each of these four constructions are all 1, meaning the value of the non-zero matrix element in the corner is simply coming from the factors $\sigma^\mu$.  The value being $2^{\mathfrak{p}}$ for the tensor with all + indices, and similarly $2^{\mathfrak{m}},2^{\mathfrak{r}},2^{\mathfrak{l}}$ for the others.  To obtain the final $t$-tensor, the final step is to couple these four tensors to the rank-$2j$ tensor.  In this operation, the values of the 6 remaining CG coefficients and the matrix elements of the 4 separate tensors determine the value of the non-zero matrix element to be
\begin{align}\label{eq:t_matrix_value}
% & \hspace{-1.5cm}
\left(t^{+_1\cdots+_{\mathfrak{p}} -_1\cdots -_{\mathfrak{m}} R_1 \cdots R_{\mathfrak{r}}  L_1 \cdots L_{\mathfrak{l}}}\right)_{c \dot {d}} 
=  
% \nonumber\\[1em]
& \langle \tfrac{\mathfrak{p}}{2}\tfrac{\mathfrak{p}}{2},\tfrac{\mathfrak{m}}{2} {-\tfrac{\mathfrak{m}}{2}}  | \tfrac{\mathfrak{p}+\mathfrak{m}}{2} \tfrac{\mathfrak{p}-\mathfrak{m}}{2} \rangle^2
\langle \tfrac{\mathfrak{p}+\mathfrak{m}}{2}\tfrac{\mathfrak{p}-\mathfrak{m}}{2},\tfrac{\mathfrak{r}}{2} {\tfrac{\mathfrak{r}}{2}}  | \tfrac{\mathfrak{p}+\mathfrak{m}+\mathfrak{r}}{2} \tfrac{\mathfrak{p}-\mathfrak{m}+\mathfrak{r}}{2} \rangle
\nonumber\\[0.5em]
& \times 
\langle \tfrac{\mathfrak{p}+\mathfrak{m}}{2}\tfrac{\mathfrak{p}-\mathfrak{m}}{2},\tfrac{\mathfrak{r}}{2} {-\tfrac{\mathfrak{r}}{2}}  | \tfrac{\mathfrak{p}+\mathfrak{m}+\mathfrak{r}}{2} \tfrac{\mathfrak{p}-\mathfrak{m}-\mathfrak{r}}{2} \rangle
\langle \tfrac{\mathfrak{p}+\mathfrak{m}+\mathfrak{r}}{2}\tfrac{\mathfrak{p}-\mathfrak{m}+\mathfrak{r}}{2},\tfrac{\mathfrak{l}}{2} {-\tfrac{\mathfrak{l}}{2}}  | j \tfrac{\mathfrak{p}-\mathfrak{m}+\mathfrak{r}-\mathfrak{l}}{2} \rangle
\nonumber\\[0.5em]
& \times 
\langle \tfrac{\mathfrak{p}+\mathfrak{m}+\mathfrak{r}}{2}\tfrac{\mathfrak{p}-\mathfrak{m}+\mathfrak{r}}{2},\tfrac{\mathfrak{l}}{2} {\tfrac{\mathfrak{l}}{2}}  | j \tfrac{\mathfrak{p}-\mathfrak{m}-\mathfrak{r}+\mathfrak{l}}{2} \rangle
% \nonumber\\
% &\qquad \times 
\, 2^{\mathfrak{p}} 2^{\mathfrak{m}} 2^{\mathfrak{r}} 2^{\mathfrak{l}} \, \delta_{c,\frac{\mathfrak{p}-\mathfrak{m}+\mathfrak{r}-\mathfrak{l}}{2}} \, \delta_{\dot {d},\frac{\mathfrak{p}-\mathfrak{m}-\mathfrak{r}+\mathfrak{l}}{2}}\nonumber\\[1em]
= & 2^{2j}\,\frac{\sqrt{(\mathfrak{p}+\mathfrak{r})!(\mathfrak{p}+\mathfrak{l})!(\mathfrak{m}+\mathfrak{r})!(\mathfrak{m}+\mathfrak{l})!}}{(2j)!} \, \delta_{c,\frac{\mathfrak{p}-\mathfrak{m}+\mathfrak{r}-\mathfrak{l}}{2}} \, \delta_{\dot {d},\frac{\mathfrak{p}-\mathfrak{m}-\mathfrak{r}+\mathfrak{l}}{2}} \nonumber\\
%%%%%
= & 2^{2j}\,\frac{\sqrt{(j+c)!(j+\dot {d})!(j-\dot {d})!(j-c)!}}{(2j)!} \, \delta_{\mathfrak{m},\mathfrak{p}-c-\dot{d}} \, \delta_{\mathfrak{r},j+c-\mathfrak{p}} \, \delta_{\mathfrak{l},j+\dot {d}-\mathfrak{p}} \,, 
\nonumber \\[1em] 
&\hspace{-1cm} \mathfrak{p},\mathfrak{m},\mathfrak{r},\mathfrak{l} \in \{0,\cdots,2j\}\,, \quad \mathfrak{p}+\mathfrak{m}+\mathfrak{r}+\mathfrak{l}=2j \,,
% \nonumber\\
% &
\quad c,\dot {d} \in \{+j,\cdots,-j\} 
\,,
\end{align}
a surprisingly simple and elegant result.  The last expression with the Kronecker deltas for $\mathfrak{p},\mathfrak{m},\mathfrak{r},\mathfrak{l}$ is useful if one considers a specific $(c,\dot d)$ matrix element of the $t$-tensor and one wants to evaluate which Lorentz tensor elements will have non-zero values at that position. One observes that for the non-zero matrix elements, the value 
% of the matrix element 
can be written as an expression that in the $\{+-R\,L\}$-basis only makes reference to the number of Lorentz indices $(\mathfrak{p},\mathfrak{m},\mathfrak{r},\mathfrak{l})$, or one that only references the spinor (matrix) indices $(c,\dot d)$.

As a corollary of Eq.~(\ref{eq:t_matrix_value}), there is a convenient way to express the matrix $C$ [Eq.~(\ref{eq:def_charge_conj_C})] as a $t$-tensor element. The matrix representation of $C$ only has non-zero elements on the off-diagonal. This means it needs to be built from $t$-tensor elements with only indices $R$ or $L$ (non-zero $\mathfrak{r},\mathfrak{l}$ in Eq.~(\ref{eq:t_index})), we refer again to Fig.~\ref{fig:t_recursion} to appreciate this visually. The spin-1/2 result $C^{(\frac{1}{2})}= {\rm i} \sigma^2$ motivates us to consider $t^{2\cdots 2}$.  We have
\begin{equation}
    t^{2\cdots 2} = \frac{1}{(2{\rm i})^{2j}}\sum_{k=0}^{2j} 
    \binom{2j}{k}
    (-1)^k t^{R_1\cdots R_{2j-k}L_1\cdots L_k}\,.
\end{equation}
Using Eq.~(\ref{eq:t_matrix_value}) with $\mathfrak{p}=\mathfrak{m}=0, \mathfrak{r}=2j-k, \mathfrak{l}=k$, it is then found that
\begin{equation}\label{eq:C_as_t2}
    C = {\rm i}^{2j} t^{2\cdots 2}\,.
\end{equation}
This relation can be of interest if one wants to directly apply the algebra of the $t$-tensors in Sec.~\ref{sec:t_gamma_tensors_algebra} in certain expressions (similar to how in certain contexts $\epsilon^{ij}=i\sigma^2$ is used).  The trade-off is the expressions do not \emph{look} explicitly covariant anymore, although they still are.
It is worth stressing that $C$ is invariant under Lorentz transformations (see comment after Eq.(\ref{eq:charge_conj_J})), and Eq.~(\ref{eq:C_as_t2}) must be read as an equality between matrix elements, as technically both sides of the equation have different spinor indices.  

We want to remark that because the traceless condition is implemented on the Lorentz indices, the elements of the $t$-tensor (with Cartesian components) do not necessarily correspond to $\mathfrak{u}$(N=$2j+1$) bases that are commonly used elsewhere.  While for spin-1/2 they do correspond to the Pauli matrices, for spin 1 the elements of the $t^{\mu\nu}$ (with at least one spatial index) do not entirely match the Gell-Mann basis for $\mathfrak{su}$(3) for instance.

%%%%%%%%%%%%%%%%%%%%%%%%%%%%%%%%%%%%%%
%%%% connection with MSH %%%%%%%%%%%%%
%%%%%%%%%%%%%%%%%%%%%%%%%%%%%%%%%%%%%%

\section{Correspondence with massive spinor-helicity formalism}
\label{sec:MSH}

The left- and right- handed spinors of Eq.~(\ref{eq:uspinors_def}) carry both a spinor and little group index.  This is reminiscent of the massive spinor-helicity (MSH) formalism developed in Ref.~\cite{Arkani-Hamed:2017jhn} and used in the construction of general $S$-matrix amplitudes.  The difference being that both indices used for the chiral spinors are $(2j+1)$-valued (without any redundancies), while MSH uses symmetrized rank-$2j$ tensor products of spin-1/2 spinors that carry a 2-valued spinor and little group index.  Of course both objects in the end carry the same number of independent indices ($2j+1$).  In this section, we sketch the correspondence between the chiral spinors and massive spinor-helicity tensor products. 
 For MSH, we follow the conventions used in Ref.~\cite{Ochirov:2018uyq}, but use slightly different typography for the indices, see App.~\ref{sec:notation} for a summary.  As in the rest of the text, we use lowercase Roman indices for spinor indices, where $a,b$ are reserved for spin-1/2.  Whereas outside of this section we typically use $\lambda$ for the little group polarization index, in this section we switch to uppercase Roman indices $A,B$ for these little group indices to avoid confusion with the notation $\lambda,\tilde \lambda$ for chiral spin-1/2 spinors, which is widely used in the spinor-helicity formalism.

Before we turn to objects with little group indices, we take Eq.~(\ref{eq:singlet_fullreduction}) and contract it with $p_{(\mu)}=p_{\mu_1}\cdots p_{\mu_{2j}}$ to arrive at
\begin{align}
 \Pi(p) = p_{(\mu)} t^{(\mu)}_{c_{2j}\dot d_{2j}}  = \prod_{i=1}^{2j} \tensor{\left[j-\tfrac{i-1}{2},j-\tfrac{i}{2},\tfrac{1}{2}\right]}{_{c_{2j-i+1}}^{c_{2j-i}a_i}} 
\tensor{\left[j-\tfrac{i-1}{2},j-\tfrac{i}{2},\tfrac{1}{2}\right]}{_{\dot d_{2j-i+1}}^{\dot d_{2j-i}\dot b_i}} p_{a_i \dot b_i},
\end{align} 
which expresses the SO(3,1) to SL$(2,\mathbb{C})$ correspondence for higher dimensional chiral representations on the left side, and we used for spin-1/2 $p_{a \dot b} \equiv p_\mu \sigma^\mu_{a\dot b}$ on the right side.

For massive spin-1/2, spinors are obtained by taking the ``square root'' of $p_{a_i \dot b_i}$, and higher-spin particles are represented as symmetric tensor products of these spin-1/2 spinors \cite{Arkani-Hamed:2017jhn}. For the spin-1/2 case, the massive Weyl spinors used in MSH are identical to the chiral spinors with the original Jacob \& Wick choice of standard boost~\cite{Jacob:1959at}, corresponding to the $\tilde p_h$ boost parameters of Eq.~(\ref{eq:ptilde_h}).  In equations, we have the following correspondence with the MSH notation of Refs.~\cite{Arkani-Hamed:2017jhn,Ochirov:2018uyq}:
\begin{subequations}
\begin{align}\label{eq:spinor_helicity_correspondence_L}
&u_{hL,a}(p,A) = \lambda_{pa}^A\equiv |p^A\rangle_a, \\
    &u_{hR}^{\dot a}(p,A) = \tilde \lambda_p^{\dot a A}\equiv |p^A]^{\dot a},\\
    &[\text{spin 1/2}]\nonumber
\end{align}    
\end{subequations}
where we use Eqs.~(\ref{eq:uspinors_def}), (\ref{eq:chiralspinor_rf}), (\ref{eq:general_boosts_t}) and (\ref{eq:general_boosts_tbar}).

To make the correspondence concrete for the higher spin case, we can contract the left chiral recursion relation shown in Eq.~(\ref{eq:singlet_fullreduction}) with the boost parameters $\tilde p_h$:
\begin{align}\label{eq:chiralspinor_MSH_correspondence}
u_{hL,c_{2j}}(p,A_{2j}) &= m^{j} \tilde p_h^{\mu_1}\cdots \tilde p_h^{\mu_{2j}} \tensor{\left(t^{\mu_{1} \cdots \mu_{2j} }\right)}{_{c_{2j}\dot d_{2j}}}\delta^{\dot d_{2j}A_{2j}}\nonumber\\
%%%%%%%%%%%%%
&= (-)^{j(2j+1)}\prod_{i=1}^{2j} \tensor{\left[j-\tfrac{i-1}{2},j-\tfrac{i}{2},\tfrac{1}{2}\right]}{_{c_{2j-i+1}}^{c_{2j-i}a_i}}
\tensor{\left[j-\tfrac{i-1}{2},j-\tfrac{i}{2},\tfrac{1}{2}\right]}{^{A_{2j-i+1}}_{A_{2j-i} A_i}} |p^{A_i}\rangle_{a_i},
\end{align} 
where $|p^{A_i}\rangle_{a_i}=
\tensor{(m^{\frac{1}{2}}\sigma^{\tilde p_h})}{_{a_i \dot b_i}}\delta^{\dot b_i A_i}$ and Eq.~(\ref{eq:CG_def}) was used.  
For an explanation regarding the mixed indices in $\delta^{\dot b_i A_i}$, see the paragraph enclosing Eqs.~(\ref{eq:uspinors_def})--(\ref{eq:chiralspinor_rf}).
On the left side of 
% this equation 
Eq.(\ref{eq:chiralspinor_MSH_correspondence}) we have the chiral spinors used in 
% the construction of 
this article, on the right hand side are the symmetric spin-1/2 tensor products used in MSH\footnote{See Ref.~\cite{Barut:1963zzb} for a proof of the symmetry in indices $a_i, A_i$ in this reduction.}.  
Both clearly carry little group and Lorentz spinor indices that give these objects the required transformation properties.  
% We need to comment on the placement of some of the indices here.  $\beta_i$ in the helicity spinor is the SU(2) little group index.  The correspondence with the spinorial index in the standard boost (written as the contraction between the $t$-tensor and the $\tilde p^\mu$ here) happens in the rest frame where as commented on earlier in Sec.~\ref{sec:boosts_spinors} the notations can be identified.  We choose the position of the little group index in the helicity spinors to conform with the conventions of Ref.~\cite{Ochirov:2018uyq}.  
An equivalent expression can be written for the right chiral sector, using $\tilde p_h^*$, $\bar t$, $|p^{A_i}]^{\dot a_i}$, and $\bar \sigma$.
This correspondence is independent of the standard boost considered and would be identically valid for the other types,
% considered, 
like canonical or light-front spinors.

Additionally, it is also possible to write the chiral spinors used here in an MSH-like notation.  This is again independent of the type of standard boost considered, as these different types of spinors just differ by rest-frame rotations (see App.~\ref{sec:hel_lf_spinors}), which do not affect any of the following results.  We introduce the following notation:
\begin{subequations}
\begin{align}
    &u_{L,c}(p,A) \equiv |j,p^A\rangle_c,\\
    &u_R^{\dot c}(p,A) \equiv |j,p^A]^{\dot c}.
\end{align}    
\end{subequations}
Then the various equations that these chiral spinors obey can be written as
\begin{subequations}
\begin{align}
    &p_{c \dot d}\equiv p_{(\mu)}t^{(\mu)} = |j,p^A\rangle_c [j,p_A|_{\dot d},
    &p^{\dot c d }\equiv p_{(\mu)}\bar t^{(\mu)} = (-)^{2j}\,|j,p^A]^{\dot c} \langle j,p_A|^{d},\\
    %%%%%%%
    &|j,p^A\rangle_c \langle j,p_A|^d = (-)^{2j}m^{2j}\delta_c^d,
    &|j,p^A]^{\dot c} [j,p_A|_{\dot d} = m^{2j}\delta_{\dot d}^{\dot c},\\
    %%%%%%%%
    &p^{\dot c d}|j,p^A\rangle_d = (-)^{2j+1}m^{2j}|j,p^A]^{\dot c},
    &p_{c \dot d}|j,p^A]^{\dot d} = m^{2j}|j,p^A\rangle _{c},\\
    %%%%%%%%%%%
    &\langle j,p^A|^c \, p_{c \dot d} = (-)^{2j}m^{2j}|j,p^A]_{\dot d},
    &[j,p^A|_{\dot c} p^{\dot c d} = -m^{2j}|j,p^A\rangle ^{d},\\
    %%%%%%%%%%%%%%%%%%%%%%
    &\langle j,p^A |^c |j,p^B\rangle_c = (-)^{2j}m^{2j}C^{AB},
    &[j,p^A |_{\dot c} |j,p^B]^{\dot c} = m^{2j}C^{AB}\,, 
    %%%%%%%%%%%%%%%%%%%%%%
\end{align}
\end{subequations}
which can be compared to Eqs.~(2.15) in Ref.~\cite{Ochirov:2018uyq}.
To conclude this section, we want to comment on the fact that while the map sketched here implies that applications can be explored in both formalism, this does not necessarily mean one approach is superior to the other, and this can be context-dependent.  At least for us, obtaining the covariant multipole classification which is outlined in this article was more natural using the chiral spinors, given the proximity with a few decades of existing phenomenology on the QCD correlators of interest.  Given the extensive literature successfully using MSH for high-energy applications, there are probably no immediate advantages to reformulating those results using the chiral spinors.  Further study should expose to what extent this correspondence can be useful in applications.

%%%%%%%%%%%%%%%%%%%%%%%%%%%%%%%%%%%%%%
%%%%% t-tensor Algebra  %%%%%%%%%%%%%%
%%%%%%%%%%%%%%%%%%%%%%%%%%%%%%%%%%%%%%

\section{Algebra of the \texorpdfstring{$t$}{t}- and \texorpdfstring{$\gamma$-}{gamma-}tensors}
\label{sec:t_gamma_tensors_algebra}

In this section, we will discuss the algebraic properties of the $t$-tensors and derive general reduction formulas for the product of $t$-tensors. Subsequently, in Sec.~\ref{sec:gen_bilinears}, we will demonstrate their practical applications in the computation of generalized bilinears. 

As established in Sec.~\ref{sec:constructions}, the expressions of chiral bispinors and generalized gamma matrices both contain $t$-tensors. Consequently, the computation of bispinor bilinears will yield products of $t$- and $\bar t$-tensors. 
In Sec.~\ref{sec:gen_bilinears} we will show that products of $t$-tensors appearing in the bispinor bilinears always exhibit an alternating pattern in $t$ and $\bar t$, such as $t\bar{t}t\bar{t}t\ldots$ or $\bar{t}t\bar{t}t\bar{t}\ldots$.  This is a consequence of their spinorial transformation properties, where only the alternating patterns transform in a simple manner (as the bilinears do).   
We prove below that such products can be reduced to a linear expression in $t$ (or $\bar{t}$). 
These reductions applied to the calculation of bilinears with arbitrary spin will enable the identification of useful and universally applicable formulas.  

Moreover, using Eq.~(\ref{eq:def_eta}) we show that the reduction of non-alternating combinations, as for instance $t \mathbb{1} t$ and $\bar t \mathbb{1} \bar t$, can be obtained (algebraically) as special cases from the alternating ones.  
Non-alternating quadratic products enter in the calculus of commutation relations between $t$- or $\bar t$-tensors and the corresponding representation of Lorentz generators. 
They are used in Sec.~\ref{sec:covariantly_independent_terms_in_quadratic_t_products} to isolate covariant independent tensor substructures that appear in the reduction of the alternating products $t\bar t$ and $\bar t t$.

\subsection{Cubic products of \texorpdfstring{$t$}{t}-tensors}
\label{sec:Cubic_products}

We start with the reduction of the cubic product $t\bar t t \to t$. We show that it is the cubic product that exhibits the most irreducible behavior.  By carrying out the {\it barring} operation [Eq.~(\ref{eq:t_chargeconj})] analogous remarks apply to the reduction $\bar t t \bar t \rightarrow \bar t$.
As detailed in Sec.~\ref{sec:t_tensors_gen}, the $t$-tensor
%\footnote{Analogous remarks apply for $\bar{t}^{\mu_1 \cdots \mu_{2j}}$.} 
contains a basis for the Hermitian matrices. Note that Hermitian conjugation retains the $t\bar t t$ structure, but with an exchange of indices: 
\begin{equation}
\left( t^{\mu_{1} \cdots \mu_{2j} }  \bar{t}^{\rho_{1} \cdots \rho_{2j} } t^{\sigma_{1} \cdots \sigma_{2j} } \right)^\dagger = t^{\sigma_{1} \cdots \sigma_{2j} }  \bar{t}^{\rho_{1} \cdots \rho_{2j} } t^{\mu_{1} \cdots \mu_{2j} } 
\,.
\end{equation}
In general, the product $t\bar t t$ in itself is no longer Hermitian.  By separating the Hermitian and skew-Hermitian terms in the expression, we identify the Hermitian part of the cubic product with the tensor symmetric under exchange of the set of indices $\{\mu_1 \cdots \mu_{2j} \}$ and $\{\sigma_1 \cdots \sigma_{2j} \}$, while the skew-Hermitian part is antisymmetric under the same exchange:
\begin{align}\label{eq:cubict_(anti)hermitian}
t^{\mu_{1} \cdots \mu_{2j} }  \bar{t}^{\rho_{1} \cdots \rho_{2j} } t^{\sigma_{1} \cdots \sigma_{2j} } = &\frac{1}{2} \left[ t^{\left(\mu\right) }  \bar{t}^{\,\left(\rho\right) } t^{\left(\sigma\right) } + \left( t^{\left(\mu\right) }  \bar{t}^{\,\left(\rho\right) } t^{\left(\sigma\right) } \right)^\dagger \right] +\frac{1}{2} \left[ t^{\left(\mu\right) }  \bar{t}^{\,\left(\rho\right) } t^{\left(\sigma\right) } - \left( t^{\left(\mu\right) }  \bar{t}^{\,\left(\rho\right) } t^{\left(\sigma\right) } \right)^\dagger \right] 
\nonumber\\
=  &\frac{1}{2} \left[ t^{\left(\mu\right) }  \bar{t}^{\,\left(\rho\right) } t^{\left(\sigma\right) } + t^{\left(\sigma\right) }  \bar{t}^{\,\left(\rho\right) } t^{\left(\mu\right) } \right] +\frac{1}{2} \left[ t^{\left(\mu\right) }  \bar{t}^{\,\left(\rho\right) } t^{\left(\sigma\right) } - t^{\left(\sigma\right) }  \bar{t}^{\,\left(\rho\right) } t^{\left(\mu\right) } \right]  
\,,
\end{align}
where we use the notational convention of Eq.~(\ref{eq:symmetrized_indices_shorthand}).
If we reinstate the spinor indices on the $t$ and $\bar t$-matrices (see Eq.~(\ref{eq:prop_ttensor}), we see that $t\bar t t$ transforms in its spinor indices as $t$ does. This corresponds to the $(j,0)\otimes(0,j) = (j,j)$ representation, which is irreducible.
It follows that we can expand the cubic product on the Hermitian basis contained in the $t$-tensor.  It is also only the reduction $t\bar t t \to t$ that preserves the alternating order of $t$ and $\bar t$ in higher order alternating monomials. In the reduction, we use complex coefficients $\mathcal{C}_{c}$  in order to generate both the Hermitian and skew-Hermitian part:
\begin{align}\label{eq:def_coefficient_t_reduction_cubic}
t^{\mu_{1} \cdots \mu_{2j} }  \bar{t}^{\rho_{1} \cdots \rho_{2j} } t^{\sigma_{1} \cdots \sigma_{2j} }& \; = \; \mathcal{C}_{c}^{\mu_{1} \cdots \mu_{2j} \rho_{1} \cdots \rho_{2j} \sigma_{1} \cdots \sigma_{2j} \alpha_{1} \cdots \alpha_{2j} } t_{\alpha_{1} \cdots \alpha_{2j}} 
% \nonumber \\
% & 
\; = \; 
\left( \operatorname{Re}\left[\mathcal{C}_{c}^{(\mu) (\rho) (\sigma) (\alpha) }\right] + {\rm i} \operatorname{Im}\left[\mathcal{C}_{c}^{(\mu) (\rho) (\sigma) (\alpha) }\right] \right) t_{(\alpha)} \,.
% \nonumber\\
% & \; = \; \left( \operatorname{Re}\left[\mathcal{C}_{c}^{\mu_{1} \cdots \mu_{2j} \rho_{1} \cdots \rho_{2j} \sigma_{1} \cdots \sigma_{2j} \alpha_{1} \cdots \alpha_{2j} }\right] + {\rm i} \operatorname{Im}\left[\mathcal{C}_{c}^{\mu_{1} \cdots \mu_{2j} \rho_{1} \cdots \rho_{2j} \sigma_{1} \cdots \sigma_{2j} \alpha_{1} \cdots \alpha_{2j} }\right] \right) t_{\alpha_{1} \cdots \alpha_{2j}} \,.
\end{align}
 Here the two terms map to the two terms written on the right-hand side of Eq.~(\ref{eq:cubict_(anti)hermitian}), meaning the (imaginary) real part generates the (skew-)Hermitian part of the cubic product.

Since $t$ and $\bar{t}$ carry Lorentz tensor indices in their role as intertwiners, it follows from the transformation properties of $t$ and $\bar t$ (Eq.~(\ref{eq:t_cov})) that $\mathcal{C}_c$ also transforms as a tensor under Lorentz transformations, as do its real and imaginary parts considered separately. The application of Eq.~(\ref{eq:t_cov}) also shows $\mathcal{C}_c$ is an invariant tensor.

Therefore, the $\mathcal{C}_c$ can be written as a polynomial of the metric tensor where terms have zero or one single Levi-Civita tensor\footnote{We remind that any pair of Levi-Civita tensors $\epsilon^{\mu\rho\sigma\alpha}\epsilon^{\nu\lambda\chi\beta}$ can be substituted by a homogeneous polynomial of degree four in the metric.}. We found above that the real part $\operatorname{Re}\left[\mathcal{C}_{c}\right]$ is symmetric in exchange of $\{(\mu)\}$ and $\{(\sigma) \}$, while the imaginary part $\operatorname{Im}\left[\mathcal{C}_{c}\right]$ is antisymmetric under the same exchange. 
We can identify other properties the tensor $\mathcal{C}_{c}$ has under permutation of indices, which are also independently fulfilled by $\operatorname{Re}\left[\mathcal{C}_{c}\right]$ and $\operatorname{Im}\left[\mathcal{C}_{c}\right]$. Under permutations \emph{within} the individual sets of indices $\{(\mu) \}$, $\{(\rho) \}$, $\{(\sigma) \}$ and $\{(\alpha)\}$, the tensor $\mathcal{C}_{c}$  is symmetric and traceless.  For the first three sets, this follows from the symmetry properties of the $t$-tensors on the left-hand side of the reduction equation, see Eqs.~(\ref{eq:t_symmetric}) and (\ref{eq:t_traceless}).  
For the last set $\{ \alpha_i \}$, this follows as they are contracted with a $t$-tensor which exhibits the same symmetric and traceless properties.

The next step is of course to detail the explicit form of the invariant tensor $\mathcal{C}_c$.  In Appendix~\ref{sec:proof_cubic_reduction}, we prove that the reduction formula for the cubic product of $t$-tensors is given by\footnote{Note that both sides of the equation are symmetric also under permutations on the set of indices $\{\mu_{1} \cdots \mu_{2j} \}$.}
\begin{equation}\label{eq:t_reduction_cubic}
t^{\mu_{1} \cdots \mu_{2j} }  \bar{t}^{\rho_{1} \cdots \rho_{2j} } t^{\sigma_{1} \cdots \sigma_{2j} } = \frac{1}{[(2j)!]^2}  
\underset{\left\{(\rho),(\sigma) \right\}}{\mathcal{S}} 
\left( \prod_{l=1}^{2j} {\mathcal{C}}^{\mu_{l} \rho_{l} \sigma_{l} \alpha_{l}} \right) t_{\alpha_{1} \cdots \alpha_{2j}} \,,
\end{equation}
where we use the symmetrization notation of Sec.~\ref{sec:notation}, and introduced the invariant rank-4 coefficient tensor
\begin{equation}\label{eq:coefficients_reduction_cubic}
{\mathcal{C}}^{\mu \rho \sigma \alpha } = g^{\mu \rho } g^{\sigma \alpha } - g^{\mu \sigma } g^{\rho \alpha } + g^{\mu \alpha } g^{\rho \sigma } + {\rm i} \epsilon^{\mu \rho \sigma \alpha } \,.
\end{equation}
It is clear that this structure exhibits all the properties mentioned in the preceding paragraph.

The  symmetrization on the right-hand side of Eq.~(\ref{eq:t_reduction_cubic}) guarantees the symmetry of the Lorentz indices on the left-hand side of that equation.  
The multiplicative structure of the coefficients, from 1 to $2j$, on the right-hand side is such that the overall structure of the algebra is self similar for any spin. This reflects the recursion properties of Eq.~(\ref{eq:t_recursion}).

The charge conjugated ({\it barred}) expression of Eq.~(\ref{eq:t_reduction_cubic}) is equivalent to
\begin{align}\label{eq:t_reduction_cubicbar}
\bar{t}^{\mu_{1} \cdots \mu_{2j} }  {t}^{\rho_{1} \cdots \rho_{2j} } \bar{t}^{\sigma_{1} \cdots \sigma_{2j} } 
= \frac{1}{[(2j)!]^2}  &  
\underset{\left\{(\rho),(\sigma) \right\}}{\mathcal{S}} 
\left( \prod_{l=1}^{2j} \overbar{\mathcal{C}}^{\mu_{l} \rho_{l} \sigma_{l} \alpha_{l}} \right) \bar{t}_{\alpha_{1} \cdots \alpha_{2j}} \,,
\end{align}
where use was made of Eq.~(\ref{eq:barring_op}), and we introduced $\overbar{\mathcal{C}}^{\mu \rho \sigma \alpha }$ as
\begin{equation}\label{eq:coefficients_reduction_cubic_bar}
\overbar{\mathcal{C}}^{\mu \rho \sigma \alpha } \equiv \left({\mathcal{C}}^{\mu \rho \sigma \alpha }\right)^* = {\mathcal{C}}^{\sigma \rho \mu \alpha } = g^{\mu \rho } g^{\sigma \alpha } - g^{\mu \sigma } g^{\rho \alpha } + g^{\mu \alpha } g^{\rho \sigma } - {\rm i} \epsilon^{\mu \rho \sigma \alpha } \,.
\end{equation}

It is worth stressing what Eq.~(\ref{eq:t_reduction_cubic}) accomplishes. Through its application, matrix (spinor index) multiplication is replaced with multiplication by ordinary C-numbers (invariant tensors). This is the main reason to carry out the reduction of products of $t$-tensors. The possibility of achieving the degree of generality and simplicity of the results and calculations that will be discussed further in the article follows directly from this trade-off.  Worth noting is that for spin $1/2$ the algebra discussed here corresponds to the ``V{\tiny{EE}}'' algebra (space-time algebra, $\sigma^\mu \vee \sigma^\nu$) discussed in Ref.~\cite{SALINGAROS19831}. 

Useful algebraic relations satisfied by the coefficients $\mathcal{C}$ and $\overbar{\mathcal{C}}$ are summarized in App.~\ref{sec:coeff_relations}.
We conclude this section noting that the two alternating cubic products of Eqs.~(\ref{eq:t_reduction_cubic}) and (\ref{eq:t_reduction_cubicbar}) can be written as  
\begin{subequations}
\label{eq:cubic_products}
\begin{align}\label{eq:cubic_products_tbartt}
{t}^{\mu_1 \cdots \mu_{2j}}  \overbar{t}^{\rho_1 \cdots \rho_{2j}} {t}^{\sigma_1 \cdots \sigma_{2j}}
& =  \frac{1}{[(2j)!]^2}  
 \underset{\left\{(\rho),(\sigma) \right\}}{\mathcal{S}} 
 \left[ \operatorname{Re}\left\{ \prod_{l=1}^{2j} {\mathcal{C}}^{\mu_{l} \rho_{l} \sigma_{l} \alpha_{l}} \right\}
 + {\rm i} \operatorname{Im}\left\{ \prod_{l=1}^{2j} {\mathcal{C}}^{\mu_{l} \rho_{l} \sigma_{l} \alpha_{l}} \right\} 
 \right] t_{\alpha_{1} \cdots \alpha_{2j}} \,,
\end{align}    
\begin{align}\label{eq:cubic_products_barttbart}
\overbar{t}^{\mu_1 \cdots \mu_{2j}}  {t}^{\rho_1 \cdots \rho_{2j}} \overbar{t}^{\sigma_1 \cdots \sigma_{2j}} 
 & = \frac{1}{[(2j)!]^2}  
 \underset{\left\{(\rho),(\sigma) \right\}}{\mathcal{S}} 
 \left[ \operatorname{Re}\left\{ \prod_{l=1}^{2j} {\mathcal{C}}^{\mu_{l} \rho_{l} \sigma_{l} \alpha_{l}} \right\}
 - {\rm i} \operatorname{Im}\left\{ \prod_{l=1}^{2j} {\mathcal{C}}^{\mu_{l} \rho_{l} \sigma_{l} \alpha_{l}} \right\} 
 \right] \overbar{t}_{\alpha_{1} \cdots \alpha_{2j}} \,,
\end{align}
\end{subequations}
a result that will be used in Sec.~\ref{sec:dirac_basis}.

\subsection{Quadratic products of \texorpdfstring{$t$}{t}-tensors}  
\label{sec:quadratic_rpoducts_t}

In this section, we show that the quadratic monomials of $t$-tensors exhibit a different and richer structure than the cubic monomials. These structures are in one-to-one correspondence with
the independent on-shell bilinears that can be constructed using generalized spinors. As we shall see in the next section, for any spin representation, the ``traceless'' form of every and each of these structures correspond to 
one covariant $\mathfrak{sl}(2,\mathbb{C})$ multipole.

\subsubsection{The tensor coefficient in the quadratic reduction}
\label{sec:quadratic_coeff}

%As $t^{0\cdots 0} = \bar{t}^{0\cdots 0} = \mathbb{1}$ (see Eq.~(\ref{eq:tzero})), 
The reduction formula for quadratic products follows from a special case of the cubic product reduction formula, where one selects all zero components in one of the $t$-tensors [see Eq.~(\ref{eq:tzero_identity})],
\begin{align}\label{eq:quad_from_cube}
    t^{\mu_1\cdots \mu_{2j}} \bar{t}^{\rho_1\cdots \rho_{2j}} = t^{\mu_1\cdots \mu_{2j}} \bar{t}^{\rho_1\cdots \rho_{2j}} t^{0\cdots 0} = t^{\mu_1\cdots \mu_{2j}} \bar{t}^{\rho_1\cdots \rho_{2j}} t^{\sigma_{1} \cdots \sigma_{2j} } \eta_{\sigma_{1}} \cdots \eta_{\sigma_{2j}} 
    \,.
\end{align}
Here the four-vector $\eta^\mu$ is defined such that\footnote{To simplify the notation, we will use $\bm J\equiv \bm J^{(j)}$  and $\mathbb{1} \equiv \mathbb{1}^{(j)}$ from here on.
}
\begin{align}\label{eq:def_eta}
 t^{\sigma_{1} \cdots \sigma_{2j} } \eta_{\sigma_{1}} \cdots \eta_{\sigma_{2j}} = \mathbb{1} \,,\\
    \eta^\mu = (1,0,0,0)\,.
\end{align}
From Eqs.~(\ref{eq:t_reduction_cubic}) and (\ref{eq:t_reduction_cubicbar})  we find that the reduction of quadratic monomials $t\bar{t}$ and $\bar{t} t$ are given by
% \footnote{See the remarks on symmetrizations following Eq.~(\ref{eq:coefficients_t_reduction_cubic}).} 
\begin{subequations}\label{eq:def_t_reduction_quadratic}
\begin{align} \label{eq:def_t_reduction_quadratic_a}  
t^{\mu_{1} \cdots \mu_{2j} }  \bar{t}^{\rho_{1} \cdots \rho_{2j} } 
= \frac{1}{(2j)!}  & \underset{\left\{(\rho)\right\}}{\mathcal{S}} \left( \prod_{l=1}^{2j} {\mathcal{C}}^{\mu_{l} \rho_{l} \sigma_{l} \alpha_{l}} \eta_{\sigma_{l}} \right) {t}_{\alpha_{1} \cdots \alpha_{2j}} 
= \frac{1}{(2j)!} \underset{\left\{(\rho)\right\}}{\mathcal{S}} \left( \prod_{l=1}^{2j} {\mathcal{Q}}^{\mu_{l} \rho_{l} \alpha_{l}} \right) {t}_{\alpha_{1} \cdots \alpha_{2j}}\,, 
\\
%%%%%%%%%%%%%%%
\bar{t}^{\mu_{1} \cdots \mu_{2j} }  {t}^{\rho_{1} \cdots \rho_{2j} } 
= \frac{1}{(2j)!}  & \underset{\left\{(\rho)\right\}}{\mathcal{S}} \left( \prod_{l=1}^{2j} \overbar{\mathcal{C}}^{\mu_{l} \rho_{l} \sigma_{l} \alpha_{l}} \eta_{\sigma_{l}} \right) \bar{t}_{\alpha_{1} \cdots \alpha_{2j}}
= \frac{1}{(2j)!} \underset{\left\{(\rho)\right\}}{\mathcal{S}} \left(\prod_{l=1}^{2j} \overbar{\mathcal{Q}}^{\mu_{l} \rho_{l} \alpha_{l}}  \right) \bar{t}_{\alpha_{1} \cdots \alpha_{2j}}\,, \label{eq:def_t_reduction_quadratic_b}
\end{align}
\end{subequations}
where from Eqs.~(\ref{eq:coefficients_reduction_cubic}) and (\ref{eq:coefficients_reduction_cubic_bar}) we have
\begin{subequations}\label{eq:coefficients_reduction_quadratic}
\begin{align}
{\mathcal{Q}}^{\mu\rho\alpha}={\mathcal{C}}^{\mu\rho\sigma\alpha}\eta_\sigma & =  g^{\mu \rho } \eta ^{\alpha } - g^{\rho \alpha } \eta^{\mu } + g^{\mu \alpha } \eta^{\rho } + {\rm i} \epsilon^{\mu \rho \sigma \alpha } \eta_{\sigma } \,,
\\
\overbar{\mathcal{Q}}^{\mu\rho\alpha}=\overbar{\mathcal{C}}^{\mu\rho\sigma\alpha}\eta_\sigma & = g^{\mu \rho } \eta ^{\alpha } - g^{\rho \alpha } \eta^{\mu } + g^{\mu \alpha } \eta^{\rho } - {\rm i} \epsilon^{\mu \rho \sigma \alpha } \eta_{\sigma }  = \left( {\mathcal{Q}}^{\mu\rho\alpha} \right)^* \,.
\end{align}    
\end{subequations}
We refer the reader to App.~\ref{sec:coeff_relations}, where useful algebraic relations among $\mathcal{Q}$ and $\overbar{\mathcal{Q}}$ are summarized.

Quadratic reductions without an alternating-bar sequence (such as $tt$ and $\overbar t \overbar t$) do not appear in the bilinear calculus. Rather, they appear in relation with covariant properties of $t$-tensors, like the calculus of the commutation relations with the Lorentz generators, see Sec.~\ref{sec:covariantly_independent_terms_in_quadratic_t_products}.
The non-alternating quadratic reductions can be algebraically obtained as special cases of the cubic ones, Eqs.~(\ref{eq:t_reduction_cubic}) and (\ref{eq:t_reduction_cubicbar})  are given by
\begin{subequations}\label{eq:def_tt_bartbart_reduction_quadratic}
\begin{align} \label{eq:def_tt_reduction_quadratic}  
t^{\mu_{1} \cdots \mu_{2j} }  {t}^{\sigma_{1} \cdots \sigma_{2j} } 
= \frac{1}{(2j)!}  & \underset{\left\{(\sigma)\right\}}{\mathcal{S}} \left( \prod_{l=1}^{2j} {\mathcal{C}}^{\mu_{l} \rho_{l} \sigma_{l} \alpha_{l}} \eta_{\rho_{l}} \right) {t}_{\alpha_{1} \cdots \alpha_{2j}} 
= \frac{1}{(2j)!} \underset{\left\{(\sigma)\right\}}{\mathcal{S}} \left( \prod_{l=1}^{2j} {\mathcal{R}}^{\mu_{l} \sigma_{l} \alpha_{l}} \right) {t}_{\alpha_{1} \cdots \alpha_{2j}}\,, 
\\
%%%%%%%%%%%%%%%
\overbar{t}^{\mu_{1} \cdots \mu_{2j} }  \overbar{t}^{\sigma_{1} \cdots \sigma_{2j} } 
= \frac{1}{(2j)!}  & \underset{\left\{(\sigma)\right\}}{\mathcal{S}} \left( \prod_{l=1}^{2j} \overbar{\mathcal{C}}^{\mu_{l} \rho_{l} \sigma_{l} \alpha_{l}} \eta_{\rho_{l}} \right) \bar{t}_{\alpha_{1} \cdots \alpha_{2j}}
= \frac{1}{(2j)!} \underset{\left\{(\sigma)\right\}}{\mathcal{S}} \left(\prod_{l=1}^{2j} \overbar{\mathcal{R}}^{\mu_{l} \sigma_{l} \alpha_{l}}  \right) \bar{t}_{\alpha_{1} \cdots \alpha_{2j}}\,, \label{eq:def_bartbart_reduction_quadratic}
\end{align}
\end{subequations}
where 
%from Eqs.~(\ref{eq:coefficients_reduction_cubic}) and (\ref{eq:coefficients_reduction_cubic_bar}) we have
\begin{subequations}\label{eq:R_coefficients_reduction_quadratic}
\begin{align}
{\mathcal{R}}^{\mu\sigma\alpha}={\mathcal{C}}^{\mu\rho\sigma\alpha}\eta_\rho & =  \eta^{\mu} g^{\rho \alpha} - \eta^{\alpha} g^{\mu \rho } + \eta^{\sigma} g^{\mu \alpha } + {\rm i} \epsilon^{\mu \rho \sigma \alpha } \eta_{\rho} \,,
\\
\overbar{\mathcal{R}}^{\mu\sigma\alpha}=\overbar{\mathcal{C}}^{\mu\rho\sigma\alpha}\eta_\rho & = \eta^{\mu} g^{\rho \alpha} - \eta^{\alpha} g^{\mu \rho } + \eta^{\sigma} g^{\mu \alpha } - {\rm i} \epsilon^{\mu \rho \sigma \alpha } \eta_{\rho}  = \left( {\mathcal{R}}^{\mu\sigma\alpha} \right)^* \,.
\end{align}    
\end{subequations}
Note that compared to $\mathcal{Q}^{\mu\rho\alpha}$, the contraction between $\mathcal{C}^{\mu\rho\sigma\alpha}$ and $\eta$ happens in a different index.
We refer the reader to App.~\ref{sec:coeff_relations}, where useful
algebraic relations among $\mathcal{Q}$ and $\overbar{\mathcal{Q}}$, as well as $\mathcal{R}$ and $\overbar{\mathcal{R}}$ are summarized.
% ; and to App.~\ref{sec:generalized_quadratic_reductions_and_CR} where some specific cases are shown that are needed for later calculations.

One observes that due to the presence of the four-vector $\eta$, the coefficients in these reductions are not Lorentz invariant (as they were for the cubic reduction), but only \emph{rotationally} invariant. 
It is worth stressing that the coefficients in all the reduction formulas are Lorentz tensors, {\rm i.e.} they transform covariantly.

\subsubsection{Quadratic products are reducible}
\label{sec:quadratic_t_products_reducible}

 If we reinstate spinor indices in the quadratic product, we have $\tensor{(t\bar t)}{_{c}^{d}}$, which shows that the quadratic product is a reducible intertwining map 
\begin{equation}\label{eq:sl2c_quad}
    (j,0)\otimes (j,0) = \overset{2j}{\underset{m=0}{\oplus}}(m,0)
    \,,
\end{equation}
between spinor and Lorentz tensor representations.
Lorentz tensors in the $(m,0)$ [integer $m$] representation are rank $2m$ anti-self-dual\footnote{For  example, see Eq.~(\ref{eq:dual_multipoles_gamma}). } tensors that have a so-called bi-index structure in their indices.  The $2m$ indices are grouped in $m$ pairs 
% of two 
[$(\mu_k \rho_k), k=1,\cdots,m$], that have the following mixed permutation symmetries:
\begin{enumerate}
    \item The tensor is anti-symmetric under interchange of indices forming a \emph{pair} (with the same numeric label in the case shown here)
\begin{subequations}
\label{eq:bi_index}
\begin{equation}
\mathcal{T}^{\cdots, \mu_k \rho_k , \cdots} = - \mathcal{T}^{\cdots, \rho_k \mu_k, \cdots} \,,
\end{equation}
\item The tensor is symmetric under interchange of \emph{pairs} of indices (with different numeric label in the case shown here) 
\begin{equation}
\mathcal{T}^{\cdots, \mu_k \rho_k, \cdots, \mu_l \rho_l , \cdots} = \mathcal{T}^{\cdots, \mu_l \rho_l , \cdots, \mu_k \rho_k, \cdots} \,.
\end{equation}    
\end{subequations}
\end{enumerate}
Similarly, for $\bar t t$ we  have that it intertwines between $(0,j) \otimes (0,j) = \overset{2j}{\underset{m=0}{\oplus}}(0,m)$ spinor and Lorentz tensor representations, where the $(0,m)$ Lorentz tensors are self-dual rank $2m$ tensors with the same bi-index structure.

The next steps help us to identify the components in this quadratic product that transform according to the $2j+1$ irreducible representations listed on the right.  The results of this section play a central role in the construction of a basis for the generalized gamma matrices.  A similar reduction has been considered by Williams in Ref.~\cite{Williams:1965rga}. While Williams employ linear combinations in spinor indices, we use the structure of the algebra in its space-time indices, which due to the presence of the invariant tensors produces more intuitive results.  Since the reduction is unique, both approaches are equivalent.

The powers of ${\mathcal{Q}}^{\mu\rho\alpha}$ and $\overbar{\mathcal{Q}}^{\mu\rho\alpha}$ tensors
% (\ref{eq:coefficients_reduction_quadratic})
in Eqs.~(\ref{eq:def_t_reduction_quadratic}) can be expanded, and the quadratic products can be written as
\begin{subequations}\label{eq:quadratic_products}
\begin{align}
\label{eq:quadratic_products_tbart}
{t}^{\mu_1 \cdots \mu_{2j}}  \bar{t}^{\rho_1 \cdots \rho_{2j}} & = \frac{1}{(2j)!}  \underset{\left\{(\rho) \right\}}{\mathcal{S}}  \sum_{m=0}^{2j}  \left[ 
%B^{2j}_{m}
\binom{2j}{m}
\frac{1}{(2j)!} \underset{\left\{(\mu \rho)\right\}}{\mathcal{S}} \left( \prod_{l=1}^{m} {\mathcal{Q}}_{\text{red}}^{\mu_{l} \rho_{l} \alpha_{l}} \  \prod_{k=m+1}^{2j} g^{\mu_k \rho_k} \eta^{\alpha_k}  \right)
\right]
 t_{\alpha_{1} \cdots \alpha_{2j}} 
 \,,
\end{align}
\begin{align}
\label{eq:quadratic_products_bartt}
\bar{t}^{\mu_1 \cdots \mu_{2j}} {t}^{\rho_1 \cdots \rho_{2j}} & = \frac{1}{(2j)!}  \underset{\left\{(\rho) \right\}}{\mathcal{S}}  \sum_{m=0}^{2j}  \left[  
%B^{2j}_{m}
\binom{2j}{m}
\frac{1}{(2j)!} \underset{\left\{(\mu \rho) \right\}}{\mathcal{S}} \left( \prod_{l=1}^{m} \overbar{\mathcal{Q}}_{\text{red}}^{\mu_{l} \rho_{l} \alpha_{l}} \  \prod_{k=m+1}^{2j} g^{\mu_k \rho_k} \eta^{\alpha_k}  \right)
\right] 
 \bar{t}_{\alpha_{1} \cdots \alpha_{2j}}
  \,,
\end{align}
\end{subequations}
where we use the notation introduced in Appendix~\ref{sec:notation}.
We emphasize that symmetrizations ${\mathcal{S}}_{\left\{(\mu\rho)\right\}}$
are done treating the pairs of indices $\mu_{r}\rho_{r}$ as one entity, as the symmetry property is inherited from that of $t_{\alpha_{1} \cdots \alpha_{m} 0 \cdots 0}$.

The reduced tensors are explicitly
\begin{subequations}\label{eq:quadratic_coefficients_Qred}
\begin{align}
\label{eq:properties_coefficients_Q_a}
{\mathcal{Q}}_{\text{red}}^{\mu \rho \alpha } & =  - g^{\rho \alpha } \eta^{\mu } + g^{\mu \alpha } \eta^{\rho } + {\rm i}\epsilon^{\mu \rho \sigma \alpha } \eta_{\sigma } 
\equiv \mathcal{C}_{\text{red}}^{\mu \rho \sigma \alpha } \eta_{\sigma}
%& \equiv \mathcal{C}^{\mu \rho \sigma \alpha } \eta_{\sigma} -  g^{\mu \rho } \eta ^{\alpha } 
\,,
\\
\label{eq:properties_coefficients_Q_b}
\overbar{\mathcal{Q}}_{\text{red}}^{\mu \rho \alpha } = \left({\mathcal{Q}}_{\text{red}}^{\mu \rho \alpha }\right){}^* &  =  - g^{\rho \alpha } \eta^{\mu } + g^{\mu \alpha } \eta^{\rho } - {\rm i} \epsilon^{\mu \rho \sigma \alpha } \eta_{\sigma } 
\equiv {\overbar{\mathcal{C}}}_{\text{red}}^{\mu \rho \sigma \alpha } \eta_{\sigma}
%\equiv {\overbar{\mathcal{C}}}^{\mu \rho \sigma \alpha } \eta_{\sigma} -  g^{\mu \rho } \eta ^{\alpha } 
\,,
\end{align}    
\end{subequations}
where
\begin{subequations}\label{eq:cubic_coefficients_Cred}
\begin{align}
\label{eq:cubic_coefficients_Cred_a}
{{\mathcal{C}}}_{\text{red}}^{\mu \rho \nu \sigma  } =  - g^{\mu \nu} g^{\rho \sigma}  + g^{\mu \sigma} g^{\rho \nu} + {\rm i} \epsilon^{\mu \rho \nu \sigma} 
\,,
\\
\label{eq:cubic_coefficients_Cred_b}
{\overbar{\mathcal{C}}}_{\text{red}}^{\mu \rho \nu \sigma  } =  - g^{\mu \nu} g^{\rho \sigma}  + g^{\mu \sigma} g^{\rho \nu} - {\rm i} \epsilon^{\mu \rho \nu \sigma} 
\,.
\end{align}    
\end{subequations}
For a summary on the algebraic relations among $\mathcal{Q}_\text{red}$, $\overbar{\mathcal{Q}}_\text{red}$, $\mathcal{C}_\text{red}$ and $\overbar{\mathcal{C}}_\text{red}$, we refer the reader to App.~\ref{sec:coeff_relations}.
% 
% We will explicitly show results for ${t}^{\mu_1 \cdots \mu_{2j}}  \bar{t}^{\rho_1 \cdots \rho_{2j}}$, Eq.~(\ref{eq:quadratic_products_tbart}). The equivalent expressions for $\bar{t}^{\mu_1 \cdots \mu_{2j}} {t}^{\rho_1 \cdots \rho_{2j}}$ can be obtained by applying the \emph{barring} operation to the results shown here, see Eq.~(\ref{eq:barring_op}). 

\subsubsection{Covariantly independent terms in the quadratic products } 
\label{sec:covariantly_independent_terms_in_quadratic_t_products}

In this section, we prove the covariant independence of each and every term in the sum of Eq.~(\ref{eq:quadratic_products}). Explicitly, we shall show that for every $m$, 
\begin{subequations}\label{eq:cov_independent_substructures_T_barT}
\begin{align}\label{eq:cov_independent_substructures_T}
    &\ProdQt_m^{(\mu\rho)} \equiv  \prod_{l=1}^{m} {\mathcal{Q}}_{\text{red}}^{\mu_{l} \rho_{l} \alpha_{l}} \prod_{r=m+1}^{2j} \eta^{\alpha_r} t_{\alpha_{1} \cdots \alpha_{2j}} =\prod_{l=1}^{m} {\mathcal{Q}}_{\text{red}}^{\mu_{l} \rho_{l} \alpha_{l}} t_{\alpha_{1} \cdots \alpha_{m} 0 \cdots 0} \,, \qquad  (0 \leq m \leq 2j) \,,\\
    %%%%%
    &\overbar{\ProdQt}_m^{(\mu\rho)} \equiv  \prod_{l=1}^{m} {\overbar{\mathcal{Q}}}_{\text{red}}^{\mu_{l} \rho_{l} \alpha_{l}} \prod_{r=m+1}^{2j} \eta^{\alpha_r} \bar t_{\alpha_{1} \cdots \alpha_{2j}} =\prod_{l=1}^{m} {\overbar{\mathcal{Q}}}_{\text{red}}^{\mu_{l} \rho_{l} \alpha_{l}} t_{\alpha_{1} \cdots \alpha_{m} 0 \cdots 0} \,,
\end{align}    
\end{subequations}
transforms under a Lorentz transformation as a covariant independent tensor.  From the anti- and self-dual nature of $\mathcal{Q}_\text{red}^{\mu\rho\alpha}$ and $ \overbar{\mathcal{Q}}_\text{red}^{\mu\rho\alpha}$ [see Eqs.~(\ref{eq:Qred_antiselfdual})], we can infer that the $\ProdQt_m^{(\mu\rho)}$, $\overbar{\ProdQt}_m^{(\mu\rho)}$ are anti-self-dual, respectively self-dual in each pair of $\mu_i\rho_i$ indices.
 
We start by writing the relation between the $t$-tensors and the spin-$j$ chiral left- and right-representations of the Lorentz generators of $\mathfrak{sl}(2,\mathbb{C})$, $\mathbb{M}^{\mu\rho}_{(j)}$ and $\overbar{\mathbb{M}}^{\mu\rho}_{(j)}$).  By explicit calculation one obtains\footnote{The explicit form of $t^{a 0\cdots 0}$ ($a=1,2,3$) follows either from constructing these components using Eq.~(\ref{eq:t_matrix_value}), or from the algorithm in App.~\ref{sec:t_algorithm} [see Eq.~(\ref{eq:t_J})].} \begin{subequations}\label{eq:sl(2,C)_generators_chiral}
\begin{align}
\label{eq:sl(2,C)_generators_left_chiral}
{\mathbb{M}}^{\mu \rho}_{(j)} & = 
\left(\begin{array}{cccc}
0 & -{\rm i} J_1 & -{\rm i} J_2 & -{\rm i} J_3 \\
{\rm i}J_1 & 0 & J_3 & - J_2 \\
{\rm i}J_2 & - J_3 & 0 & J_1 \\
{\rm i}J_3 & J_2 & - J_1 & 0
\end{array}\right) = ({\rm i}\, j) {\mathcal{Q}}_{\text{red}}^{\mu \rho \alpha_1} \prod_{s=2}^{2j} \eta^{\alpha_s}  {t}_{\alpha_{1} \cdots \alpha_{2j}} = ({\rm i}\, j) {\ProdQt}_1^{(\mu\rho)}
\,,  \\[0.5em]
\label{eq:sl(2,C)_generators_right_chiral}
\overbar{\mathbb{M}}^{\mu \rho}_{(j)} & = 
\left(\begin{array}{cccc}
0 & {\rm i} J_1 & {\rm i} J_2 & {\rm i} J_3 \\
-{\rm i}J_1 & 0 & J_3 & - J_2 \\
-{\rm i}J_2 & - J_3 & 0 & J_1 \\
-{\rm i}J_3 & J_2 & - J_1 & 0
\end{array}\right) = ({\rm i}\, j) \overbar{\mathcal{Q}}_{\text{red}}^{\mu \rho \alpha_1} \prod_{s=2}^{2j} \eta^{\alpha_s}  \bar{t}_{\alpha_{1} \cdots \alpha_{2j}} = ({\rm i}\, j) \overbar{\ProdQt}_1^{(\mu\rho)}
\,,
\end{align} 
\end{subequations}
where, ${\mathbb{M}}^{\mu \rho}$ and $\overbar{\mathbb{M}}^{\mu \rho}$ both obey the Lorentz algebra
\begin{subequations}\label{eq:sl(2,C)_generators_CR}
\begin{align}
\left[ \mathbb{M}^{\mu \rho} , \mathbb{M}^{\nu \lambda} \right] & =  {\rm i} \left( g^{\mu \nu} \mathbb{M}^{\rho \lambda} - g^{\rho \nu} \mathbb{M}^{\mu \lambda} + g^{\mu \lambda} \mathbb{M}^{\rho \nu} - g^{\rho \lambda} \mathbb{M}^{\mu \nu}  \right)  
\,, \\
\left[ \overbar{\mathbb{M}}^{\mu \rho} , \overbar{\mathbb{M}}^{\nu \lambda} \right] & = {\rm i} \left( g^{\mu \nu} \overbar{\mathbb{M}}^{\rho \lambda} - g^{\rho \nu} \overbar{\mathbb{M}}^{\mu \lambda} + g^{\mu \lambda} \overbar{\mathbb{M}}^{\rho \nu} - g^{\rho \lambda} \overbar{\mathbb{M}}^{\mu \nu}  \right) 
\,,
\end{align} 
\end{subequations}
and $J_i$ ($1\leq i \leq 3$) are the spin-$j$ representations of the three generators of rotations.  Here and in the following, we dropped the explicit reference to the spin representation [label $(j)$]  on $\mathbb{M}$ and $J_i$.

The transformation properties of the $\ProdQt_m^{(\mu\rho)}$ under the Lorentz group can be deduced from the commutators with the group generators, i.e. the $\mathbb{M}^{\mu\nu}$.
Using Eqs.~(\ref{eq:CR_for_tt}) and (\ref{eq:algebra_quadratic_coeff_b}), we find 
\begin{align}\label{eq:CR_M_Tm}
{\left[{\mathbb{M}}^{\nu \sigma}, \ProdQt_m^{(\mu \rho)}\right] = {\rm i} \sum_{n=1}^m \left( \right.} & g^{\mu_n \sigma}  \ProdQt_m^{\mu_1 \rho_1, \ldots, \mu_{n-1} \rho_{n-1}, \nu \rho_n, \mu_{n+1} \rho_{n+1}, \ldots, \mu_m \rho_m} 
\nonumber\\
& 
-g^{\mu_n \nu} \ProdQt_m^{\mu_1 \rho_1, \ldots, \mu_{n-1} \rho_{n-1}, \sigma \rho_n, \mu_{n+1} \rho_{n+1}, \ldots, \mu_m \rho_m} 
\nonumber\\
& +g^{\rho_n \sigma} \ProdQt_m^{\mu_1 \rho_1, \ldots, \mu_{n-1} \rho_{n-1}, \mu_{n} \nu, \mu_{n+1} \rho_{n+1}, \ldots, \mu_m \rho_m} 
\nonumber\\
& \left. 
-g^{\rho_n \nu} \ProdQt_m^{\mu_1 \rho_1, \ldots, \mu_{n-1} \rho_{n-1}, \mu_{n} \sigma, \mu_{n+1} \rho_{n+1}, \ldots, \mu_m \rho_m}  \right) \,,
\end{align}
which shows that the $\ProdQt_m^{(\mu\rho)}$ transform within themselves under Lorentz transformations. Thus, they are independent tensors that carry the bi-index structure discussed in Sec.~\ref{sec:quadratic_t_products_reducible}. 
Therefore, Eq.~(\ref{eq:quadratic_products_tbart}) can be written as the following sum of $2j$ covariantly independent bi-indexed tensors
\begin{equation}\label{eq:quadratic_products_tbart_with_T}
{t}^{\mu_1 \cdots \mu_{2j}}  \bar{t}^{\rho_1 \cdots \rho_{2j}} = \frac{1}{(2j)!} \underset{\left\{(\rho) \right\}}{\mathcal{S}}  \sum_{m=0}^{2j} \left[ \binom{2j}{m}
\frac{1}{(2j)!} \underset{\left\{(\mu \rho)\right\}}{\mathcal{S}} \left( \ProdQt_m^{(\mu\rho)} \prod_{k=m+1}^{2j} g^{\mu_k \rho_k} \right)
\right]
\,.
\end{equation}
% and similarly for Eq.~(\ref{eq:quadratic_products_bartt}).
% 
We have explicitly shown the results for ${t}^{\mu_1 \cdots \mu_{2j}}  \bar{t}^{\rho_1 \cdots \rho_{2j}}$, Eq.~(\ref{eq:quadratic_products_tbart}). The equivalent expressions for $\bar{t}^{\mu_1 \cdots \mu_{2j}} {t}^{\rho_1 \cdots \rho_{2j}}$ [Eq.~(\ref{eq:quadratic_products_bartt})] can be obtained by applying the \emph{barring} operation to the outcomes shown here, see Eq.~(\ref{eq:barring_op}).

\subsubsection{Orthogonalization of the covariantly independent terms}%in the quadratic products } 
\label{sec:orthogonal_basis_quadratic_t}

In anticipation of Sec.~\ref{sec:dirac_basis}, where we enumerate a complete basis of generalized Dirac matrices, we construct an orthogonal basis out of  the covariantly independent tensors $\ProdQt_m^{(\mu\rho)}$ introduced in the previous section. After this orthogonalization, the basis elements correspond to irreducible representations of the Lorentz group.  

In  Sec.~\ref{sec:t_tensors_gen}, it was shown that the $t$-tensors contain a basis for $(2j+1) \times (2j+1)$ Hermitian matrices, 
% of order $2j+1$, 
{\emph{i.e.}}, a complete basis for $\mathfrak{u}(2j+1)$. In the $\ProdQt_m^{(\mu\rho)}$, the contractions of the $t$-tensor with the tensors ${\mathcal{Q}}_{\text{red}}^{\mu \rho \alpha}$ account for a convenient repackaging of the matrices in the $t$-tensors.  The explicit expressions of the ${\mathcal{Q}}_{\text{red}}^{\mu \rho \alpha}$,
\begin{align}
\label{eq:Qred_explicit} 
& {\mathcal{Q}}_{\text{red}}^{\mu \rho  0}  = 
\left(
\begin{array}{cccc}
 0 & 0 & 0 & 0 \\
 0 & 0 & 0 & 0 \\
 0 & 0 & 0 & 0 \\
 0 & 0 & 0 & 0 \\
\end{array}
\right) 
\,, \
{\mathcal{Q}}_{\text{red}}^{\mu \rho 1} = 
\left(
\begin{array}{cccc}
 0 & 1 & 0 & 0 \\
 -1 & 0 & 0 & 0 \\
 0 & 0 & 0 & \rm i \\
 0 & 0 & - \rm i & 0 \\
\end{array}
\right) 
\,, 
% \nonumber \\
% & 
\ {\mathcal{Q}}_{\text{red}}^{\mu \rho 2} = 
\left(
\begin{array}{cccc}
 0 & 0 & 1 & 0 \\
 0 & 0 & 0 & - \rm i \\
 -1 & 0 & 0 & 0 \\
 0 & \rm i & 0 & 0 \\
\end{array}
\right) 
\,, \
{\mathcal{Q}}_{\text{red}}^{\mu \rho 3} = 
\left(
\begin{array}{cccc}
 0 & 0 & 0 & 1 \\
 0 & 0 & \rm i & 0 \\
 0 & -\rm i & 0 & 0 \\
 -1 & 0 & 0 & 0 \\
\end{array}
\right) \,,
\end{align}  
show that each $\mu\rho$ (antisymmetric) index pair only has support for one specific value of $\alpha$, where for each $\alpha$-value one real ($\pm 1$) and one imaginary ($\pm \text{i}$) independent element appears (up to antisymmetrization in $(\mu\rho))$.  Consequently, the covariant tensors $\ProdQt_m^{(\mu\rho)}$ contain \emph{all} independent Hermitian matrices in $\operatorname{Re}\left\{\prod_{l=1}^{m} {\mathcal{Q}}_{\text{red}}^{\mu_{l} \rho_{l} \alpha_{l}}\right\} t_{\alpha_{1} \cdots \alpha_{m} 0 \cdots 0}$, and all the anti-Hermitian ones in ${\rm i}\operatorname{Im}\left\{\prod_{l=1}^{m} {\mathcal{Q}}_{\text{red}}^{\mu_{l} \rho_{l} \alpha_{l}}\right\} t_{\alpha_{1} \cdots \alpha_{m} 0 \cdots 0}$. See Eqs.~(\ref{eq:sl(2,C)_generators_chiral}) for the explicit case $m=1$, where both the Hermitian $J_i$ and anti-Hermitian $\text{i}J_i$ appear.
In other words, the set of elements $\ProdQt_m^{(\mu\rho)}$ ($0 \leq m \leq 2j$) contains a basis for $(2j+1) \times (2j+1)$ complex matrices
% all square complex matrices of order $2j+1$ 
when considered as a vector space over $\mathbb{R}$\footnote{Considered as a vector space over $\mathbb{C}$, the set of Hermitian (or anti-Hermitian) matrices by itself is already a complete basis for complex matrices.}. 
We will show in Sec.~\ref{sec:quadratic_t_vs_multipoles} that the orthogonal basis of covariant tensors that we construct below corresponds (up to a numerical factor) to the basis of covariant multipoles of $\mathfrak{sl}(2,\mathbb{C})$, {\emph i.e.} is a basis for any operator acting on chiral spin-$j$ spinors.  

% To single out the content of $\ProdQt_m^{(\mu\rho)}$
In order to identify within the set $\ProdQt_m^{(\mu\rho)}$ ($1 \leq m \leq 2j$) the objects 
that truly transform as the irreducible Lorentz group representations, we remove from every term ($\ProdQt_m^{(\mu\rho)}$) the content of lower rank elements of the basis set. 
To identify this lower rank content, we study the elimination of  covariant traces, {\emph i.e.}  we constrain the tensor to vanish when a pair of its space-time indices is contracted. 
The bi-indexed symmetry of Eq.~(\ref{eq:bi_index}) allows us to choose any two indices from two different bi-index pairs. 
% The result of 
Taking the simplest non-trivial traces of $\ProdQt_m^{(\mu\rho)}$ gives the result
\begin{subequations}
\begin{align}\label{eq:g_trace_T_1}
\ProdQt_{m}^{(\mu\rho)} g_{\mu_{m-1} \mu_{m}}  & = - \ProdQt_{m-2}^{(\mu\rho)} \, g^{\rho_{m-1} \rho_{m}}  \,,\\
\label{eq:g_trace_T_2}
\ProdQt_{m}^{(\mu\rho)} g_{\mu_{m-1} \rho_{m}}  & =  \ProdQt_{m-2}^{(\mu\rho)} \, g^{\mu_{m} \rho_{m-1}} \,.
\end{align}    
\end{subequations}
Henceforth, this results leads us to consider the projection of $\ProdQt_{m}^{(\mu\rho)}$ onto an invariant fourth-rank subspace given by 
\begin{equation}
\operatorname{Re}\left\{{\mathcal{C}}_{\text{red}}^{\mu_{m-1} \rho_{m-1} \mu_{m} \rho_{m}}\right\} = - g^{\mu_{m-1} \mu_{m}}g^{\rho_{m-1} \rho_{m}} + g^{\mu_{m-1} \rho_{m}}g^{\mu_{m} \rho_{m-1}}
\,,
\end{equation}
which is the real part of the ${\mathcal{C}}_{\text{red}}$-tensor and exhibits the same bi-index symmetry in its indices as the $\ProdQt_m^{(\mu\rho)}$.  
For completeness, we must also consider the projection given by the Levi-Civita tensor (${\rm i} \epsilon^{\mu_{m-1} \rho_{m-1} \mu_{m} \rho_{m}}$), which is the only other independent fourth-rank invariant satisfying the same bi-indexed symmetry and appears as the imaginary part of the ${\mathcal{C}}_{\text{red}}$-tensor.
Contractions between these two invariant tensors and the $\ProdQt_m^{(\mu\rho)}$ and between themselves evaluate to
\begin{subequations}\label{eq:Cred_trace_T}
\begin{align}\label{eq:Real_Cred_trace_T}
\operatorname{Re}\left\{{\mathcal{C}}^{\text{red}}_{\mu_{m-1} \rho_{m-1} \mu_{m} \rho_{m}}\right\} \, \ProdQt_{m}^{(\mu\rho)}   & = 8 \ProdQt_{m-2}^{(\mu\rho)}   \,, \\
\label{eq:Im_Cred_trace_T}
%{\rm i} \epsilon_{\mu_{m-1} \rho_{m-1} \mu_{m} \rho_{m}}
\text{i}\operatorname{Im}\left\{{\mathcal{C}}^{\text{red}}_{\mu_{m-1} \rho_{m-1} \mu_{m} \rho_{m}}\right\}\, \ProdQt_{m}^{(\mu\rho)}  & =  8 \ProdQt_{m-2}^{(\mu\rho)}  \,, \\
\operatorname{Re}\left\{{\mathcal{C}}^{\text{red}}_{\mu_{1} \rho_{1} \mu_{2} \rho_{2}} \right\} 
\operatorname{Re}\left\{{\mathcal{C}}^{\text{red},\mu_{1} \rho_{1} \mu_{2} \rho_{2}}\right\}  &= 24\,,\\
\left(\rm i \operatorname{Im}\left\{{\mathcal{C}}_{\text{red}}^{\mu_{1} \rho_{1} \mu_{2} \rho_{2}}\right\}\right)\left(\rm i\operatorname{Im}\left\{{\mathcal{C}}^{\text{red},\mu_{1} \rho_{1} \mu_{2} \rho_{2}}\right\}\right)&=24\,,\\
\label{eq:ReCred_ImCred}
\operatorname{Re}\left\{{\mathcal{C}}^{\text{red}}_{\mu_{1} \rho_{1} \mu_{2} \rho_{2}}\right\} \operatorname{Im}\left\{{\mathcal{C}}_{\text{red}}^{\mu_{1} \rho_{1} \mu_{2} \rho_{2}}\right\} & = 0  \,.
\end{align}    
\end{subequations}
Thus, the next lower rank element of the basis contained in $\ProdQt_{m}^{(\mu\rho)}$ is $\ProdQt_{m-2}^{(\mu\rho)}$.  In order to remove this contribution we can combine Eqs.~(\ref{eq:Cred_trace_T}) by using contractions with any linear combination of $\operatorname{Re}\left\{{\mathcal{C}}^{\text{red}}\right\}$ and $\operatorname{Im}\left\{{\mathcal{C}}^{\text{red}}\right\}$ (with two non-zero coefficients).   The final result, however, will not depend on the exact choice of linear combination. We opt to use the simplest combination ${\mathcal{C}}_{\text{red}}$ as the coefficient 8 in both Eqs.~(\ref{eq:Real_Cred_trace_T}) and (\ref{eq:Im_Cred_trace_T}) show that the subtraction can be written as proportional to $\mathcal{C}_\text{red}$:
\begin{equation}
\label{eq:tildeT}
\widetilde{\ProdQt}_{m}^{(\mu\rho)}  = {\ProdQt}_{m}^{(\mu\rho)} - \frac{1}{2m-1} \frac{1}{(m-2)! 2!!} \underset{\left\{(\mu\rho) \right\}}{\mathcal{S}} \ProdQt_{m-2}^{(\mu\rho)}  \, {\mathcal{C}}_{\text{red}}^{\mu_{m-1} \rho_{m-1} \mu_{m} \rho_{m}} \,, 
\end{equation}
where the coefficient in front of the second term is such that ${\mathcal{C}}^{\text{red}}_{\mu_{a} \rho_{a} \mu_{b} \rho_{b}} \widetilde{\ProdQt}_{m}^{(\mu\rho)}$ ($1 \leq a,b \leq m$ and $\ a \neq b$) does not contain any $\ProdQt_{m-2}^{(\mu\rho)}$ term.  Contractions with the ${\mathcal{C}}_{\text{red}}$-tensor treats the pairs of indices $\mu_a\rho_a$ ($\mu_b\rho_b$) as one entity, and ${\mathcal{C}}_{\text{red}}^{\mu_a\rho_a\mu_b\rho_b}$ is symmetric under exchanges of these pair of indices. In other words, it resembles a metric that realizes the bi-index symmetry. For this reason, we refer to the contraction with ${\mathcal{C}}_{\text{red}}$-tensor as taking a ${\mathcal{C}}_{\text{red}}$-\emph{trace}. Similarly, the operation of removing a lower rank  ${\ProdQt}_{n}^{(\mu\rho)}$-tensor ($n < m$), like in Eq.~(\ref{eq:tildeT}), we refer to it as removing a ${\mathcal{C}}_{\text{red}}$-\emph{trace}. 

Although $\widetilde{\ProdQt}_{m}^{(\mu\rho)}$ does not contain $\ProdQt_{m-2}^{(\mu\rho)}$, it does contain $\ProdQt_{m-4}^{(\mu\rho)}$, which can be seen by explicit calculation 
\begin{equation}
\label{eq:Cred_trace_tildeT}
{\mathcal{C}}^{\text{red}}_{\mu_{m-1} \rho_{m-1} \mu_{m} \rho_{m}} \, \widetilde{\ProdQt}_{m}^{(\mu\rho)}  =   \frac{1}{(2m-1)(2m-3)} \frac{1}{(m-4)! (4)!!} \underset{\left\{(\mu\rho) \right\}}{\mathcal{S}} \ProdQt_{m-4}^{(\mu\rho)}  \, {\mathcal{C}}_{\text{red}}^{\mu_{m-3} \rho_{m-3} \mu_{m-2} \rho_{m-2}}  \,.   
\end{equation}
The term proportional to $\ProdQt_{m-4}^{(\mu\rho)}$ appearing here, results from contracting ${\mathcal{C}}^{\text{red}}_{\mu_{m-1} \rho_{m-1} \mu_{m} \rho_{m}}$ with the second term on the right-hand side of Eq.~(\ref{eq:tildeT}). 
This procedure can be iterated to remove all contributions from lower-$m$ terms, where it is clear the jump in $m$-values happens in intervals of 2.  

To summarize this construction, the complete orthogonal basis, for which we use the notation $\mathcal{T}_{m}^{(\mu\rho)}$, is obtained as follows:
\begin{enumerate}

\item As the ${\mathcal{C}}_{\text{red}}$-trace removals lead to subtractions with $m-2,m-4,\ldots$, two independent towers of tensors appear. The first is labeled with even integer $m$ and starts with 
\begin{equation}
\label{eq:Tau_0}
\mathcal{T}_{0} =  t_{0 \cdots 0} =  \mathbb{1} \,.   
\end{equation}
The second is labeled with odd integer $m$ and starts with 
\begin{equation}
\label{eq:Tau_1}
\mathcal{T}_{1}^{\mu\rho} =  {\ProdQt}_{1}^{\mu\rho} = {\mathcal{Q}}_{\text{red}}^{\mu \rho \alpha}  t_{\alpha 0 \cdots 0} \,.   
\end{equation}

\item After removing all the lower rank tensor
% order multipole 
contributions, the $m$-th term of the orthogonal basis can be written as
\begin{align}
\label{eq:Tau_m}
\mathcal{T}_{m}^{(\mu\rho)} = & {\ProdQt{}}_{m}^{(\mu\rho)} + \sum_{n=1}^{\lfloor \frac{m}{2} \rfloor} (-)^{n} \frac{(2m-2n-1)!!}{(2m-1)!!} \frac{1}{(m-2n)! (2n)!!} 
% \nonumber \\
% & \hspace{2cm}
% \times 
\underset{\left\{(\mu\rho) \right\}}{\mathcal{S}} \ProdQt{}_{m-2n}^{(\mu\rho)} \prod_{l = m-2n+1, m-2n+3, \cdots}^{m-1} {\mathcal{C}}_{\text{red}}^{\mu_{l} \rho_{l} \mu_{l+1} \rho_{l+1}}
  \,,   
\end{align}
where, the floor function ${\lfloor \frac{m}{2} \rfloor}$ represents the nearest integer that is less than or equal to $\frac{m}{2}$.
For all $m \geq 2$ these satisfy
\begin{equation}
\label{eq:Cred_Tau_m}
{\mathcal{C}}^{\text{red}}_{\mu_{a} \rho_{a} \mu_{b} \rho_{b}} \mathcal{T}_{m}^{\cdots \mu_a\rho_a \cdots \mu_b\rho_b \cdots} = 0  \,, \quad 1 \leq a,b \leq m , \,  \text{with} \ a \neq b \,.   
\end{equation}

\item If $m$ is even all the subtracted terms contain ${\ProdQt}_{n}^{(\mu\rho)}$ with even $n$ ($< m$). The last subtractions end when we reach 
\begin{equation}
\label{eq:trace_T0}
(-)^{\lfloor \frac{m}{2} \rfloor}  \frac{(m-1)!!}{(2m-1)!!}  \frac{1}{(m)!!} \underset{\left\{(\mu\rho) \right\}}{\mathcal{S}} \ProdQt_{0}^{(\mu\rho)}  \prod_{l = 1, 3, \cdots}^{m-1} {\mathcal{C}}_{\text{red}}^{\mu_{l} \rho_{l} \mu_{l+1} \rho_{l+1}}  \,,   
\end{equation}
where $\ProdQt_{0}^{(\mu\rho)}=\ProdQt_{0}=\mathbb{1}$.
If $m$ is odd, all the ${\ProdQt}_{n}^{(\mu\rho)}$ terms in the subtractions have odd $n$ ($< m$), and we keep subtracting terms until we reach 
\begin{equation}
\label{eq:trace_T1}
(-)^{\lfloor \frac{m}{2} \rfloor}  \frac{(m)!!}{(2m-1)!!}  \frac{1}{(m-1)!!} \underset{\left\{(\mu\rho) \right\}}{\mathcal{S}} \ProdQt_{1}^{(\mu\rho)}  \prod_{l = 2, 4, \cdots}^{m-1} {\mathcal{C}}_{\text{red}}^{\mu_{l} \rho_{l} \mu_{l+1} \rho_{l+1}}  \,.   
\end{equation}

\end{enumerate}

As an example, we provide explicit expressions for the first five terms of this construction, which allow to construct all needed covariant multipoles up to the spin-2 case.
The first two elements of the basis are given by Eq.~(\ref{eq:Tau_0}) and (\ref{eq:Tau_1}). They correspond to the identity matrix, respectively the right-handed chiral representation of the generators of the Lorentz group (up to a factor $(\text{i}j)$).
The following three elements of the basis are
\begin{subequations}
\begin{align}
\label{eq:Tau2_quadrupole}
\mathcal{T}_{2}^{\mu_1 \rho_1 , \mu_2 \rho_2} &= \ProdQt_{2}^{(\mu\rho)}  - \frac{1}{3} \mathcal{C}_{\text{red}}^{\mu_1 \rho_1 \mu_2 \rho_2} \mathbb{1}   \,,\\
%%%%%%%%%%%%%%%%%%%%%%%%%%%%%%%%%
\label{eq:Tau3_octupole}
\mathcal{T}_{3}^{\mu_1 \rho_1 , \mu_2 \rho_2 , \mu_3 \rho_3} &= \ProdQt_{3}^{(\mu\rho)} - \frac{1}{5}\left( \ProdQt_{1}^{\mu_1\rho_1} \mathcal{C}_{\text{red}}^{\mu_2 \rho_2 \mu_3 \rho_3} + \ProdQt_{1}^{\mu_2\rho_2} \mathcal{C}_{\text{red}}^{\mu_3 \rho_3 \mu_1 \rho_1} + \ProdQt_{1}^{\mu_3\rho_3} \mathcal{C}_{\text{red}}^{\mu_1 \rho_1 \mu_2 \rho_2} \right) \,,\\
%%%%%%%%%%%%%%%%%%%%%%%%%%%
\label{eq:Tau4_hexadecapole}
\mathcal{T}_{4}^{\mu_1 \rho_1 , \mu_2 \rho_2 , \mu_3 \rho_3} &= \ProdQt{}_{4}^{(\mu\rho)} 
- \frac{1}{7} \left( \ProdQt{}_{2}^{\mu_1 \rho_1 \mu_2 \rho_2} \mathcal{C}_{\text{red}}^{\mu_3 \rho_3 \mu_4 \rho_4} + \ProdQt{}_{2}^{\mu_1 \rho_1 \mu_3 \rho_3} \mathcal{C}_{\text{red}}^{\mu_4 \rho_4 \mu_2 \rho_2} 
\right. \nonumber \\
& \hspace{2.2cm} \left. 
+ \ProdQt{}_{2}^{\mu_1 \rho_1 \mu_4 \rho_4} \mathcal{C}_{\text{red}}^{\mu_2 \rho_2 \mu_3 \rho_3}  + \ProdQt{}_{2}^{\mu_3 \rho_3 \mu_4 \rho_4} \mathcal{C}_{\text{red}}^{\mu_1 \rho_1 \mu_2 \rho_2} 
\right. \nonumber \\
& \hspace{2.2cm} \left. 
+ \ProdQt{}_{2}^{\mu_4 \rho_4 \mu_2 \rho_2} \mathcal{C}_{\text{red}}^{\mu_1 \rho_1 \mu_3 \rho_3} + \ProdQt{}_{2}^{\mu_2 \rho_2 \mu_3 \rho_3} \mathcal{C}_{\text{red}}^{\mu_1 \rho_1 \mu_4 \rho_4} \right)
\nonumber \\
& \hspace{.5cm} + \frac{1}{35}\left( \mathcal{C}_{\text{red}}^{\mu_1 \rho_1 \mu_2 \rho_2} \mathcal{C}_{\text{red}}^{\mu_3 \rho_3 \mu_4 \rho_4} + \mathcal{C}_{\text{red}}^{\mu_1 \rho_1 \mu_3 \rho_3} \mathcal{C}_{\text{red}}^{\mu_4 \rho_4 \mu_2 \rho_2} + \mathcal{C}_{\text{red}}^{\mu_1 \rho_1 \mu_4 \rho_4} \mathcal{C}_{\text{red}}^{\mu_2 \rho_2 \mu_3 \rho_3} \right) \mathbb{1}  
 \,.
\end{align}    
\end{subequations}
The analogous construction for the left-handed chiral representations is obtained  by using the barred tensors instead (both $\overbar{\ProdQt}_{m}^{(\mu\rho)}$ and $\overbar{\mathcal{C}}_\text{red}$).  

We finish this section by counting the independent matrices in $\mathcal{T}_{m}^{(\mu\rho)}$. First, we determine the number of independent matrices in 
\begin{equation}
    \ProdQt_m^{(\mu\rho)} = \prod_{l=1}^{m} {\mathcal{Q}}_{\text{red}}^{\mu_{l} \rho_{l} \alpha_{l}} t_{\alpha_{1} \cdots \alpha_{m} 0 \cdots 0}
\,.
\end{equation}
Because the indices of the $t$-tensor are contracted with ${\mathcal{Q}}_{\text{red}}^{\mu_{l} \rho_{l} \alpha_{l}}$ they only attain the values $\alpha_i=1,2,3$, see Eq.~(\ref{eq:Qred_explicit}). Since $t_{\alpha_{1} \cdots \alpha_{m} 0 \cdots 0}$ ($1 \leq \alpha_i \leq 3$) 
is a symmetric three-dimensional $m$th-rank tensor\footnote{Note that only the full $t_{\alpha_{1} \cdots \alpha_{m} 0 \cdots 0}$ is covariantly traceless, {\emph i.e.} when $0 \leq \alpha_i \leq 3$.}, it contains $\binom{m+2}{m}=\frac{(m+2)!}{m! 2!}$ independent (Hermitian) matrices. The contraction with the ${\mathcal{Q}}_{\text{red}}^{\mu_{l} \rho_{l} \alpha_{l}}$ ($1 \leq l \leq m$) doubles this number as $\ProdQt_m^{(\mu\rho)}$ contains $\binom{m+2}{m}$ Hermitian matrices and the same amount of anti-Hermitian matrices, see the discussion at the beginning of this section. Therefore there are a total of $2\binom{m+2}{m}$ independent matrices in $\ProdQt_m^{(\mu\rho)}$ when considered as a vector space over $\mathbb{R}$.
When we construct the $\mathcal{C}_{\text{red}}$-traceless $\mathcal{T}_{m}^{(\mu\rho)}$ tensor, we remove from $\ProdQt_m^{(\mu\rho)}$ the $2\binom{m}{m-2}$ independent matrices in $\ProdQt_{m-2}^{(\mu\rho)}$, which contains contributions from all $\mathcal{T}_{n}^{(\mu\rho)} \, (n=m-2,m-4,\ldots$). Thus, the number of independent matrices in $\mathcal{T}_{m}^{(\mu\rho)}$ is
\begin{equation}\label{eq:independent_matrices_Tau_m}
    2\binom{m+2}{m} - 2\binom{m}{m-2} = 2(2m+1) \,, \qquad m \geq 1
\,.
\end{equation}
The total number of independent matrices in our orthogonal basis is
\begin{equation}\label{eq:independent_matrices_in_basis_Tau_m}
    \sum_{m=0}^{2j} 2(2m+1)-1 = 2(2j+1)^2-1
\,,
\end{equation}
where the subtraction by one accounts for the fact that the anti-Hermitian ${\rm i}\mathcal{T}_0={\rm i}\mathbb{1}$ is not included in the construction or counting.  When considering the basis as one over $\mathbb{C}$, the Hermitian and anti-Hermitian matrices are not independent, and the counting is reduced by a factor of 2 in Eq.~(\ref{eq:independent_matrices_Tau_m}).  The $2m+1$ result then matches the dimensionality of the $(m,0)$ irrep listed in Eq.~(\ref{eq:sl2c_quad}) and the total independent elements in all $\mathcal{T}_m^{(\mu\rho)}$ is $(2j+1)^2$.  We remind that the anti-self-dual nature of the representations is reflected in the presence of the $Q_\text{red}$ tensors in the definition of the $\mathcal{T}_m^{(\mu\rho)}$, see Eq.~(\ref{eq:Qred_antiselfdual}).

\subsection{Covariant \texorpdfstring{$\mathfrak{sl}(2,\mathbb{C})$}{sl(2,C)} multipoles} 
% {Quadratic products as towers of \texorpdfstring{$\mathfrak{sl}(2,\mathbb{C})$}{sl(2,C)} multipoles } 
\label{sec:quadratic_t_vs_multipoles}

For a recent and detailed study on the topic of $\mathfrak{sl}(2,\mathbb{C})$ multipoles, see Ref.~\cite{Cotogno:2019vjb}. In this publication, the authors construct a basis of covariant multipoles, which they use to expand operators.
The $\mathfrak{sl}(2,\mathbb{C})$ multipole of order $m$ is defined by the rank-$2m$ bi-indexed Lorentz tensor 
\begin{equation}\label{eq:sl2C_multipoles}
{\mathcal{M}}_{m}^{\mu_1 \rho_1 , \cdots , \mu_{m} \rho_{m}} 
= \frac{1}{m!} \underset{\left\{\left(\mu \rho\right)\right\}}{\mathcal{S}} \prod_{r=1}^m \mathbb{M}^{\mu_{r} \rho_{r}}  - \left({\text{Traces}}\right)
\,,
%\nonumber
\end{equation}
and $\mathcal{M}_0=\mathbb{1}$. Here, ${\mathbb{M}}^{\mu \rho}$ corresponds to the representation of Lorentz generators for a given spin $j$, and  the traces guarantee the tracelessness property of the multipoles.

In Ref.~\cite{Cotogno:2019vjb}, three kind of possible constraints over the multipoles are considered. They correspond to the following contractions with  ${\mathcal{C}}_{\text {red }}^{\mu \nu \lambda \sigma}$
\begin{subequations}\label{eq:multipoles_traces}
\begin{align}\label{eq:multipoles_traces_a}
\mathcal{M}^{\cdots, \mu \nu, \cdots, \rho \sigma, \cdots} {\mathcal{C}}^{\text {red }}_{\mu \nu \rho \sigma} & =0 \,,
\\
\label{eq:multipoles_traces_b}
\mathcal{M}^{\cdots, \mu \mu^{\prime}, \cdots, \nu \nu^{\prime}, \cdots, \rho \sigma, \cdots} {\mathcal{C}}^{\text {red }}_{\mu \nu \rho \sigma} & =0 \,, 
\\
\label{eq:multipoles_traces_c}
\mathcal{M}^{\cdots, \mu \mu^{\prime}, \cdots, \nu \nu^{\prime}, \cdots, \rho \rho^{\prime}, \cdots, \sigma \sigma^{\prime}, \cdots} {\mathcal{C}}^{\text {red }}_{\mu \nu \rho \sigma} & =0
\,.
\end{align}    
\end{subequations}
In Appendix~\ref{sec:generalized_quadratic_reductions_and_CR}, we show that the $m$-th $\mathfrak{sl}(2,\mathbb{C})$ multipole is related to our orthogonal basis $\mathcal{T}_{m}^{(\mu\rho)}$ by the simple proportionality
% The $m$th multipoles are given by
\begin{subequations}\label{eq:sl(2,C)_multipole_tTensors}
\begin{align}
\label{eq:sl(2,C)_multipole_tTensors_a}
{\mathcal{M}}_{m}^{\mu_1 \rho_1 ,\cdots, \mu_m \rho_m} &  = \frac{{\rm i}^m}{2^m} m! \binom{2j}{m}  {\mathcal{T}}_m^{\mu_1 \rho_1 ,\cdots, \mu_m \rho_m} \,,
\\
\label{eq:sl(2,C)_multipole_tTensors_b}
\overbar{\mathcal{M}}_{m}^{\mu_1 \rho_1 ,\cdots, \mu_m \rho_m} & = \frac{{\rm i}^m}{2^m} m! \binom{2j}{m} \overbar{\mathcal{T}}_m^{\mu_1 \rho_1 ,\cdots \mu_m \rho_m}
\,.
\end{align} 
\end{subequations}
Therefore, like the $\mathcal{T}_m^{(\mu\rho)}$, the $\mathfrak{sl}(2,\mathbb{C})$ multipoles are covariantly independent and must be ${\mathcal{C}}_{\text{red}}$-{\it traceless} in the sense of Eq.~(\ref{eq:Cred_Tau_m}). Furthermore, the counting of independent Lorentz tensor elements in $\mathcal{T}_m^{(\mu\rho)}$ shown in Sec.~\ref{sec:orthogonal_basis_quadratic_t} establishes that it is not possible to impose any other non-trivial constraint. Otherwise we will have less than $(2j+1)^2$ independent elements and the set of tensors $\mathcal{T}_m^{(\mu\rho)}$ ($0\leq m \leq 2j$) will not contain a complete basis for the $(2j+1) \times (2j+1)$ complex matrices.
% square matrices of order $2j+1$.
% 
This indicates that only the constraint of Eq.~(\ref{eq:multipoles_traces_a}) can be applied and 
% we are of the opinion that 
Eqs.~(\ref{eq:multipoles_traces_b}) and (\ref{eq:multipoles_traces_c}) should not be imposed.

We have proved that for any spin $j$ there is a one-to-one correspondence between the independent substructures that appear in the reduction formula of quadratic products of $t$-tensors and the independent multipoles in the $2j+1$ dimensional representation of $\mathfrak{sl}(2,\mathbb{C})$. 
Moreover, in Eq.~(\ref{eq:sl(2,C)_multipole_tTensors}) we provide a natural way to identify the independent tensor structures in the quadratic product of Eq.~(\ref{eq:quadratic_products_tbart}) with the $\mathfrak{sl}(2,\mathbb{C})$ multipole expansion. 
As we will show below, this leads to an expansion of the invariant tensor structures of bilinears in terms of $\mathfrak{sl}(2,\mathbb{C})$ multipoles which is valid for any spin.

It is not the purpose of this work to establish the mathematical origin of these relations. Nevertheless, it is worth pointing out that this is an example of a general algorithm for generating a bigger Lie algebra that contains the elements of the original Lie algebra. The matrices generated in this manner form what is called the universal enveloping algebra \cite{Bekaert:904799}. In the present case, we start with a $(2j+1)\times(2j+1)$ dimensional representation of the six $\mathfrak{sl}(2,\mathbb{C})$ generators of chiral boosts and rotations and generate, by repeated application of the anticommutation operation, the $(2j+1)$ multipoles corresponding to  the (fundamental representation)  $\mathfrak{sl}(2j+1,\mathbb{C})$\footnote{
The reason for the existence of the map of Eq.~(\ref{eq:sl(2,C)_multipole_tTensors}), which connects between the rank-$2j$ $t$-tensor and the $2j+1$ covariant multipoles $\mathcal{M}_m^{(\mu\rho)}$ can be seen from the following statement.  
The algebra $\mathfrak{sl}(N,\mathbb{C})$  (by definition $N \times N$ complex traceless matrices) is equivalent to the complexification of the $\mathfrak{su}(N)$ algebra (by definition $N \times N$ Hermitian traceless matrices). This follows from the isomorphism that a complex linear combination of Hermitian traceless matrices spans the vector space of complex traceless matrices.}   sector of its universal enveloping algebra\footnote{ As we showed in Sec.~\ref{sec:t_tensors_gen}, the $t$-tensor contains the generators of $\mathfrak{u}(N)$. An algorithm to obtain the $\mathfrak{u}(N)$ generators, detailed in Appendix.~\ref{sec:t_algorithm}, follows the same procedure as was done here for $\mathfrak{sl}(2,\mathbb{C})$. We can start with a $(2j+1)\times(2j+1)$ representation of the three $\mathfrak{su}(2)$ generators (plus the identity) and obtain, from their symmetrized products, the generators of $\mathfrak{u}(N)$ that make up the $t$-tensor.}.
Due to the Cayley-Hamilton theorem applied to the $\bm J$, this sector of the universal enveloping algebra still is finite-dimensional, where the number of elements (equivalent to the highest order multipole +1) depends on the considered representation.

The universal enveloping algebra has the important feature that for the elements of the original group there is a one-to-one map to an element of the universal enveloping algebra. Furthermore, the product of group elements (composite transformations) is, on the one hand, obtained through the exponential map with a specific {\it ordered} succession of commutators (Lie brackets) of the generators of the original Lie algebra [$\mathfrak{sl}(2,\mathbb{C})$ in our case], whereas on the other hand, in the context of the universal enveloping algebra, the equivalent composite transformation can be obtained  by matrix multiplication of the multipoles, which is an associative operation.  These features expose two of the advantages of the universal enveloping algebra framework, hence the efficiency of working with the basis of covariant multipoles.

\subsection{Dirac basis and the \texorpdfstring{$\gamma^{(\mu)}$-algebra}{gamma matrix algebra}  }
\label{sec:dirac_basis}

The goal in this section is to find the irreducible covariantly independent tensors containing the $[2(2j+1)]^2$ independent matrices, that, when considered as a vector space over $\mathbb{C}$, equip us with a basis for operators. In Sec.~\ref{sec:bilinears_general} we will use this basis to calculate the bispinor bilinears for any spin, which can then be used to expand matrix elements of operators and parametrize the non-perturbative structure of hadrons. 
For spin $j$, the $t$-tensors contain $(2j+1)^2$ independent matrices, which are a basis for Hermitian matrices of rank $2j+1$. It follows from the 2 by 2 block matrix structure of the gamma matrices 
% in the Weyl representation 
that there are $4(2j+1)^2$ elements in the Dirac basis.
% \footnote{We remind the reader that we always use the Weyl representation.}.  
These include the earlier introduced gamma matrices, the identity $\mathbb{1}$, and $\gamma_5$, see Eqs.~(\ref{eq:gamma_mu}) and (\ref{eq:gamma5}).  The result of the construction is a generalization of the spin-1/2 gamma matrix basis, which can be used to exhaust all possible independent bispinor bilinears.
As in the spin-$1/2$ case, the independent elements of the generalized Dirac basis are obtained from products of $\gamma$-tensors. They are grouped based on their transformation properties under the Lorentz group, including parity and time reversal. 

We will find products of $\gamma$-tensors in general calculations using the bispinor calculus. 
Thus, the strategy, familiar from the spin-$1/2$ case, is to analyze the reducibility of $\gamma$-tensors products.
Analogous to the spin-$1/2$ case, we only need to study linear, quadratic, cubic, and quartic powers of $\gamma$-tensors. Powers higher than four can be reduced to linear combinations of the previous cases. 
% We will show  that the matrices generated in this manner form a complete basis.

From the SL$(2,\mathbb{C})$ group-theoretical viewpoint, the Dirac basis corresponds to the reduction of the exterior product of two bispinors, which transform as~\cite{Williams:1965rga,Gomez-Avila:2013qaa}
\begin{equation}\label{eq:diracbasis_grouptheory}
    [(j,0)\oplus (0,j)]\otimes [(j,0)\oplus (0,j)] = 2(j,j) \oplus 2(0,0) \overset{2j}{\underset{m=1}{\oplus}} [(m,0)\oplus (0,m)]
    \,,
\end{equation}
where the last terms are collected in pairs that are invariant under parity.

The results shown below are independent of the representation used for the Dirac matrices. However, for the sake of clarity, it is useful to choose the Weyl representation.
We start with the \emph{anti-diagonal} block matrices, corresponding to $2(2j+1)^2$ independent Dirac matrices that are exhausted by $\gamma^{\mu_1\cdots \mu_{2j}}$ and $\gamma^{\mu_1\cdots \mu_{2j}}\gamma_5$.
Then, we study the \emph{diagonal} block matrices, corresponding to the other $2(2j+1)^2$ independent Dirac matrices, which are exhausted by $\gamma^{\mu_1\cdots \mu_{2j}}\gamma^{\rho_1\cdots \rho_{2j}}$ and $\gamma_5$.

\begin{enumerate}

\item The $\gamma$-tensors (generalized gamma matrices) 
\begin{equation}\label{eq:gamma}
\gamma^{\mu_1 \cdots \mu_{2j}}
=\left(\begin{array}{cc}
0 & t^{\mu_1 \cdots \mu_{2j}} \\
\bar{t}^{\mu_1 \cdots \mu_{2j}} & 0
\end{array}\right) 
\,,
\end{equation} 
are symmetric and traceless in their Lorentz indices as $t$ and $\bar t$ are symmetric and traceless, see Sec.~\ref{sec:chiral_reps}.
Notice that they populate the \emph{anti-diagonal} blocks of the Dirac matrices.
Inheriting the number of independent matrices from the $t$-tensor building blocks, these gamma matrices contain $(2j+1)^2$ independent $2(2j+1)\times2(2j+1)$ matrices of the Dirac basis.

 \item  The rest of the matrices needed to span the \emph{anti-diagonal} blocks of the Dirac basis are found in the cubic product of gamma matrices. 
 % The rest of the matrices needed to span the Dirac matrices with  \emph{anti-diagonal} blocks are found in the cubic product of gamma matrices.
 Using Eqs.~(\ref{eq:cubic_products}) this product can be expanded as linear combinations of $\gamma^{\mu_1 \cdots \mu_{2j}}$ and $\gamma^{\mu_1 \cdots \mu_{2j}} \gamma_5$  
\begin{align}\label{eq:gamma_gamma_gamma}
\gamma^{(\mu)} \gamma^{(\rho)} \gamma^{(\sigma)} 
& =\left(\begin{array}{cc}
0 & {t}^{(\mu)} \bar{t}^{(\rho)} t^{(\sigma)}  \\
\bar{t}^{(\mu)} {t}^{(\rho)} \bar{t}^{(\sigma)} & 0
\end{array}\right) 
\nonumber \\
& 
= \frac{1}{[(2j)!]^2}  \underset{\left\{(\rho),(\sigma)\right\}}{\mathcal{S}} \left( \operatorname{Re}\left\{ \prod_{l=1}^{2j} {\mathcal{C}}^{\mu_{l} \rho_{l} \sigma_{l}}{}_{ \alpha_{l}} \right\} \gamma^{\alpha_{1} \cdots \alpha_{2j}}
+ {\rm i} \operatorname{Im}\left\{ \prod_{l=1}^{2j} {\mathcal{C}}^{\mu_{l} \rho_{l} \sigma_{l}}{}_{ \alpha_{l}} \right\} \gamma^{\alpha_{1} \cdots \alpha_{2j}} \gamma_5
\right)  
\,.
\end{align} 

The cubic products again populate the \emph{anti-diagonal} blocks of the Dirac matrices. The terms proportional to $\gamma_{\alpha_{1} \cdots \alpha_{2j}}$ do not produce any new element to the Dirac basis. They are linear combinations of products of metric tensors and a gamma matrix carrying different permutations of indices. This term corresponds to a rank-$6j$ tensor.  

The new elements of the generalized Dirac basis are found in $\gamma_{\alpha_{1} \cdots \alpha_{2j}} \gamma_5$, which can not be written as a linear combination of $\gamma_{\alpha_{1} \cdots \alpha_{2j}}$. 
Furthermore, owning to their transformation under parity inversion, $\gamma_5$ is a pseudoscalar and $\gamma^{(\mu)}$ is a proper tensor [Eqs.~(\ref{eq:parity_transf_gamma5}) and (\ref{eq:parity_transf_gamma})]. It follows that $\gamma^{\mu_{1} \cdots \mu_{2j}} \gamma_5$ is a pseudotensor, which was already stated in Eq.~(\ref{eq:parity_transf_gamma_gamma5}).

Thus, we have that
% the independent elements  
\begin{equation}\label{eq:gamma_gamma_5}
\gamma^{\mu_1 \cdots \mu_{2j}}\gamma_5
=\left(\begin{array}{cc}
0 & t^{\mu_1 \cdots \mu_{2j}} \\
-\bar{t}^{\mu_1 \cdots \mu_{2j}} & 0
\end{array}\right) 
\,,
\end{equation}
transforms as a rank-$2j$ totally symmetric pseudotensor. It is also covariantly traceless, $g_{\mu_k\mu_l}\gamma^{\cdots\mu_k\cdots \mu_{l}\cdots}\gamma_5=0$, and contains $(2j+1)^2$ independent  $2(2j+1)\times2(2j+1)$ matrices of the Dirac basis.  
Hence, we find that the Dirac subspace corresponding to \emph{anti-diagonal} block matrices is spanned by the $2(2j+1)^2$ matrices in $\gamma^{\mu_{1} \cdots \mu_{2j}}$ and $\gamma^{\mu_{1} \cdots \mu_{2j}} \gamma_5$. These correspond with the $2(j,j)$ terms in Eq.~(\ref{eq:diracbasis_grouptheory}).

\item We now turn our attention to the block \emph{diagonal} Dirac matrices, starting with the quadratic product of gamma matrices. 
In order to gain further insight on the quadratic product of $\gamma$-tensors, it is useful to write it as follows
\begin{equation}
\label{eq:quadratic_products_gamma}
\gamma^{\mu_1 \cdots \mu_{2j}} \gamma^{\rho_1 \cdots \rho_{2j}} 
= \frac{1}{2} \left[ \gamma^{\mu_1 \cdots \mu_{2j}} , \gamma^{\rho_1 \cdots \rho_{2j}} \right] + \frac{1}{2} \left\{ \gamma^{\mu_1 \cdots \mu_{2j}} , \gamma^{\rho_1 \cdots \rho_{2j}} \right\}
\,,
\end{equation}
where 
\begin{subequations}
\label{eq:(anti)commutator_gamma}
\begin{align}
%\label{eq:commutator_gamma}
\left[ \gamma^{\mu_1 \cdots \mu_{2j}} , \gamma^{\rho_1 \cdots \rho_{2j}} \right]  &=  \left(\begin{array}{cc}
{t}^{(\mu)} \bar{t}^{(\rho)} - {t}^{(\rho)} \bar{t}^{(\mu)} & 0 \\
0 & \bar{t}^{(\mu)} {t}^{(\rho)} - \bar{t}^{(\rho)} {t}^{(\mu)}
\end{array}\right) 
\,,\\
\left\{ \gamma^{\mu_1 \cdots \mu_{2j}} , \gamma^{\rho_1 \cdots \rho_{2j}} \right\} & =  \left(\begin{array}{cc}
{t}^{(\mu)} \bar{t}^{(\rho)} + {t}^{(\rho)} \bar{t}^{(\mu)} & 0 \\
0 & \bar{t}^{(\mu)} {t}^{(\rho)} + \bar{t}^{(\rho)} {t}^{(\mu)}
\end{array}\right)
\,.
\end{align}    
\end{subequations}
Using Eqs.~(\ref{eq:quadratic_products},\ref{eq:quadratic_products_tbart_with_T}), (\ref{eq:covariant_coefficients_Qred_permutations} and (\ref{eq:cov_independent_substructures_T_barT}) we find  
\begin{subequations}
\label{eq:quadratic_products_tbart_index_antisymm}
\begin{align}
\frac{1}{2} \left( {t}^{(\mu)} \bar{t}^{(\rho)} - {t}^{(\rho)} \bar{t}^{(\mu)} \right)  & = \frac{1}{(2j)!^2}  \underset{\left\{(\rho) \right\}}{\mathcal{S}} \underset{\left\{(\mu \rho)\right\}}{\mathcal{S}} \sum_{\underset{(1,3,\cdots)}{m=\text{odd}}}^{2j}  
\binom{2j}{m} \ProdQt{}_{m}^{(\mu\rho)}
% \prod_{l=1}^{m} {\mathcal{Q}}_{\text{red}}^{\mu_{l} \rho_{l} \alpha_{l}} 
\prod_{k=m+1}^{2j} g^{\mu_k \rho_k} 
% \eta^{\alpha_k} t_{\alpha_{1} \cdots \alpha_{2j} } 
 \, , 
 \\
 \frac{1}{2} \left( \bar{t}^{(\mu)} {t}^{(\rho)} - \bar{t}^{(\rho)} {t}^{(\mu)} \right)  & = \frac{1}{(2j)!^2}  \underset{\left\{(\rho) \right\}}{\mathcal{S}} \underset{\left\{(\mu \rho)\right\}}{\mathcal{S}} \sum_{\underset{(1,3,\cdots)}{m=\text{odd}}}^{2j} 
\binom{2j}{m} \overbar{\ProdQt{}}_{m}^{(\mu\rho)}
% \prod_{l=1}^{m} \overbar{{\mathcal{Q}}}_{\text{red}}^{\mu_{l} \rho_{l} \alpha_{l}} 
\prod_{k=m+1}^{2j} g^{\mu_k \rho_k} \,,
% \eta^{\alpha_k} \bar{t}_{\alpha_{1} \cdots \alpha_{2j} }
\end{align}    
\end{subequations}
and
\begin{subequations}
\label{eq:quadratic_products_tbart_index_symm}
\begin{align}
\frac{1}{2} \left( {t}^{(\mu)} \bar{t}^{(\rho)} + {t}^{(\rho)} \bar{t}^{(\mu)} \right)  & = \frac{1}{(2j)!^2}  \underset{\left\{(\rho) \right\}}{\mathcal{S}} \underset{\left\{(\mu \rho)\right\}}{\mathcal{S}} \sum_{\underset{(0,2,\cdots)}{m=\text{even}}}^{2j}  
\binom{2j}{m} {\ProdQt{}}_{m}^{(\mu\rho)}
% \prod_{l=1}^{m} {\mathcal{Q}}_{\text{red}}^{\mu_{l} \rho_{l} \alpha_{l}} 
\prod_{k=m+1}^{2j} g^{\mu_k \rho_k} 
% \eta^{\alpha_k} t_{\alpha_{1} \cdots \alpha_{2j} }
 \, , 
 \\
 \frac{1}{2} \left( \bar{t}^{(\mu)} {t}^{(\rho)} + \bar{t}^{(\rho)} {t}^{(\mu)} \right)  & = \frac{1}{(2j)!^2}  \underset{\left\{(\rho) \right\}}{\mathcal{S}} \underset{\left\{(\mu \rho)\right\}}{\mathcal{S}} \sum_{\underset{(0,2,\cdots)}{m=\text{even}}}^{2j} 
\binom{2j}{m} \overbar{\ProdQt{}}_{m}^{(\mu\rho)}
% \prod_{l=1}^{m} \overbar{{\mathcal{Q}}}_{\text{red}}^{\mu_{l} \rho_{l} \alpha_{l}} 
\prod_{k=m+1}^{2j} g^{\mu_k \rho_k} 
\,.
\end{align}    
\end{subequations}

Thus, the quadratic product of $\gamma$-tensors can be separated into two independent towers
\begin{subequations}
\label{eq:botwh_commutator_gamma}
\begin{equation}
\label{eq:commutator_gamma}
 \left[ \gamma^{(\mu)} , \gamma^{(\rho)} \right] = \frac{2}{(2j)!^2}  \underset{\left\{(\rho) \right\}}{\mathcal{S}} \underset{\left\{(\mu \rho)\right\}}{\mathcal{S}} \sum_{\underset{(1,3,\cdots)}{m=\text{odd}}}^{2j}    
\binom{2j}{m} \prod_{k=m+1}^{2j} g^{\mu_k \rho_k} 
% \eta^{\alpha_k}
\left(\begin{array}{cc}
{\ProdQt{}}_{m}^{(\mu\rho)}
% \prod_{l=1}^{m} {\mathcal{Q}}_{\text{red}}^{\mu_{l} \rho_{l} \alpha_{l}} t_{\alpha_{1} \cdots \alpha_{2j}} 
& 0 \\
0 & \overbar{\ProdQt{}}_{m}^{(\mu\rho)}
% \prod_{l=1}^{m} \overbar{{\mathcal{Q}}}_{\text{red}}^{\mu_{l} \rho_{l} \alpha_{l}} \bar{t}_{\alpha_{1} \cdots \alpha_{2j}}
\end{array}\right) 
\,,
\end{equation}
\begin{equation}
\label{eq:anti_commutator_gamma}
 \left\{ \gamma^{(\mu)} , \gamma^{(\rho)} \right\} = \frac{2}{(2j)!^2}  \underset{\left\{(\rho) \right\}}{\mathcal{S}} \underset{\left\{(\mu \rho)\right\}}{\mathcal{S}} \sum_{\underset{(0,2,\cdots)}{m=\text{even}}}^{2j}   
\binom{2j}{m} \prod_{k=m+1}^{2j} g^{\mu_k \rho_k} 
% \eta^{\alpha_k}
\left(\begin{array}{cc}
{\ProdQt{}}_{m}^{(\mu\rho)}
% \prod_{l=1}^{m} {\mathcal{Q}}_{\text{red}}^{\mu_{l} \rho_{l} \alpha_{l}} t_{\alpha_{1} \cdots \alpha_{2j}} 
& 0 \\
0 & \overbar{\ProdQt{}}_{m}^{(\mu\rho)} 
% \prod_{l=1}^{m} \overbar{{\mathcal{Q}}}_{\text{red}}^{\mu_{l} \rho_{l} \alpha_{l}} \bar{t}_{\alpha_{1} \cdots \alpha_{2j}}
\end{array}\right) 
\,.
\end{equation}    
\end{subequations}
The case $m=0$ only appears in $\left\{ \gamma^{\mu_1 \cdots \mu_{2j}} , \gamma^{\rho_1 \cdots \rho_{2j}} \right\}$, for which we have
\begin{equation}
\label{eq:anti_commutator_gamma_m_0}
 \left\{ \gamma^{\mu_1 \cdots \mu_{2j}} , \gamma^{\rho_1 \cdots \rho_{2j}} \right\} 
 \ \underset{m=0}{\longrightarrow} \ \frac{2}{(2j)!^2}  \underset{\left\{(\rho) \right\}}{\mathcal{S}} \underset{\left\{(\mu \rho)\right\}}{\mathcal{S}} 
\prod_{k=1}^{2j} g^{\mu_k \rho_k} \mathbb{1}  
 \,,
\end{equation}
where we use the empty product convention $\prod_{l=1}^{0} (\cdots) \equiv 1$, see Eq.~(\ref{eq:empty_prod_convention}).
It is straightforward to show that the quadratic product can be expanded as 
% the linear combination 
\begin{align}\label{eq:gamma_gamma}
\gamma^{\mu_1 \cdots \mu_{2j}} \gamma^{\rho_1 \cdots \rho_{2j}}
%& =\left(\begin{array}{cc}
%{t}^{\mu_1 \cdots \mu_{2j}}\bar{t}^{\rho  \cdots \rho_{2j}} & 0 \\
%0 &\bar{t}^{\mu_1 \cdots \mu_{2j}}t^{\rho_1 \cdots \rho_{2j}} 
%\end{array}\right) 
%\\
&  = \frac{1}{(2j)!^2}  \underset{\left\{(\rho) \right\}}{\mathcal{S}} \underset{\left\{(\mu \rho)\right\}}{\mathcal{S}}  \left[
\prod_{k=1}^{2j} g^{\mu_k \rho_k} \mathbb{1}  
+ \sum_{m=1}^{2j}  \binom{2j}{m} \prod_{k=m+1}^{2j} g^{\mu_k \rho_k} \eta^{\alpha_k} 
\right.
\nonumber\\
& \left. 
% \hspace{.5cm} 
\times \left( \operatorname{Re}\left\{ \prod_{l=1}^{m} {\mathcal{Q}}_{\text{red}}^{\mu_{l} \rho_{l} \alpha_{l}} \right\} \gamma_{\alpha_{1} \cdots \alpha_{2j}} \beta^{-1}
+ {\rm i} \operatorname{Im}\left\{ \prod_{l=1}^{m} {\mathcal{Q}}_{\text{red}}^{\mu_{l} \rho_{l} \alpha_{l}} \right\} \gamma_{\alpha_{1} \cdots \alpha_{2j}} \gamma_5 \beta^{-1}
\right)  
\right]
\,,
\end{align} 
which involves the Dirac matrices $\mathbb{1}$, $\gamma^{\mu_1 \cdots \mu_{2j}} \beta^{-1}$ and $\gamma^{\mu_1 \cdots \mu_{2j}} \gamma_5 \beta^{-1}$, where at least one of the indices  $\mu_{a}$ is non-zero.  Here, $\mathbb{1}$ is one of the two $(0,0)$ singlets that appears in Eq.~(\ref{eq:diracbasis_grouptheory}), while the other matrices make up the $\overset{2j}{\underset{m=1}{\oplus}} [(m,0)\oplus (0,m)]$ terms.  The excluded $\gamma_5$ is the remaining singlet that needs to be accounted for in Eq.~(\ref{eq:diracbasis_grouptheory}) and will appear in the quartic product of $\gamma$ tensors.
Therefore, we find that the quadratic product of  $\gamma$-tensors contain $2(2j+1)^2-1$ independent elements of the Dirac basis. There are $(2j+1)^2$ independent matrices in $\gamma^{\mu_1 \cdots \mu_{2j}} \beta^{-1}$, including $\mathbb{1}$. The other $(2j+1)^2-1$ independent matrices are $\gamma^{\mu_1 \cdots \mu_{2j}} \gamma_5 \beta^{-1}$, where at least one of the indices  $\mu_{a}$ is not zero (meaning $\gamma_5$ by itself is not included).

Moreover, we find that these $2(2j+1)^2-1$ Dirac matrices can be accommodated into $2j+1$ independent covariant tensor matrices, corresponding to the different terms in the summation over $m$ (including $\mathbb{1}$). 
Each $0 \leq m \leq 2j$ corresponds to a rank-$4j$ tensor, which is totally symmetric (and invariant) in the $2(2j-m)$ indices associated with $\prod g^{\mu_k \rho_k}$, whereas for the other $2m$ indices it is anti-symmetric within each $\mu_r\rho_s$ pair and symmetric for exchanges among the $m$ pairs.
We are interested in the objects that result from stripping out the metric tensors $\prod g^{\mu_k \rho_k}$ in Eqs.~(\ref{eq:botwh_commutator_gamma}), {\emph{i.e.}} 
\begin{equation}
\label{eq:quadratic_reductions_gamma}
 \mathsf{G}_m^{\mu_1 \rho_1 \cdots \rho_{m} \mu_{m}} = 
\left(\begin{array}{cc}
{\ProdQt{}}_{m}^{(\mu\rho)}
% \prod_{l=1}^{m} {\mathcal{Q}}_{\text{red}}^{\mu_{l} \rho_{l} \alpha_{l}} t_{\alpha_{1} \cdots \alpha_{m} 0 \cdots 0} 
& 0 \\
0 & \overbar{\ProdQt{}}_{m}^{(\mu\rho)}
% \prod_{l=1}^{m} \overbar{{\mathcal{Q}}}_{\text{red}}^{\mu_{l} \rho_{l} \alpha_{l}} \bar{t}_{\alpha_{1} \cdots \alpha_{m} 0 \cdots 0}
\end{array}\right) 
\,.
\end{equation}
Using the results of Sec.~\ref{sec:orthogonal_basis_quadratic_t} we can again turn this set into the following tower of orthogonal elements of the generalized Dirac basis
\begin{equation}
\label{eq:multipoles_gamma}
 \mathcal{G}_m^{\mu_1 \rho_1 \cdots \rho_{m} \mu_{m}} = 
\left(\begin{array}{cc}
\mathcal{T}_m^{\mu_1 \rho_1 \cdots \rho_{m} \mu_{m}} & 0 \\
0 & \overbar{\mathcal{T}}_m^{\mu_1 \rho_1 \cdots \rho_{m} \mu_{m}}
\end{array}\right) 
\,, \quad 0 \leq m \leq 2j
\,,
\end{equation}
see Eqs.~(\ref{eq:Tau_m}) and the surrounding discussion. 
For $m >0$, the $\mathcal{G}_m$ are the elements of the basis which can be matched one-by-one to the irreducible (extended by parity) $[(m,0)\oplus (0,m)]$ representations in Eq.~(\ref{eq:diracbasis_grouptheory}). The rank-$2m$ tensors $\mathcal{G}_m^{\mu_1 \rho_1 \cdots \rho_{m} \mu_{m}}$ generalize the familiar identity matrix and anti-symmetric tensor ($\sigma^{\mu\nu}$) that form part of the Dirac basis for spin-$1/2$. In our notation, the Lorentz generators for the bispinor representation of any spin $j$ are always equal to $({\rm i} j) \mathcal{G}_1$, see Eq.~(\ref{eq:sl(2,C)_generators_chiral}). 

According to the dimensionality of the irreps, each $\mathcal{G}_m^{(\mu\rho)}$ contains $2(2m+1)$ independent matrices.  We can verify this by using the (anti-)self-dual nature of the $\mathcal{T}_m^{(\mu\nu)},\overbar{\mathcal{T}}_m^{(\mu\nu)}$ in each bi-index pair to write
\begin{align}
\label{eq:dual_multipoles_gamma}
 \frac{\rm i}{2}\tensor{\epsilon}{^{\mu_1\rho_1}_{\alpha\beta}}\mathcal{G}_m^{\alpha \beta \cdots \rho_{m} \mu_{m}} &= 
\left(\begin{array}{cc}
-\mathcal{T}_m^{\mu_1\rho_1 \cdots \rho_{m} \mu_{m}} & 0 \\
0 & \overbar{\mathcal{T}}_m^{\mu_1\rho_1 \cdots \rho_{m} \mu_{m}}
\end{array}\right) 
\nonumber \\[0.8em]
&
=\mathcal{G}_m^{\mu_1\rho_1 \cdots \rho_{m} \mu_{m}}\, \gamma_5 = \gamma_5 \, \mathcal{G}_m^{\mu_1\rho_1 \cdots \rho_{m} \mu_{m}} 
\,,  \quad 1 \leq m \leq 2j
\,,
\end{align}
familiar from the spin-1/2 case, where $i\mathcal{G}_1^{\mu\nu}=\sigma^{\mu\nu}$. 
We can then form linear combinations
\begin{equation}\label{eq:chiral_decomposition_of_Dirac_cov_multipoles}
    \frac{1}{2}\left(\mathcal{G}_m^{\mu_1\rho_1 \cdots \rho_{m} \mu_{m}} \pm \frac{\rm i}{2}\tensor{\epsilon}{^{\mu_1\rho_1}_{\alpha\beta}}\mathcal{G}_m^{\alpha\beta \cdots \rho_{m} \mu_{m}}\right) = \frac{1}{2}(1\pm\gamma_5)\mathcal{G}_m^{\mu_1\rho_1 \cdots \rho_{m} \mu_{m}} \,,
\end{equation}
to isolate the top left or bottom right block of the $\mathcal{G}_m^{(\mu\nu)}$. Each of these blocks inherits their number of independent elements from those of $\mathcal{T}_m^{(\mu\nu)}$ ($-$ in the sum), respectively $\overbar{\mathcal{T}}_m^{(\mu\nu)}$ ($+$).  Each of these contain $2m+1$ independent matrices, as discussed in Sec.~\ref{sec:orthogonal_basis_quadratic_t}, which brings us to the desired $2(2m+1)$ independent matrices in each multipole $\mathcal{G}_m^{(\mu\nu)}$.
    
\item  The quartic power of $\gamma$-tensors is responsible for the appearance of the last independent Dirac matrix, which is $\gamma_5$. 
The reduction of the quartic power of $\gamma$-tensors produces
\begin{align}\label{eq:gamma_gamma_gamma_gamma}
\gamma^{(\mu)} \gamma^{(\rho)} \gamma^{(\sigma)} \gamma^{(\nu)} 
& \propto
\underset{\left\{(\rho),(\sigma)\right\}}{\mathcal{S}}
\left( \operatorname{Re}\left\{ \prod_{l=1}^{2j} {\mathcal{C}}^{\mu_{l} \rho_{l} \sigma_{l}}{}_{\alpha_{l}} \right\} \gamma^{\alpha_{1} \cdots \alpha_{2j}}  
+ {\rm i} \operatorname{Im}\left\{ \prod_{l=1}^{2j} {\mathcal{C}}^{\mu_{l} \rho_{l} \sigma_{l}}{}_{ \alpha_{l}} \right\} \gamma^{\alpha_{1} \cdots \alpha_{2j}} \gamma_5
\right) \gamma^{\nu_{1} \cdots \nu_{2j}}
\nonumber\\
& 
% \hspace{-2.5cm} 
\hspace{-.5cm}
\propto \underset{\left\{(\mu), (\rho),(\sigma),(\nu)\right\}}{\mathcal{S}}  
\operatorname{Re}\left\{ \prod_{l=1}^{2j} {\mathcal{C}}^{\mu_{l} \rho_{l} \sigma_{l}}{}_{\alpha_{l}} \right\}
\left( 
\prod_{k=1}^{2j} g^{\alpha_k \nu_k} \mathbb{1} 
+ \underset{\left\{(\alpha \nu)\right\}}{\mathcal{S}} \sum_{m=1}^{2j}  \binom{2j}{m} \prod_{k=m+1}^{2j} g^{\alpha_k \nu_k} \eta^{\xi_k} 
\right.\nonumber \\
& \left. 
\hspace{1.cm}  
 \times \left[ \operatorname{Re}\left\{ \prod_{r=1}^{m} {\mathcal{Q}}^{\alpha_{r} \nu_{r} \xi_{r}} \right\}\gamma_{\xi_{1} \cdots \xi_{2j}} \beta^{-1} + {\rm i} \operatorname{Im}\left\{ \prod_{r=1}^{m} {\mathcal{Q}}^{\alpha_{r} \nu_{r} \xi_{r}} \right\}\gamma_{\xi_{1} \cdots \xi_{2j}} \gamma_5 \beta^{-1} \right]
\right)
\nonumber\\
& 
% \hspace{-2.2cm}
% \hspace{.5cm}
- \underset{\left\{(\mu), (\rho),(\sigma),(\nu)\right\}}{\mathcal{S}}  
{\rm i} \operatorname{Im}\left\{ \prod_{l=1}^{2j} {\mathcal{C}}^{\mu_{l} \rho_{l} \sigma_{l}}{}_{ \alpha_{l}} \right\}
\left( 
\prod_{k=1}^{2j} g^{\alpha_k \nu_k} \mathbb{1} 
+ \underset{\left\{(\alpha \nu)\right\}}{\mathcal{S}} \sum_{m=1}^{2j}  \binom{2j}{m} \prod_{k=m+1}^{2j} g^{\alpha_k \nu_k} \eta^{\xi_k} 
\right.\nonumber \\
& \left. 
\hspace{1.cm}  
 \times 
\left[ \operatorname{Re}\left\{ \prod_{r=1}^{m} {\mathcal{Q}}^{\alpha_{r} \nu_{r} \xi_{r}} \right\}\gamma_{\xi_{1} \cdots \xi_{2j}} \beta^{-1} + {\rm i} \operatorname{Im}\left\{ \prod_{r=1}^{m} {\mathcal{Q}}^{\alpha_{r} \nu_{r} \xi_{r}} \right\}\gamma_{\xi_{1} \cdots \xi_{2j}} \gamma_5 \beta^{-1} \right] 
\right)
\gamma_5
\,.
\end{align}
The first line carries out the reduction of the cubic product $\gamma^{(\mu)} \gamma^{(\rho)} \gamma^{(\sigma)}$. In the following lines we further reduce the quadratic products $\gamma^{(\alpha)} \gamma^{(\nu)}$, 
% together with  $\{\gamma^{(\mu)},\gamma_5\}=0$, see Eq.~(\ref{eq:anti_CR_gamma_gamma5}).
and use Eq.~(\ref{eq:anti_CR_gamma_gamma5}).
Thus, we are lead to the result that the term with $m=0$ in the sum of Eq.~(\ref{eq:gamma_gamma_gamma_gamma}) is given by
\begin{align}\label{eq:gamma_gamma_gamma_gamma_m_0}
%\hspace{-1.5cm} 
\gamma^{(\mu)} \gamma^{(\rho)} \gamma^{(\sigma)} \gamma^{(\nu)}
%|_{m=0}
\  \underset{m=0}{\longrightarrow} \ 
\underset{\left\{(\mu), (\rho),(\sigma),(\nu)\right\}}{\mathcal{S}}  \left( 
\operatorname{Re}\left\{ \prod_{l=1}^{2j} {\mathcal{C}}^{\mu_{l} \rho_{l} \sigma_{l} \nu_{l}} \right\}
\mathbb{1}
- {\rm i} \operatorname{Im}\left\{ \prod_{l=1}^{2j} {\mathcal{C}}^{\mu_{l} \rho_{l} \sigma_{l} \nu_{l}} \right\}
\gamma_5
\right)
\,,
\end{align}
which means that we must add $\gamma_5$ as another independent element of the Dirac basis.
On the other hand, when $m \neq 0$ the quartic product reduces to a linear combination of  $\gamma_{\xi_{1} \cdots \xi_{2j}}\beta^{-1}$ and $\gamma_{\xi_{1} \cdots \xi_{2j}} \gamma_5 \beta^{-1}$, with at least one of the indices  $\xi_{a}$ not zero, which we have already included in the basis with the quadratic product. 
 
\end{enumerate}

Thus, in summary, the elements of the generalized Dirac basis are,   
\begin{equation}
\label{eq:Dirac_basis}
\mathbb{1} \,, \, \gamma_5 \,, \, \gamma^{\mu_{1} \cdots \mu_{2j}} \,, \, \gamma^{\mu_{1} \cdots \mu_{2j}} \gamma_5 \,, \, \mathcal{G}_m^{\mu_1 \rho_1 \cdots \rho_{m} \mu_{m}} \, (\text{or } \ \mathsf{G}_m^{\mu_1 \rho_1 \cdots \rho_{m} \mu_{m}}) \,, \quad 1 \leq m \leq 2j  \,, 
\end{equation}
 where, $\mathcal{G}_m^{\mu_1 \rho_1 \cdots \rho_{m} \mu_{m}}$ [$\mathsf{G}_m^{\mu_1 \rho_1 \cdots \rho_{m} \mu_{m}}$] are given by Eq.~(\ref{eq:multipoles_gamma}) [Eq.~(\ref{eq:quadratic_reductions_gamma})]. 
Although we have listed $\mathbb{1}$ separately for visual proposes, from our perspective it is more natural to include it as $\mathbb{1} = \mathcal{G}_0 = \mathsf{G}_0$.

The quintic and higher products of gammas are completely reducible to linear combinations of these Dirac matrices.
As can be inferred from the block structure of Dirac's matrices in the Weyl representation, we have exhausted the elements of the Dirac basis. This concludes the construction of the generalized Dirac basis, we have in total $4(2j+1)^2$ independent matrices arranged into irreducible covariant tensors. 

%%%%%%%%%%%%%%%%%%%%%%%%%
%%%%% Bilinears %%%%%%%%%
%%%%%%%%%%%%%%%%%%%%%%%%%

\section{Generalized bilinears} 
\label{sec:gen_bilinears}

\subsection{General setup}
\label{sec:bilinears_general}

In this section we will detail the calculations for the bilinears of bispinors of arbitrary spin, where we make use of the generalized Dirac matrices discussed in the previous section. These Dirac matrices form a complete basis as demonstrated in Sec.~\ref{sec:dirac_basis}.  The bilinears will Lorentz transform under separate subspaces, meaning they transform independently from each other with simple transformation laws.  
% By using the covariant algebra developed in Sec.~\ref{sec:t_gamma_tensors_algebra}, we will arrive at expressions which can be used to obtain covariant expressions for the bilinears for any spin and boost parameterization. 
Using the covariant algebra developed in Sec.~\ref{sec:t_gamma_tensors_algebra}, we will arrive at general expressions which can be particularize to obtain the bilinears for any spin value and boost parameterization.
For the spin-$1/2$ case, our results agree with those derived earlier in \cite{Lorce:2017isp}, which were obtained using a different approach.
%, based on a different construction.

The general form of the bilinears is
\begin{align}\label{eq:bilinear_general}
\bar{u}(p_f,\lambda_f)  \; \Gamma \; u(p_i,\lambda_i)  
\,,
\end{align}
where $\Gamma$ stands for any of the elements that form the generalized Dirac basis discussed in Sec.~\ref{sec:dirac_basis} and the bispinor and its adjoint can correspond to any choice of spinors (canonical, helicity, light-front).  The bispinors were introduced in the Weyl representation for the canonical spinors in Eqs.~(\ref{eq:bispinor_def}) and (\ref{eq:adjoint_bispinor_def}), and in Eqs.~(\ref{eq:bispinors_construction}) for the helicity and light-front spinors.  
The focus here is on the bilinears of positive energy bispinors.  Using Eqs.~(\ref{eq:bispinor_chargeconj}) and (\ref{eq:C_as_t2}) bilinears with negative energy bispinors can be rewritten using positive energy bispinors and reduced in a similar manner as presented here.  As the first applications that we have in mind are the use of these bilinears in the decomposition of QCD operators for composite systems of any spin, the positive energy bilinears of Eq.~(\ref{eq:bilinear_general}) are the objects of primary interest.

We consider bilinears where the masses can be different
\begin{equation}
   [ p_i^2 = m_i^2 ] \; \neq \; [ p_f^2 = m_f^2]\,,
\end{equation}
and introduce the standard average and relative four-momentum variables
\begin{subequations}
\begin{align}\label{eq:P_Delta}
    &P = \frac{\overbar m}{2}\left(\frac{p_f}{m_f}+\frac{p_i}{m_i}\right) \,, 
    &\Delta = \overbar m \left( \frac{p_f}{m_f}-\frac{p_i}{m_i} \right) \,,\\
    %%%%%%%%%%%%%%%
    \label{eq:pf_pi}
    &p_f = \frac{m_f}{\overbar m}\left(P+\frac{\Delta}{2} \right) \,,
    &p_i = \frac{m_f}{\overbar m}\left(P-\frac{\Delta}{2} \right) \,,
\end{align}    
\end{subequations}
where the average mass is
\begin{equation}
    \overbar m = \tfrac{1}{2}(m_i+m_f) \,.
\end{equation}

In order to obtain useful expressions for the bilinears, which still hold for any spin representation and spinor choice, we use Eqs.~(\ref{eq:bispinor_def}),(\ref{eq:adjoint_bispinor_def}) and (\ref{eq:boosts_t}) to  write
\begin{align} \label{eq:bilinear_gen_setup}
\bar{u}(p_f,\lambda_f) \; \Gamma \; u(p_i,\lambda_i)  
&= 
%{1\over {m_f}^{2j}} 
\overset{\circ}{u}^{\dagger}(\lambda_f) \left(\begin{array}{cc}
0 &  \Pi(\tilde{p}_f) \\
\overbar{\Pi}(\tilde{p}_f^*) & 0
\end{array} \right) 
%\left( \Gamma \right)
\Gamma
\left(\begin{array}{cc}
 \Pi(\tilde{p}_i) & 0 \\
0 & \overbar{\Pi}(\tilde{p}_i^*)
\end{array} \right) 
%{1\over {m_i}^{2j}}  
\overset{\circ}{u}(\lambda_i) 
\nonumber \\[1em]
%%
%%%%%%%%%%%%%%%%%%%%%%%%%%%%%
%%
&= 
% { 1 \over (m_f m_i)^{2j}} 
\overset{\circ}{u}^{\dagger}(\lambda_f)
\left(\begin{array}{cc}
0 &  {t}^{\beta_{1} \cdots \beta_{2j}} \tilde{p}_{f,\beta_1} \cdots \tilde{p}_{f,\beta_{2j}} \\
  \bar{t}^{\beta_{1} \cdots \beta_{2j} } (\tilde{p}_{f,\beta_1} \cdots \tilde{p}_{f,\beta_{2j}})^* & 0
\end{array} \right) \nonumber \\
&\hspace{1cm} \times\Gamma
\left(\begin{array}{cc}
 {t}^{{\alpha}_{1} \cdots \alpha_{2j}} \tilde{p}_{i,\alpha_1} \cdots \tilde{p}_{i,\alpha_{2j}} & 0 \\
0 & \bar{t}^{\alpha_{1} \cdots \alpha_{2j}} (\tilde{p}_{i,\alpha_{1}} \cdots \tilde{p}_{i,\alpha_{2j}})^*
\end{array} \right)  
\overset{\circ}{u}(\lambda_i)
\,. 
\end{align}
These expressions use general $\tilde p$ boost parameters, so are valid for any considered choice of standard boost and spinors.
The overall structure of the matrix multiplications, which we want to reduce to an expression linear in the $t$-tensor, is not affected by the contractions of all the $\tilde p$, which happen in the Lorentz indices. Therefore, we factorize the parameters into a diagonal matrix,
\begin{equation}
    \mathcal{P}^{i}_{\alpha_{1} \cdots \alpha_{2j}} = 
\left(\begin{array}{cc}
\tilde{p}_{i,\alpha_{1}} \cdots \tilde{p}_{i,\alpha_{2j}} & 0 \\
0 & \left(\tilde{p}_{i,\alpha_{1}} \cdots \tilde{p}_{i,\alpha_{2j}}\right)^*
\end{array} \right) 
\,, 
\end{equation}
and similarly for the final state set of parameters. In this manner, Eq.~(\ref{eq:bilinear_gen_setup}) can be written as
\begin{align}\label{eq:bilinear_gen_reduction}
\bar{u}(p_f,\lambda_f) \; \Gamma \; u(p_i,\lambda_i) 
= & \overset{\circ}{u}^{\dagger}(\lambda_f)\,
\mathcal{P}^{f}_{\beta_{1} \cdots \beta_{2j}}
\left(\begin{array}{cc}
0 &  {t}^{\beta_{1} \cdots \beta_{2j}} \\
  \bar{t}^{ \beta_{1} \cdots  \beta_{2j} } & 0
\end{array} \right) 
\Gamma
\left(\begin{array}{cc}
 {t}^{{\alpha}_{1} \cdots \alpha_{2j}} & 0 \\
0 & \bar{t}^{ \alpha_{1} \cdots  \alpha_{2j}}
\end{array} \right)  
\mathcal{P}^{i}_{\alpha_{1} \cdots \alpha_{2j}}
\,\overset{\circ}{u}(\lambda_i)\,.
\end{align}
The appearance of the non-covariant $\mathcal{P}^{f/i}$ reflects relativistic spin effects, related to the fact that the spinors do no transform simply under Lorentz transformations, see Eq.~(\ref{eq:spinor_transf}).

We are now in the position to give useful general expressions that the bilinears take for each independent element of the $\Gamma$ basis.  
% These expressions will be valid for any spin representation and boost parameterization.  
% 
Following the discussion at the end of Sec.~\ref{sec:boosts_spinors}, we remind the reader that the expressions can be evaluated in two manners
\begin{enumerate}
    \item (``Full $\tilde p$'') One uses the $\tilde p$ for the corresponding spinor choice (Eq.~(\ref{eq:ptilde canonical}) for canonical, Eq.~(\ref{eq:ptilde_h}) or (\ref{eq:ptilde_h'}) for helicity, Eq.~(\ref{eq:ptilde_LF}) for light-front) and one uses the (simple) standard rest-frame spinors $\overset{\circ}{u}(\lambda)$ of Eq.~(\ref{eq:spinors_RF}).
    \item (``Canonical-Melosh'') One uses the real-valued canonical $\tilde p_c$ \textbf{but} then one has to use the Melosh rotated rest-frame spinors of Eq.~(\ref{eq:spinor_RF_meloshrotated}).
\end{enumerate}
% Both approaches will yield identical results but the second approach has the advantage of the real-valued $\tilde p_c$, which allows to simplify the expressions of some of the bilinears further, as will be shown in Sec.~\ref{sec:bitensor_bilinears}.
Both approaches will yield identical results but with the second approach we can take advantage of the real-valued parameters $\tilde p_c$, which allows to simplify the expressions of some of the bilinears further.
% , as will be shown in Sec.~\ref{sec:bitensor_bilinears}.
% 
As with the generalized Dirac basis obtained in Sec.~\ref{sec:dirac_basis}, we start with bilinears corresponding to the \emph{anti-diagonal} block matrices of the Dirac basis, {\it i.e.} $\gamma^{\mu_1\cdots \mu_{2j}}$ and $\gamma^{\mu_1\cdots \mu_{2j}}\gamma_5$.
Then, we study the bilinears corresponding to block \emph{diagonal} matrices, which are associated with $\gamma_5$ and the $2j+1$ covariantly independent bi-indexed tensors $\mathbb{1}$, $\mathsf{G}_m^{\mu_1 \rho_1 \cdots \rho_{m} \mu_{m}}$ ($\mathcal{G}_m^{\mu_1 \rho_1 \cdots \rho_{m} \mu_{m}}$), with $1 \leq m \leq 2j$.

\subsection{Tensor and pseudo-tensor bilinears}
\label{sec:tensor_bilinears}

For the tensor bilinear, we substitute $\Gamma=\gamma^{ {\mu}_{1} \cdots {\mu}_{2j} }$, in Eq.~(\ref{eq:bilinear_gen_reduction})
\begin{align}%\label{eq:tensor_bilinear}
\bar{u}
(p_f,\lambda_f)  
\gamma^{ {\mu}_{1} \cdots {\mu}_{2j} }
u(p_i,\lambda_i)  
& = \overset{\circ}{u}^\dagger (\lambda_f)
\mathcal{P}^{f}_{\beta_{1} \cdots \beta_{2j}}
\left(\begin{array}{cc}
  {t}^{(\beta)} \bar{t}^{(\mu)} {t}^{(\alpha)}   &  0  \\
  0  &  \bar{t}^{(\beta)} {t}^{(\mu)} \bar{t}^{(\alpha)} 
\end{array} \right) 
\mathcal{P}^{i}_{\alpha_{1} \cdots \alpha_{2j}}
\overset{\circ}{u}(\lambda_i)  
\nonumber\\
& = 
\overset{\circ}{\phi}^\dagger (\lambda_f) \left[ 
\left(\prod^{2j}_{l=1} \tilde{p}_{f,{\beta}_{l}} \, \tilde{p}_{i,{\alpha}_{l}} \right) 
{t}^{(\beta)} \bar{t}^{(\mu)} {t}^{(\alpha)} 
+ \left(\prod^{2j}_{l=1} \tilde{p}_{f,{\beta}_{l}} \, \tilde{p}_{i,{\alpha}_{l}} \right)^* 
\bar{t}^{(\beta)} {t}^{(\mu)} \bar{t}^{(\alpha)} 
\right]
\overset{\circ}{\phi}(\lambda_i) 
\,,
\end{align}
where the expression was reduced from bispinor to chiral spinor form. The rest frame chiral spinors $\overset{\circ}{\phi}(\lambda)$ were introduced in Eqs.~(\ref{eq:chiralspinor_rf}).
Making use of the reduction formulae for the cubic products of $t$-tensors, 
Eqs.~(\ref{eq:t_reduction_cubic}) and (\ref{eq:t_reduction_cubicbar}), this can be written as
\begin{align}\label{eq:tensor_bilinear_setup}
\bar{u}(p_f,\lambda_f)  
\gamma^{ {\mu}_{1} \cdots {\mu}_{2j} }
u(p_i,\lambda_i)  
& = \frac{1}{[(2j)!]^2}  \underset{\left\{(\mu), (\alpha)\right\}}{\mathcal{S}} 
\overset{\circ}{\phi}^\dagger (\lambda_f)
\left[ \left( \prod_{l=1}^{2j} \tilde{p}_{f,{\beta}_{l}} \, \tilde{p}_{i,{\alpha}_{l}} {\mathcal{C}}^{\beta_{l} \mu_{l} \alpha_{l} \xi_{l}} \right) t_{\xi_{1} \cdots \xi_{2j}}
\right.
\nonumber\\
& \left. \hspace{4.cm} + \left( \prod_{l=1}^{2j} \left(\tilde{p}_{f,{\beta}_{l}} \, \tilde{p}_{i,{\alpha}_{l}}\right)^*  \overbar{\mathcal{C}}^{\beta_{l} \mu_{l} \alpha_{l} \xi_{l}} \right) \bar{t}_{\xi_{1} \cdots \xi_{2j}}
\right]
\overset{\circ}{\phi}(\lambda_i) 
\nonumber\\
&   = m_f^j m_i^j    \left[  \left( \prod_{l=1}^{2j} {\mathcal{C}}^{\tilde p_f \mu_{l} \tilde p_i \xi_{l}} \right) 
   \langle \lambda_f| t_{\xi_{1} \cdots \xi_{2j}} | \lambda_i \rangle
 + \left( \prod_{l=1}^{2j} {\mathcal{C}}^{\tilde p_f \mu_{l} \tilde p_i \xi_{l}} \right)^* \langle \lambda_f| \bar t_{\xi_{1} \cdots \xi_{2j}} | \lambda_i \rangle
 \right] \,,
\end{align}
where we use the notation
\begin{equation}\label{eq:Ctilde}
    \mathcal{C}^{\tilde p_f \mu_l \tilde p_i \xi _l} \equiv \mathcal{C}^{\beta_{l} \mu_{l} \alpha_{l} \xi_{l}}\tilde p_{f,\beta_l} \tilde p_{i,\alpha_l}
    \,,
\end{equation}
and
\begin{align}\label{eq:t_tensor_matrix_elements_bw_rest_spinors}
    &\langle \lambda_f| t_{\xi_{1} \cdots \xi_{2j}} | \lambda_i \rangle = \frac{\overset{\circ}{\phi}^\dagger(\lambda_f)t_{\xi_{1} \cdots \xi_{2j}} \overset{\circ}{\phi}(\lambda_i)}{m_f^j m_i^j}
    \,,
\end{align}
denotes the normalized contraction of a $t$-tensor with the \emph{appropriate} rest-frame spinor (see remarks at the end of Sec.~\ref{sec:bilinears_general}).
Through the contractions with symmetric objects the product of the $2j$ copies of ${\mathcal{C}}$ is symmetric in the three set of indices $\beta_l$, $\alpha_l$, and $\xi_l$. Thus, the initial expression in Eq.~(\ref{eq:tensor_bilinear_setup}) is guaranteed to be symmetric in the remaining index set $\mu_l$ and all symmetrizations are trivial.

We now introduce the average and difference between $\tilde p_f$ and $\tilde p_i$ as
\begin{subequations}\label{eq:P_Delta_tildes}
\begin{align} 
\label{eq:P_tilde}
& \widetilde{P} = \frac{1}{2}\left( \tilde{p}_f + \tilde{p}_i \right) \,,
\\
\label{eq:Delta_tilde}
& \widetilde\Delta =  \tilde{p}_f - \tilde{p}_i
\,.
\end{align}
\end{subequations}
Similar to $\tilde p_f$ and $\tilde p_i$, also $\widetilde P$ and $\widetilde \Delta$ are \textit{not} four-vectors.  They only appear (contracted) in the final expressions for the spinors and bilinears and we never need to consider any transformation properties. Thus, for all purposes we can still do ordinary vector calculus with them, treating them here just as vectors. We have 
\begin{equation}\label{eq:parameters_to_relative_average}
\tilde{p}^{f}_{\beta} \tilde{p}^{i}_{\alpha} = \left( \widetilde{P}_{\beta} \widetilde{P}_{\alpha} -\frac{1}{4} \widetilde{\Delta}_{\beta} \widetilde{\Delta}_{\alpha} \right) + \frac{1}{2}\left( \tilde{\Delta}_{\beta} \tilde{P}_{\alpha} - \tilde{\Delta}_{\alpha} \tilde{P}_{\beta} \right)
\,,
\end{equation}
which allows us to write Eq.~(\ref{eq:Ctilde}) as
\begin{align}\label{eq:contraction_C_parameters}
\mathsf{V}^{\mu \xi} (\tilde{P},\tilde{\Delta}) \equiv \mathcal{C}^{\tilde p_f \mu \tilde p_i \xi}  = 2\left( \widetilde{P}^\mu \widetilde{P}^\xi - \frac{1}{4}\widetilde{\Delta}^\mu \widetilde{\Delta}^\xi \right) - \left( \widetilde{P}^2  - \frac{1}{4}\widetilde{\Delta}^2 \right) g^{\mu \xi} + {\rm i} \varepsilon^{ \mu \xi \widetilde{P} \widetilde{\Delta}}
%\\
%& = 2\left( \widetilde{P}^\mu \widetilde{P}^\xi - \frac{1}{4}\widetilde{\Delta}^\mu \widetilde{\Delta}^\xi \right) - \left( \tilde{p}_f \cdot \tilde{p}_i \right) g^{\mu \xi} + {\rm i} \varepsilon^{ \mu \xi \widetilde{P} \widetilde{\Delta}}
\,.
\end{align}
When elastic currents $m_f=m_i$ in the Breit frame ($\bm{p}_f=-\bm{p}_i $) are considered, $\tilde{p}_f \cdot \tilde{p}_i=\widetilde{P}^2  - \widetilde{\Delta}^2/4 = 1$.  This choice of frame simplifies Eq.~(\ref{eq:contraction_C_parameters}) and all expressions that depend on it considerably. 

Using the notation of Eq.~(\ref{eq:contraction_C_parameters}), we have for the tensor bilinear $\Gamma=\gamma^{ {\mu}_{1} \cdots {\mu}_{2j} }$,
\begin{align}\label{eq:tensor_bilinear}
\bar{u}(p_f,\lambda_f) \, 
\gamma^{ {\mu}_{1} \cdots {\mu}_{2j} }\,
u(p_i,\lambda_i)  &  =  m_f^j m_i^j\left\{ \left[ \prod_{l=1}^{2j}  \mathsf{V}^{\mu_{l} \xi_{l}} (\tilde{P},\tilde{\Delta}) \right] 
\langle \lambda_f| t_{\xi_{1} \cdots \xi_{2j}} | \lambda_i \rangle
+ \left[ \prod_{l=1}^{2j}  \mathsf{V}^{\mu_{l} \xi_{l}} (\tilde{P},\tilde{\Delta}) \right]^* 
\langle \lambda_f| \bar{t}_{\xi_{1} \cdots \xi_{2j}} | \lambda_i \rangle\right\}
\,,
\end{align}
while for the pseudotensor $\Gamma=\gamma^{ {\mu}_{1} \cdots {\mu}_{2j} } \gamma_5$, we find
\begin{align}\label{eq:pseudotensor_bilinear}
\bar{u}(p_f,\lambda_f) \, 
\gamma^{ {\mu}_{1} \cdots {\mu}_{2j} } \gamma_5\,
u(p_i,\lambda_i)  
& = \overset{\circ}{u}^\dagger (\lambda_f)
\mathcal{P}^{f}_{\beta_{1} \cdots \beta_{2j}}
\left(\begin{array}{cc}
  - {t}^{(\beta)} \bar{t}^{(\mu)} {t}^{(\alpha)}   &  0  \\
  0  &  \bar{t}^{(\beta)} {t}^{(\mu)} \bar{t}^{(\alpha)} 
\end{array} \right) 
\mathcal{P}^{i}_{\alpha_{1} \cdots \alpha_{2j}}
\overset{\circ}{u}(\lambda_i)  
\nonumber\\
& \hspace{-1.cm} = m_f^j m_i^j\left\{ - \left[ \prod_{l=1}^{2j}  \mathsf{V}^{\mu_{l} \xi_{l}} (\tilde{P},\tilde{\Delta}) \right]  
\langle \lambda_f| t_{\xi_{1} \cdots \xi_{2j}} | \lambda_i \rangle
+ \left[ \prod_{l=1}^{2j}  \mathsf{V}^{\mu_{l} \xi_{l}} (\tilde{P},\tilde{\Delta}) \right]^* 
\langle \lambda_f| \bar{t}_{\xi_{1} \cdots \xi_{2j}} | \lambda_i \rangle\right\}
\,,
\end{align}
where the only difference with the tensor bilinear of Eq.~(\ref{eq:tensor_bilinear}) is the sign of the first term, originating from the extra $\gamma_5$.

We finish this section with some remarks about the flexibility of our framework, 
% in the spirit of the remarks at 
see the end of Sec.~\ref{sec:bilinears_general}. 
Equations~(\ref{eq:tensor_bilinear}) and (\ref{eq:pseudotensor_bilinear}) were obtained by writing the momentum dependent spinors as a boost, represented by a matrix in spinor space, acting on a rest-frame spinor.  Then the combination of boost and Dirac matrices are simplified by using the reduction formulas for the $t$-tensors.  Up to here the formulas are valid for both ``Full $\tilde p$'' and ``Canonical-Melosh'' approaches.  

In the ``Full $\tilde p$'' approach, all the information about the motion of the particles appears in the boost parameters $\tilde p$, which are spinor choice dependent and in general complex valued. The rest-frame spinors are the simplest possible [Eq.~(\ref{eq:spinors_RF})] and select the element on row $\lambda_f$ and column $\lambda_i$ in the $t$-tensor matrices.   From Sec.~\ref{sec:t_tensors_gen}, we know that in the light-front spherical basis these matrices only contain one non-zero element and are thus extremely sparse, simplifying the numerical evaluation.

In the ``Canonical-Melosh'' approach, the motion of the particles is contained both in the rest-frame spinors $\langle \lambda_f|\cdots|\lambda_i\rangle$, which are now the \emph{Melosh-rotated} spinors of Eq.~(\ref{eq:spinor_RF_meloshrotated}), and also in the boost parameters $\tilde p= \tilde p_c$, which are real valued. 
% In the ``Canonical-Melosh'' approach, the motion of the particles is contained in both the rest-frame spinors and the canonical boosts. The rest-frame spinors $\langle \lambda_f|\cdots|\lambda_i\rangle$, which are now the \emph{Melosh-rotated} spinors of Eq.~(\ref{eq:spinor_RF_meloshrotated}). The boost parameters are the canonical, with parameters $\tilde p= \tilde p_c$, which are real valued.
The latter property allows to further simplify the expressions.  The (pseudo)tensor bilinear can be written as
\begin{subequations}\label{eq:tensor_bilinear_can_melosh}
\begin{align}\label{eq:tensor_bilinear_canonical}
\bar{u}(p_f,\lambda_f) \, 
\gamma^{ {\mu}_{1} \cdots {\mu}_{2j} }\,
u(p_i,\lambda_i)  
&  =  m_f^j m_i^j \left\{  
\operatorname{Re}\left[ \prod_{l=1}^{2j}  \mathsf{V}^{\mu_{l} \xi_{l}} (\widetilde{P}_{c},\widetilde{\Delta}_{c})   \right] 
\langle \tilde\lambda_f| (t+\bar{t})_{\xi_{1} \cdots \xi_{2j}}  | \tilde\lambda_i \rangle 
\right.\nonumber
\\
& \hspace{2cm} \left. + {\rm i} \operatorname{Im}\left[ \prod_{l=1}^{2j}  \mathsf{V}^{\mu_{l} \xi_{l}} (\widetilde{P}_{c},\widetilde{\Delta}_{c})   \right]
\langle \tilde\lambda_f| (t-\bar{t})_{\xi_{1} \cdots \xi_{2j}}  | \tilde\lambda_i \rangle 
\right\}
\,,
\end{align}    
\begin{align}\label{eq:pseudotensor_bilinear_canonical}
    \bar{u}(p_f,\lambda_f) \, 
\gamma^{ {\mu}_{1} \cdots {\mu}_{2j} } \gamma_5\,
u(p_i,\lambda_i)  
&  =  -m_f^j m_i^j \left\{ 
\operatorname{Re}\left[ \prod_{l=1}^{2j}  \mathsf{V}^{\mu_{l} \xi_{l}} (\widetilde{P}_{c},\widetilde{\Delta}_{c})   \right]
\langle \tilde\lambda_f| (t-\bar{t})_{\xi_{1} \cdots \xi_{2j}}  | \tilde\lambda_i \rangle 
\right.\nonumber
\\ 
& \hspace{2cm} \left. + {\rm i} \operatorname{Im}\left[ \prod_{l=1}^{2j}  \mathsf{V}^{\mu_{l} \xi_{l}} (\widetilde{P}_{c},\widetilde{\Delta}_{c})   \right] 
\langle \tilde\lambda_f| (t+\bar{t})_{\xi_{1} \cdots \xi_{2j}}  | \tilde\lambda_i \rangle \right\}
\,,
\\[1em]
&
% \hspace{2.5cm} 
[\text{Canonical-Melosh}]\nonumber
\end{align}
\end{subequations}
where we wrote $\langle \tilde \lambda_f|$, $|\tilde \lambda_i\rangle$ to emphasize the Melosh-rotated nature of these rest-frame spinors, and include a subindex $c$ to remind us that canonical boost parameters must be used.  
From these expressions, it is straightforward to read off the behavior of these bilinears under parity. $\operatorname{Re} [ \prod_{l=1}^{2j}  \mathsf{V}^{\mu_{l} \xi_{l}}(\widetilde{P}_{c},\widetilde{\Delta}_{c}) ]$ ($\operatorname{Im} [ \prod_{l=1}^{2j}  \mathsf{V}^{\mu_{l} \xi_{l}} (\widetilde{P}_{c},\widetilde{\Delta}_{c}) ]$) contains an even (odd) number of Levi-Civita tensors, thus behaves as a tensor (pseudo-tensor) under parity. Meanwhile, parity switches the positions of the left- and right-chiral spinor in the bispinor, which in Eqs.~(\ref{eq:tensor_bilinear_can_melosh}) results in interchanging the $t$ and $\bar t$ matrix element.  
Therefore, bilinear (\ref{eq:tensor_bilinear_canonical}) transforms as a proper tensor. For bilinear (\ref{eq:pseudotensor_bilinear_canonical}) there is a sign change, thus it transforms as a pseudotensor.  
Note that Eq.~(\ref{eq:t_tbar_sign_relations}) also restricts the number of spatial indices on the elements of the $t$-tensors that contribute in each of the terms, i.e. even (odd) in the first (second) term of Eq.~(\ref{eq:tensor_bilinear_canonical}) and vice versa for the pseudotensor epression.  These agree with the discussed behavior under parity.

It is worth noting that both approaches have their advantages in evaluating these bilinears.  
% The \emph{Canonical-Melosh} has no difference in the Lorentz-tensor structure, which is entirely written using invariant tensors and real parameters $\widetilde P, \widetilde \Delta$, which have the same form for every type considered. The difference is in the $t$-matrix elements, which are evaluated between the Melosh-rotated rest-frame spinors and differ for every type of spin. These are, however, straightforward to calculate, but the Melosh rotation depends on the momentum of the particles in the bilinear. 
For every type of spinor the \emph{Canonical-Melosh} has no difference in the Lorentz-tensor structure, which is entirely written using invariant tensors and the same (real) canonical parameters $\widetilde P, \widetilde \Delta$. The difference between spinors type appears in the $t$-matrix elements, which are evaluated between the Melosh-rotated rest-frame spinors. The Melosh rotation, although straightforward to calculate, depends on the momentum of the particles in the bilinear.  
% In the \emph{Full $\tilde p$} approach, one has to use Eqs.~(\ref{eq:tensor_bilinear}),(\ref{eq:pseudotensor_bilinear}).  For those expressions, the $t$-matrix elements are identical for any type of spin considered and one can use Eq.~(\ref{eq:t_index}) to determine which Lorentz elements of the $t$-tensors contribute for a certain spin transition.  These elements, through the $\widetilde P, \widetilde \Delta$ that contracts the matrix elements, then determine which elements of the bilinear will be non-zero.
In the \emph{Full $\tilde p$} approach Eqs.~(\ref{eq:tensor_bilinear}), (\ref{eq:pseudotensor_bilinear}) are used. 
For those expressions, the $t$-matrix elements remain the same for any spin types. Equation~(\ref{eq:t_index}) can be used to identify which specific Lorentz components of the $t$-tensors contribute to a given spin transition. 
These components, combined with $\widetilde P$ and $\widetilde \Delta$, determine the non-zero matrix elements of the bilinear.
% In both approaches the contraction in Lorentz indices is best evaluated in the $\{+-R\,L\}$-basis, as it was shown in Sec.~\ref{sec:t_tensors_gen} that the $t$-tensors are particularly simple and sparse in that basis.
In both approaches one can take advantage of the $\{+-R\,L\}$-basis for the contraction in Lorentz indices, as it was shown in Sec.~\ref{sec:t_tensors_gen} that the $t$-tensors are particularly simple and sparse in that basis.

\subsection{Scalar and pseudo-scalar bilinears}
\label{sec:scalar_pseudoscalar_bilinears}

The scalar bilinear corresponds to the special case $m=0$ of the covariantly independent bi-indexed elements $\mathsf{G}_m^{\mu_1 \rho_1 \cdots \rho_{m} \mu_{m}}$, which we present in the next section. 
We single it out here, however, because it is the simplest multipole (monopole) and it is present for all spin representations.
Substituting $\Gamma=\mathbb{1}$  in Eq.~(\ref{eq:bilinear_gen_reduction}), we have
\begin{align}\label{eq:scalar_bilinear}
\bar{u}{(p_f,s_f)}  
%\Gamma
u{(p_i,s_i)}  
& = \overset{\circ}{u}^\dagger(\lambda_f) \mathcal{P}^{f}_{\beta_{1} \cdots \beta_{2j}}
\left(\begin{array}{cc}
  0  &  {t}^{(\beta)} \bar{t}^{(\alpha)}   \\
 \bar{t}^{(\beta)} {t}^{(\alpha)}  &  0 
\end{array} \right) 
 \mathcal{P}^{i}_{\alpha_{1} \cdots \alpha_{2j}} \overset{\circ}{u}(\lambda_i) 
\nonumber\\
& = \overset{\circ}{\phi}^\dagger(\lambda_f) 
\left[
\left(\prod^{2j}_{l=1} \tilde{p}_{f,{\beta}_{l}} \, \tilde{p}_{i,{\alpha}_{l}}^* \right)
% (\tilde{p}_{f,{\beta}_{1}} \cdots \tilde{p}_{f,{\beta}_{2j}} ) 
{t}^{(\beta)} \bar{t}^{(\alpha)} 
% (\tilde{p}_{i,{\alpha}_{1}} \cdots \tilde{p}_{i,{\alpha}_{2j}})^* 
% \right.\nonumber \\
% &\hspace{2cm} \left. 
% + (\tilde{p}_{f,{\beta}_{1}} \cdots \tilde{p}_{f,{\beta}_{2j}})^*
+ \left(\prod^{2j}_{l=1} \tilde{p}_{f,{\beta}_{l}}^* \, \tilde{p}_{i,{\alpha}_{l}} \right)
\bar{t}^{(\beta)} {t}^{(\alpha)} 
% (\tilde{p}_{i,{\alpha}_{1}} \cdots \tilde{p}_{i,{\alpha}_{2j}}) 
\right] 
\overset{\circ}{\phi}(\lambda_i)  
\,,
\end{align}
where rest frame chiral spinors $\overset{\circ}{\phi}(\lambda)$ are given in Eqs.~(\ref{eq:chiralspinor_rf}).
Making use of the reduction formulae for the quadratic products of $t$-tensors 
Eqs.~(\ref{eq:def_t_reduction_quadratic}), and the notation of Eq.~(\ref{eq:t_tensor_matrix_elements_bw_rest_spinors}), we find
\begin{align}\label{eq:scalar_bilinear_setup}
\bar{u}(p_f,\lambda_f)  
u(p_i,\lambda_i)  &  = m_f^j m_i^j    \left\{  \left( \prod_{l=1}^{2j} {\mathcal{Q}}^{\tilde p_f  \tilde p_i^* \xi_{l}} \right) 
   \langle \lambda_f| t_{\xi_{1} \cdots \xi_{2j}} | \lambda_i \rangle
 + \left( \prod_{l=1}^{2j} {\mathcal{Q}}^{\tilde p_f \tilde p_i^* \xi_{l}} \right)^* \langle \lambda_f| \bar t_{\xi_{1} \cdots \xi_{2j}} | \lambda_i \rangle
 \right\}
 \,.
\end{align}

For the pseudo-scalar $\Gamma=\gamma_5$  in Eq.~(\ref{eq:bilinear_gen_reduction}), we have
\begin{align}\label{eq:pseudo-scalar_bilinear}
\bar{u}{(p_f,s_f)}  
\gamma_5
u{(p_i,s_i)} 
& =  \overset{\circ}{u}^\dagger(\lambda_f) \mathcal{P}^{f}_{\beta_{1} \cdots \beta_{2j}} 
\left(\begin{array}{cc}
  0  &  {t}^{(\beta)} \bar{t}^{(\alpha)}   \\
   - \bar{t}^{(\beta)} {t}^{(\alpha)}  &  0 
\end{array} \right) 
 \mathcal{P}^{i}_{\alpha_{1} \cdots \alpha_{2j}} \overset{\circ}{u}(\lambda_i)   
\nonumber \\
& = m_f^j m_i^j    \left\{  \left( \prod_{l=1}^{2j} {\mathcal{Q}}^{\tilde p_f  \tilde p_i^* \xi_{l}} \right) 
   \langle \lambda_f| t_{\xi_{1} \cdots \xi_{2j}} | \lambda_i \rangle
 - \left( \prod_{l=1}^{2j} {\mathcal{Q}}^{\tilde p_f \tilde p_i^* \xi_{l}} \right)^* \langle \lambda_f| \bar t_{\xi_{1} \cdots \xi_{2j}} | \lambda_i \rangle
 \right\},
\end{align}
where the only difference with the scalar bilinear of Eq.~(\ref{eq:scalar_bilinear}) is the sign of the second term, originating from the extra $\gamma_5$. 

% In the ``Canonical-Melosh'' approach, using Eqs.~(\ref{eq:P_Delta_tildes}) and the fact that the canonical boost-parameters are real valued, we introduce
We take advantage of the ``Canonical-Melosh'' approach, where the (canonical) boost-parameters are real valued, to simplify further.  
Using Eqs.~(\ref{eq:parameters_to_relative_average}) we introduce
\begin{equation}
\label{eq:contraction_Q_parameters}
\mathsf{S}^{\xi} (\widetilde{P}_{c},\widetilde{\Delta}_{c}) = \mathcal{Q}^{\tilde p_f^c  \tilde p_i^c \xi}  = \left( \widetilde{P}_c^2  - \frac{1}{4}\widetilde{\Delta}_c^2 \right) \eta^{\xi}  - \left( \widetilde{P}_c^0  \widetilde{\Delta}_c^\xi - \widetilde{\Delta}_c^0 \widetilde{P}_c^\xi \right) + {\rm i} \varepsilon^{0 \widetilde{\Delta}_c \widetilde{P}_c \xi}
\,,
\end{equation}
where the first term on the right side is zero when evaluated in the Breit frame (see remark in the previous section).
The (pseudo)scalar bilinear can be written as
\begin{subequations}\label{eq:scalar_bilinear_can_melosh}
\begin{align}\label{eq:scalar_bilinear_canonical}
\bar{u}(p_f,\lambda_f) 
u(p_i,\lambda_i)  
&  =  m_f^j m_i^j \left\{ \operatorname{Re} \left[ \prod_{l=1}^{2j}  \mathsf{S}^{\xi_{l}} (\widetilde{P}_{c},\widetilde{\Delta}_{c}) \right]
\langle \tilde\lambda_f| (t+\bar{t})_{\xi_{1} \cdots \xi_{2j}}  | \tilde\lambda_i \rangle 
\right. \nonumber \\ 
& \left. \hspace{2.cm}
+ {\rm i} \operatorname{Im} \left[ \prod_{l=1}^{2j}  \mathsf{S}^{\xi_{l}} (\widetilde{P}_{c},\widetilde{\Delta}_{c}) \right] 
\langle \tilde\lambda_f| (t-\bar{t})_{\xi_{1} \cdots \xi_{2j}}  | \tilde\lambda_i \rangle 
\right\}
\,,
\end{align}    
\begin{align}\label{eq:pseudoscalar_bilinear_canonical}
    \bar{u}(p_f,\lambda_f) \, 
 \gamma_5\,
u(p_i,\lambda_i)  
&  =  m_f^j m_i^j \left\{ \operatorname{Re} \left[ \prod_{l=1}^{2j}  \mathsf{S}^{\xi_{l}} (\widetilde{P}_{c},\widetilde{\Delta}_{c}) \right]
\langle \tilde\lambda_f| (t-\bar{t})_{\xi_{1} \cdots \xi_{2j}}  | \tilde\lambda_i \rangle 
\right. \nonumber \\ 
& \left. \hspace{2.cm}
+ {\rm i} \operatorname{Im} \left[ \prod_{l=1}^{2j}  \mathsf{S}^{\xi_{l}} (\widetilde{P}_{c},\widetilde{\Delta}_{c}) \right] 
\langle \tilde\lambda_f| (t+\bar{t})_{\xi_{1} \cdots \xi_{2j}}  | \tilde\lambda_i \rangle 
\right\}
\,,
\\[1em]
&
% \hspace{2.5cm} 
[\text{Canonical-Melosh}]\nonumber
\end{align}
\end{subequations}
where we wrote $\langle \tilde \lambda_f|\cdots|\tilde \lambda_i\rangle$ to emphasize the Melosh-rotated nature of these rest-frame spinors, and include a subindex $c$ to emphasize that (real-valued) canonical boost parameters must be used.  From these expressions, it is straightforward to read off the behavior of these bilinears under parity.  
$\operatorname{Re} [ \prod_{l=1}^{2j}  \mathsf{S}^{\xi_{l}}(\widetilde{P}_{c},\widetilde{\Delta}_{c}) ]$ ($\operatorname{Im} [ \prod_{l=1}^{2j}  \mathsf{S}^{\xi_{l}} (\widetilde{P}_{c},\widetilde{\Delta}_{c}) ]$) contains an even (odd) number of Levi-Civita tensors, thus behaves as a proper tensor (pseudo-tensor) under parity. Meanwhile, parity switches the positions of the left- and right-chiral spinor in the bispinor, which results in interchanging the $t$ and $\bar t$ matrix element. 
%Parity switches the positions of the left- and right-chiral spinor in the bispinor, which in Eqs.~(\ref{eq:scalar_bilinear_can_melosh}) results in interchanging the $t$ and $\bar t$ matrix element.  
For bilinear (\ref{eq:scalar_bilinear_canonical}) the expression does not change sign under a parity inversion, meaning it transforms as a scalar. For bilinear (\ref{eq:pseudoscalar_bilinear_canonical}) there is a sign change, thus it transforms as a pseudo-scalar.

\subsection{ Bilinears of bi-indexed tensors}
\label{sec:bitensor_bilinears}

For the sake of simplicity of the resulting expressions, in this section we provide a general expression for bi-indexed-tensor bilinears in terms of the Dirac basis elements $\mathsf{G}_m^{(\mu \rho)}$ instead of the multipoles $\mathcal{G}_m^{(\mu \rho)}$. The bilinears in terms of the covariant multipoles can be obtained using the equations (\ref{eq:multipoles_gamma}) and (\ref{eq:Tau_m}) together with the results shown here.  

Each of the $2j$ covariantly independent bi-indexed-tensor bilinears related to $\Gamma = \mathsf{G}_m^{(\mu \rho)}$  ($1 \leq m \leq 2j$) corresponds to
\begin{align}\label{eq:bi_indexed_bilinear_def}
\bar{u}{(p_f,s_f)}  
\mathsf{G}_m^{(\mu \rho)}
u{(p_i,s_i)}  = & \overset{\circ}{\phi}^\dagger(\lambda_f) 
\left[ 
% (\tilde{p}_{f,{\beta}_{1}} \cdots \tilde{p}_{f,{\beta}_{2j}} )
\left(\prod^{2j}_{l=1} \tilde{p}_{f,{\beta}_{l}} \, \tilde{p}_{i,{\alpha}_{l}}^* \right)
{t}^{(\beta)} \, 
\bar{t}_{\xi_{1} \cdots \xi_{m} 0 \cdots 0} \left(\prod_{l=1}^{m} \overbar{\mathcal{Q}}_{\text{red}}^{\mu_{l} \rho_{l} \xi_{l}}\right) 
\bar{t}^{(\alpha)} 
\right. \nonumber \\
& \hspace{1.cm} \left. 
% + (\tilde{p}_{f,{\beta}_{1}} \cdots \tilde{p}_{f,{\beta}_{2j}})^*
\; + \left(\prod^{2j}_{l=1} \tilde{p}_{f,{\beta}_{l}}^* \, \tilde{p}_{i,{\alpha}_{l}} \right)
\bar{t}^{(\beta)} \,
t_{\xi_{1} \cdots \xi_{m} 0 \cdots 0} \left(\prod_{l=1}^{m} {\mathcal{Q}}_{\text{red}}^{\mu_{l} \rho_{l} \xi_{l}} \right) 
{t}^{(\alpha)} 
\right]
\overset{\circ}{\phi}(\lambda_i)  
\,,
\end{align}
where rest-frame chiral spinors $\overset{\circ}{\phi}(\lambda)$ are given in Eqs.~(\ref{eq:chiralspinor_rf}).
In order to simplify Eq.~(\ref{eq:bi_indexed_bilinear_def}) we use twice the reduction formulas for the quadratic products of $t$-tensors Eqs.~(\ref{eq:def_t_reduction_quadratic}). As suggested by the grouping in Eq.~(\ref{eq:bi_indexed_bilinear_def}), we perform a first reduction on the quadratic product ${t}^{(\beta)} \, \bar{t}_{\xi_{1} \cdots \xi_{m} 0 \cdots 0}$  ($\bar{t}^{(\beta)} \, {t}_{\xi_{1} \cdots \xi_{m} 0 \cdots 0}$), see Sec.~\ref{sec:quadratic_t_products_reducible} for details on a method of reducing these products. The resulting $t$-tensor ($\bar{t}$-tensor) is combined with the remaining $\bar{t}^{(\alpha)}$ (${t}^{(\alpha)}$) and subsequently also reduced.
% and the notation of Eq.~(\ref{eq:t_tensor_matrix_elements_bw_rest_spinors}), 
We find  
\begin{align}\label{eq:bi_indexed_bilinear_reduction}
\bar{u}{(p_f,s_f)}  
\mathsf{G}_m^{(\mu \rho)}
u{(p_i,s_i)}  = & \left[ \prod_{l=1}^{m} \overbar{\mathcal{Q}}_{\text{red}}^{\mu_{l} \rho_{l}}{}_{\xi_{l}} {\mathcal{Q}}^{\tilde{p}_{f} \xi_{l}}{}_{\sigma_{l}} {\mathcal{Q}}^{\sigma_{l} \tilde{p}_{i}^*  \nu_{l}} \right]
\left[ \prod_{l=m+1}^{2j} {\mathcal{Q}}^{\tilde{p}_{f} \tilde{p}_{i}^* \nu_{l}}  \right]
 \langle \lambda_f|  t_{\nu_{1} \cdots \nu_{2j}} | \lambda_i \rangle
\nonumber\\
& + \left[ \prod_{l=1}^{m} \overbar{\mathcal{Q}}_{\text{red}}^{\mu_{l} \rho_{l}}{}_{\xi_{l}} {\mathcal{Q}}^{\tilde{p}_{f} \xi_{l}}{}_{\sigma_{l}} {\mathcal{Q}}^{\sigma_{l} \tilde{p}_{i}^*  \nu_{l}} \right]^*
\left[\prod_{l=m+1}^{2j} {\mathcal{Q}}^{\tilde{p}_{f} \tilde{p}_{i}^* \nu_{l}}  \right]^*
 \langle \lambda_f| \bar t_{\nu_{1} \cdots \nu_{2j}} | \lambda_i \rangle
 \,.
\end{align}

In the ``Canonical-Melosh'' approach, using Eqs.~(\ref{eq:parameters_to_relative_average}) and the fact that the canonical boost parameters are real valued, we introduce
%where, $\mathsf{B}^{\mu_{l} \rho_{l} \nu_{l}}$ is given by
\begin{align}\label{eq:bi_indexed_bilinear_B}
\mathsf{B}^{\mu \rho \nu} (\widetilde{P}_{c},\widetilde{\Delta}_{c}) = & \;\overbar{\mathcal{Q}}_{\text{red}}^{\, \mu \rho}{\,}_{\xi} {\mathcal{Q}}^{\tilde{p}_{f}^c \xi}{}_{\sigma} {\mathcal{Q}}^{\sigma \tilde{p}_{i}^c{}^*  \nu} = \overbar{\mathcal{Q}}_{\text{red}}^{\, \mu \rho}{\,}_{\xi} {\mathcal{Q}}^{\tilde{p}_{f}^c \xi}{}_{\sigma} {\mathcal{Q}}^{\sigma \tilde{p}_{i}^c  \nu} 
\nonumber\\
 = &  \; 
%\left[ 
2 \left( \tilde{P}^{\mu}_c \tilde{P}^{\nu}_c - \frac{1}{4} \tilde{\Delta}^{\mu}_c \tilde{\Delta}^{\nu}_c \right) \eta^{\rho} - 2 \left( \tilde{P}^{\rho}_c \tilde{P}^{\nu}_c - \frac{1}{4} \tilde{\Delta}^{\rho}_c \tilde{\Delta}^{\nu}_c \right) \eta^{\mu} 
%\right.
\nonumber\\
& 
%\left. \hspace{.5cm} 
+ 2 \left( \tilde{P}^{\mu}_c \tilde{P}^{0}_c - \frac{1}{4} \tilde{\Delta}^{\mu}_c \tilde{\Delta}^{0}_c \right) g^{\rho \nu} - 2 \left( \tilde{P}^{\rho}_c \tilde{P}^{0}_c - \frac{1}{4} \tilde{\Delta}^{\rho}_c \tilde{\Delta}^{0}_c \right) g^{\mu \nu} 
%\right.
\nonumber\\
& 
%\left.  \hspace{.5cm} 
+ \left( \tilde{P}^2_c - \frac{1}{4} \tilde{\Delta}^2_c \right) \left( \eta^{\mu} g^{\rho \nu} - \eta^{\rho} g^{\mu \nu} \right) + \left( \tilde{P}^{\mu}_c \tilde{\Delta}^{\rho}_c - \tilde{P}^{\rho}_c \tilde{\Delta}^{\mu}_c \right) \eta^{\nu} 
%\right.
\nonumber\\
& 
%\left. \hspace{.5cm} 
- 2 \left( \tilde{P}^{0}_c \tilde{\Delta}^{\rho}_c - \tilde{P}^{\rho} \tilde{\Delta}^{0}  \right) g^{\mu \nu} + 2 \left( \tilde{P}^{0}_c \tilde{\Delta}^{\mu} - \tilde{P}^{\mu}_c \tilde{\Delta}^{0}_c \right) g^{\rho\nu} 
%\right.
\nonumber\\
& 
%\left. \hspace{.5cm}  
+ {\rm i} {\epsilon}^{\tilde{\Delta}_c \tilde{P}_c \rho \nu} \eta^{\mu} - {\rm i} {\epsilon}^{\tilde{\Delta}_c \tilde{P}_c \mu \nu} \eta^{\rho} + {\rm i} {\epsilon}^{0 \tilde{P}_c \mu \rho} \tilde{P}^{\nu}_c - \frac{1}{4} {\rm i} {\epsilon}^{0 \tilde{\Delta}_c \mu \rho} \tilde{\Delta}^{\nu}_c 
%\right.
\nonumber\\
& 
%\left. \hspace{.5cm} 
+ \left( \tilde{P}^2_c - \frac{1}{4} \tilde{\Delta}^2_c \right) {\rm i}\epsilon^{0 \mu \rho \nu} + \left( 2\tilde{P}^0_c - \tilde{\Delta}^0_c \right) {\rm i}\epsilon^{\tilde{P}_c \mu \rho \nu} + \left( \tilde{P}^0_c - \frac{1}{2} \tilde{\Delta}^0_c \right) {\rm i}\epsilon^{\tilde{\Delta}_c \mu \rho \nu}  
%\right]
\,,
\end{align}
to write the bi-indexed bilinears as
\begin{align}\label{eq:bi_indexed_bilinear}
\bar{u}(p_f,\lambda_f) 
\mathsf{G}_m^{(\mu \rho)}
u(p_i,\lambda_i)  
&  =  m_f^j m_i^j \left\{ \operatorname{Re} \left[  \prod_{l=1}^{m}  \mathsf{B}^{\mu_{l} \rho_{l} \xi_{l}} (\widetilde{P}_{c},\widetilde{\Delta}_{c}) \prod_{l=m+1}^{2j}  \mathsf{S}^{\xi_{l}} (\widetilde{P}_{c},\widetilde{\Delta}_{c}) \right] 
\langle \tilde\lambda_f| (t+\bar{t})_{\xi_{1} \cdots \xi_{2j}}  | \tilde\lambda_i \rangle 
\right. \nonumber\\
& \hspace{1.5cm} \left. 
\; + {\rm i} \operatorname{Im} \left[  \prod_{l=1}^{m}  \mathsf{B}^{\mu_{l} \rho_{l} \xi_{l}} (\widetilde{P}_{c},\widetilde{\Delta}_{c}) \prod_{l=m+1}^{2j}  \mathsf{S}^{\xi_{l}} (\widetilde{P}_{c},\widetilde{\Delta}_{c}) \right]
\langle \tilde\lambda_f| (t-\bar{t})_{\xi_{1} \cdots \xi_{2j}}  | \tilde\lambda_i \rangle 
\right\}
\,,   
\\[1em]
&\hspace{2.5cm} [\text{Canonical-Melosh}]\nonumber
\end{align}
where $\mathsf{S}^{\xi} (\widetilde{P}_{c},\widetilde{\Delta}_{c})$ is given by Eq.~(\ref{eq:contraction_Q_parameters}). We wrote $\langle \tilde \lambda_f|\cdots|\tilde \lambda_i\rangle$ to emphasize the Melosh-rotated nature of these rest-frame spinors, and include a subindex $c$ to remind us that (real-valued) canonical boost parameters must be used.  For $m=0$ Eq.~(\ref{eq:scalar_bilinear_canonical}) is recovered.  From these expressions, it is straightforward to read off the behavior of these bilinears under parity.  
The term {\small$\operatorname{Re} [\prod_{l=1}^{m}  \mathsf{B}^{\mu_{l} \rho_{l} \xi_{l}} (\widetilde{P}_{c},\widetilde{\Delta}_{c}) \prod_{l=m+1}^{2j}  \mathsf{S}^{\xi_{l}} (\widetilde{P}_{c},\widetilde{\Delta}_{c})]$} contains an even number of Levi-Civita pseudotensors, while {\small$\operatorname{Im} [\prod_{l=1}^{m}  \mathsf{B}^{\mu_{l} \rho_{l} \xi_{l}} (\widetilde{P}_{c},\widetilde{\Delta}_{c}) \prod_{l=m+1}^{2j}  \mathsf{S}^{\xi_{l}} (\widetilde{P}_{c},\widetilde{\Delta}_{c})]$} contains an odd number of Levi-Civita pseudotensors, thus under parity they behave as a tensor and pseudo-tensor respectively. 
Meanwhile, parity switches the positions of the left- and right-chiral spinor in the bispinor, which results in interchanging the $t$ and $\bar t$ matrix element. 
Therefore, the bi-indexed bilinears behave as proper tensors under a parity inversion.

%%%%%%%%%%%%%%%%%%%%%%%%%%%%%%%%%%%%%%
%%%%% On-Shell Identities %%%%%%%%%%%%
%%%%%%%%%%%%%%%%%%%%%%%%%%%%%%%%%%%%%%

%\newpage

\section{On-shell identities} 
\label{sec:onshell_dentities}

It is well-known in the spin-1/2 case that for the on-shell case, not all the bilinears considered in Sec.~\ref{sec:gen_bilinears} are independent.  These are the famous Gordon identities~\cite{Gordon:1928}.  As is shown here, the same is true for the general spin case.  This allows to reduce (or choose) the number of independent bilinears that need to be considered in the decomposition of operator matrix elements.

We start from the generalized Dirac equation,
\begin{align}\label{gen Dirac eq}
\left( \gamma^{\mu_1 \dots \mu_{2j}} (p_{\mu_1} \dots p_{\mu_{2j}}) - m^{2j} \right) u(p,\lambda) = \left({\sh{p}}^{(j)} - m^{2j}\right) u(p,\lambda) = 0
\,,
\end{align}
where, ${\sh{p}}^{(j)}=\gamma^{\mu_1 \dots \mu_{2j}} p_{\mu_1} \dots p_{\mu_{2j}}$.
The Dirac equation leads to generalized on-shell identities
\begin{equation}
\bar u(p',\lambda') \left( \Gamma \right)  u(p,\lambda) = \frac{1}{2 \bar{m}^{2j}} \bar u(p',\lambda') \left( \left\{ {\shl{P}}^{(j)} , \Gamma \right\} + \frac{1}{2} \left[ {\shl{\Delta}}^{(j)} , \Gamma \right]  \right) u(p,\lambda)
\,,
\end{equation}
and
\begin{equation}
0 = \bar u(p',\lambda') \left( \frac{1}{2} \left\{ {\shl{\Delta}}^{(j)} , \Gamma \right\} + \left[ {\shl{P}}^{(j)} , \Gamma \right]  \right) u(p,\lambda)
\,.
\end{equation}
Here, $\Gamma$ stands for any of the independent elements of the generalized Dirac basis in Eq.~(\ref{eq:Dirac_basis}):
% the independent generalized gamma matrices that form a basis for spin-1 bilinears in the bi-spinor representation,
\begin{align}
\Gamma = \mathbb{1}: & \ \text{scalar}, \nonumber\\
\Gamma =  \gamma^{\mu_1 \dots \mu_{2j}}: & \  \text{symmetric traceless proper rank-$2j$ tensor}, \nonumber\\
\Gamma = \mathcal{G}_m^{\mu_1 \rho_1 \cdots \rho_{m} \mu_{m}}: & \ \text{bi-indexed traceless rank-$2m$  tensors}  \ (1 \leq m \leq 2j),  \nonumber\\
\Gamma = \gamma^{\mu_1 \dots \mu_{2j}} \gamma_5: & \  \text{symmetric traceless rank-$2j$ pseudotensor},\nonumber \\
\Gamma = \gamma_5: & \ \text {pseudoscalar} ,
 \nonumber
\end{align}
 we introduced the notation ${\shl{P}}^{(j)}=\gamma^{\mu_1 \dots \mu_{2j}} P_{\mu_1 \dots \mu_{2j}}$, ${\shl{\Delta}}^{(j)}=\gamma^{\mu_1 \dots \mu_{2j}} \Delta_{\mu_1 \dots \mu_{2j}}$, where
\begin{align}
P_{\mu_1 \dots \mu_{2j}} & = \frac{\bar{m}^{2j}}{2}\left(\frac{p^{\prime}_{\mu_1} \dots p^{\prime}_{\mu_{2j}}}{{m^{\prime}}^{2j}}+\frac{p_{\mu_1} \dots p_{\mu_{2j}}}{m^{2j}}\right) \,,
\\
\Delta_{\mu_1 \dots \mu_{2j}} & = \bar{m}^{2j} \left(\frac{p^{\prime}_{\mu_1} \dots p^{\prime}_{\mu_{2j}}}{{m^{\prime}}^{2j}} - \frac{p_{\mu_1} \dots p_{\mu_{2j}}}{m^{2j}} \right) \,, 
\\
\bar{m}^{2j} & = \frac{1}{2}\left({m^{\prime}}^{2j}+m^{2j}\right) \,.
\end{align} 
These statisfy the following symmetry and orthogonality properties: 
\begin{align}
P^{\mu_1 \dots \mu_{2j}} {\left(p^{\prime}, p\right)}  = & \; P^{\mu_1 \dots \mu_{2j}} {\left(p, p'\right)} \,,
\\
\Delta^{\mu_1 \dots \mu_{2j}}\left(p^{\prime}, p\right) = & \; - \Delta^{\mu_1 \dots \mu_{2j}}\left(p, p^{\prime}\right) \,,
\\
P^{\mu_1 \dots \mu_{2j}} \Delta_{\mu_1 \dots \mu_{2j}}  = & \; 0
\,.
\end{align} 
If the initial and final masses are equal $m^{\prime} = m$, the definitions of $P_{\mu_1 \dots \mu_{2j}}$ and $\Delta_{\mu_1 \dots \mu_{2j}}$ reduce to
\begin{align}
\bar{m}^{2j} &= m^{2j} \,, \\
P_{\mu_1 \dots \mu_{2j}}  &= \frac{1}{2} \left(p^{\prime}_{\mu_1} \dots p^{\prime}_{\mu_{2j}} + p_{\mu_1} \dots p_{\mu_{2j}} \right) \,, \\
\Delta_{\mu_1 \dots \mu_{2j}}  &=  \left(p^{\prime}_{\mu_1} \dots p^{\prime}_{\mu_{2j}} - p_{\mu_1} \dots p_{\mu_{2j}} \right)
\,.
\end{align}

The existence of on-shell identities 
means that not all terms within the full set of bilinears are independent, and the identities can be used to reduce the set into an independent one. More specifically, the on-shell identities mean that we can always replace bilinears of $\gamma^{(\mu)},\gamma^{(\mu)}\gamma_5$ with bilinears of commutators and anticommutators of two gamma matrices, which correspond to linear combinations of bilinears of $\mathcal{G}_m^{(\mu\rho)}$.  This implies one can always work with block-\emph{diagonal} elements of the Dirac basis, {\it i.e.} $ \mathbb{1} \,, \, \gamma_5 \,, \, \mathcal{G}_m^{\mu_1 \rho_1 \cdots \rho_{m} \mu_{m}}  \, (1 \leq m \leq 2j)$, which can be identified with specific multipoles.

\section{Discussion and outlook}
\label{sec:discussion}

If we want to summarize the constructions presented here in one sentence, we would describe it as repackaging expressions with general SL(2,$\mathbb C$) chiral (bi)spinors into Lorentz tensors.  Working with the Lorentz tensors simplifies the construction of covariant (or invariant) expressions.  For the spin-1/2 case this is something that we all have learned when studying the Dirac calculus, here we exposed details and interconnections for the general higher-spin case.  The $t$-tensors, and their derived objects ($\gamma$-matrices, the $\mathcal{T}_m,\mathcal{G}_m$ multipoles etc.), allow for the mapping between these different (repackaged) tensor representations of the Lorentz group's double cover SL(2,$\mathbb{C}$).  This property is reflected in  the $t$-tensors and generalized Dirac basis carrying both spinorial (matrix) and Lorentz (tensor) indices.  For general spin $j$, these objects contain a convenient packaging of all the Clebsch-Gordan coefficients needed to build the larger spin representations, but in such a way that they still transform covariantly.  The mapping to Lorentz tensors facilitates the composition of invariants and allow us to circumvent spinor calculations, which can be less intuitive to deal with.

Our results in Sec.~\ref{sec:onshell_dentities} show that on-shell bilinears for any spin-$j$ can be parametrized exclusively using a complete set of covariant $\mathfrak{sl}(2,\mathbb{C})$ multipoles.  This is a very appealing result, which is valid for any spin.  
Our final expressions allow simple switching between different types of spinors through a change in the $\tilde p$ parameter or implementing Melosh rotations on the rest-frame spinors.  The multipole structure allows immediate identification of similar structures across the different spin cases.  The use of chiral (bi)spinors also means we are using objects containing only the physical degrees of freedom and no extra constraints need to be imposed.  The algorithm to construct the $t$-tensors presented in Sec.~\ref{sec:t_tensors_gen}, in combination with the bilinear expressions presented in Sec.~\ref{sec:gen_bilinears}, demonstrate that the whole framework is straightforward to implement in calculations or numerical evaluations.

Of course, a framework only becomes practical when it can be applied to relevant applications.  We want to use our results to have a fresh look at the matrix elements of QCD operators used in non-perturbative phenomenology.  Local operators encode the form factors (electromagentic, weak, gravitational, generalized) of hadronic and nuclear bound states, while bilocal light-ray operators do the same for partonic distribution functions (collinear, transverse, generalized, $\ldots$).  The use of multipole bilinears will aid to parametrize these operators in a unified and systematic manner for any spin. This should make comparisons between the different spin cases more straightforward.  These studies will be the topic of follow-up work.  The goal there will not necessarily be to replace existing parametrizations and decompositions, as any decomposition in the $\mathfrak{sl}(2,\mathbb{C})$ multipoles will result in distributions that are a linear combination of existing ones.  Instead, our aim is to provide a more straightforward identification for the physical content of each distribution.  The resulting expressions will apply to any spin, which is advantageous from both a theoretical and practical standpoint.   Our expressions hold in any frame, but to make the connection with phenomenology, using preferential frames like the Breit or Drell-Yan frame will put to zero several of the parameters and yield simplified expressions.

Our study focused on matrix element with the same spin content in the initial and final state.  A more general treatment of matrix elements for nuclear and hadronic amplitudes of interest would also include transition matrix element between particles of different spin, or different number of particles in the initial and final state (e.g. breakup matrix elements).  The treatment of these cases should also be possible through extensions of the framework presented here.  While not discussed in this work, the $t$-tensors exist for more general $(j,j')$ representations~\cite{Barut:1963zzb,Weinberg:1964cn}, although Weinberg shows that for causal fields they can be written as derivatives of $(j-j',0)$ or $(0,j'-j)$ representations.  These tensors (and derived objects from their algebra) can still, however, be useful in transitions between states of different spin.  Additionally, Williams~\cite{Williams:1965rga} has also introduced objects that intertwine between one whole-valued $(m,0)$ or $(0,m)$ spinorial representation and bi-indexed Lorentz tensors.  With all these ingredients it is possible to build structures that are linear in the spinors of all involved particles and carry a definite Lorentz tensor structure.  Separately, a dedicated study of the $(j,j')$ representations could still be interesting for its own sake.

The bilinears for which expressions were constructed still carry dependence on the little group polarization indices of the spinors.  Consequently, they have combined Lorentz and little group covariance properties.  These little group characteristics reflect themselves in the appearance of the 
$\tilde p^\mu$ parameters in the final expressions.  Observables (cross sections, asymmetries etc.) do not have this explicit little group dependence anymore because the $S$-matrix amplitude and its complex conjugate are contracted in their little group indices with the density matrices of initial and final state particles, which also carry little group indices.  Explicit appearance of little group indices can be avoided now, by attaching the objects that carry the little group indices in the $S$-matrix amplitude (spinors, polarization vectors, etc.) to the density matrix and as such introduce a covariant density matrix and covariant ``amputated'' amplitudes that only carry Lorentz indices, see Ref.~\cite{Zwanziger:1965}.  To quote the well-known spin-1/2 example for the covariant spin-density matrix:
\begin{equation}
    \sum_{\lambda,\lambda'}\rho(\bm S)_{\lambda \lambda'}\, u(p,\lambda)\bar{u}(p,\lambda') = (\slashed{p}+m)\left(\frac{1+\gamma_5\slashed{S}}{2}\right),
\end{equation}
where $\bm S$ is the polarization vector characterizing the rest-frame spin-1/2 ensemble, and $S^\mu = \tensor{\left(\Lambda_\text{SB}[p]\right)}{^\mu_\nu}\overset{\circ}{S}^\nu\;[\overset{\circ}{S}^\mu=(0,\bm S)]$ its covariant generalization.  Therefore, to accommodate polarized targets or beams in this formalism, it could be interesting to analyze the covariant density matrices for these chiral spinors too, building on the earlier work by Zwanziger~\cite{Zwanziger:1965}.

% While we have referred to the bilinears as having nice covariant transformation properties, the truth is that they do not transform completely covariant due to relativistic spin effects.  This is apparent in the appearance of the non-covariant $\tilde p^\mu$ in the final bilinear expressions and is related to the appearance of Wigner rotations associated with spinor boosts.  The final $S$-matrix amplitudes, which contribute to observables, remain invariant and relativistic spin effects are eliminated through cancellations between different bilinears or through density matrices that exhibit similar transformation properties, such as Wigner rotations. To quote the well-known spin-1/2 example for the covariant spin-density matrix:
% \begin{equation}
%     \sum_{\lambda,\lambda'}\rho(\bm S)_{\lambda \lambda'}\, u(p,\lambda)\bar{u}(p,\lambda') = (\slashed{p}+m)\left(\frac{1+\gamma_5\slashed{S}}{2}\right),
% \end{equation}
% where $\bm S$ is the polarization vector characterizing the rest-frame spin-1/2 ensemble, and $S^\mu = \tensor{\left(\Lambda_\text{SB}[p]\right)}{^\mu_\nu}\overset{\circ}{S}^\nu\;[\overset{\circ}{S}^\mu=(0,\bm S)]$ its covariant generalization.  Therefore it could be interesting to analyze density matrices (and their covariant extensions) for these chiral spinors too, building on earlier work by Zwanziger~\cite{Zwanziger:1965}.

As outlined in Sec.~\ref{sec:MSH}, we note that the $(j,0)\oplus(0,j)$ spinor construction used in this work can be mapped to the massive spinor-helicity formalism of Ref.~\cite{Arkani-Hamed:2017jhn}. 
This connection opens the door to applying our covariant multipole decomposition directly within modern on-shell amplitude computations, bridging non-perturbative QCD applications and high-energy scattering methods.
Finally, all our results were derived
using the 1+3D flat space-time geometry of the Standard Model. We did not consider any
other cases here, but extensions in different dimensions and/or non-flat geometries could
be an interesting topic of study.

%%%%%%%%%%%%%%%%%%%%%%%%%%%%%%%%%%%%%%
%%%%% Appendices  %%%%%%%%%%%%%%%%%%%%
%%%%%%%%%%%%%%%%%%%%%%%%%%%%%%%%%%%%%%

\appendix

%%%%%%%%%%%%%%%%%%%%%%%%%%%%%%%%%%%%%%
%%%%% Other spinors (LF, helicity)%%%%
%%%%%%%%%%%%%%%%%%%%%%%%%%%%%%%%%%%%%%

\section{Conventions and notation}
\label{sec:notation}

Throughout this article, we use the following standard conventions: 
\begin{itemize}

\item The index conventions are
\begin{enumerate}
    \item Lowercase Greek letters for space-time Lorentz four-vector indices.
    \item Lowercase Roman letters at the start of the alphabet ($a,b,\ldots, f$) for SL(2,$\mathbb{C}$) left-chiral spinorial indices. For the right-chiral representation we use the same convention with dotted indices. For spin-$j$ representations, these indices take values from $-j$ to $j$.  In most of the text, we reserve the use of $a,b$ for the spin-1/2 case, and use $c,d,\ldots$ for the general spin case.  The exception is App.~\ref{sec:proof_cubic_reduction}, see specific comments there.
    \item We use $\lambda$ for the SU(2) little group polarization index of the chiral spinors.  In Sec.~\ref{sec:MSH}, we use uppercase Roman letters $A,B$ for these indices to make the connection with the MSH notation more immediate.
    \item We use lowercase Roman letters in the middle of the alphabet $i,j,\ldots,n,r,s$ as general counters in summations and products or as Lorentz indices in expressions where only spatial indices should be considered.  The complex number $\mathrm i^2=-1$ is distinguished from the index $i$.
    \item In Sec.~\ref{sec:t_gamma_tensors_algebra}, we use $\mathfrak{p},\mathfrak{m},\mathfrak{l},\mathfrak{r}$ to indicate the number of plus, minus, left and right indices in the $t$-tensor matrix elements.  
    \item For bi-spinors we do not write any spinor index explicitly, as they contain spinors from both chiralities. We denote with $u(p,\lambda)$ the canonical bi-spinor (no label). For the helicity spinors we use $u_h(p,\lambda)$, $u_{h'}(p,\lambda)$, and for light-front $u_{LF}(p,\lambda)$.
\end{enumerate}

\item The Minkowski metric is
\begin{equation}\label{eq:metric}
    g^{\mu\nu} = \text{diag}(1,-1,-1,-1)\,,
\end{equation}
and the covariant Levi-Civita tensor has sign $\epsilon^{0123}=1$.  

\item  The relation between a general element $g$ of the Lie group $G$ and the generators of the group, which form a basis $X_i$ for the associated Lie algebra $\mathfrak{g}$, is given by the exponential map
\begin{equation} \label{eq:exp_map}
    g = e^{-{\rm i}\sum_i t_iX_i}\,.
\end{equation}
The real parameters $t_i$ uniquely characterize the group element and we always use a minus sign in the exponential map (both for rotations and boosts).
\item The construction uses left- and right-handed chiral spinors (defined in Sec.~\ref{sec:irreps_Lorentz}), and bispinors as parity conserving direct sum representations of a left- and right-handed spinor.  For the bispinors we work in the Weyl representation, with the left-handed (right-handed) spinor occupies the top (bottom) rows of the bispinor, see Sec.~\ref{sec:bispinors}.
\item For any representation of spin $j$, $\gamma_5$ is defined in the Weyl representation as 
\begin{equation}\label{eq:gamma5_notation}
 \gamma_5 \equiv \left(\begin{array}{cc}
-\mathbb{1}^{(j)} & 0 \\
0  & \mathbb{1}^{(j)} 
\end{array} \right)
\,, 
\end{equation}
where $\mathbb{1}^{(j)}$ is the identity matrix of rank $2j+1$. For spin 1/2, this corresponds to 
\begin{align}
    &\gamma_5 = {\rm i}\gamma^0\gamma^1\gamma^2\gamma^3 \,,\\
    &\operatorname{Tr}[\gamma^\mu \gamma^\nu \gamma^\rho \gamma^\sigma \gamma_5] = -4{\rm i}\epsilon^{\mu\nu\rho\sigma}
    \,. 
\end{align}

\item We use light-front and transverse spherical components
\begin{align}\label{eq:coords_LCRL}
    &a^\pm = a^0\pm a^3 \,,\nonumber\\
    &a^{R} = a^1 +{\rm i}a^2 \,,\nonumber\\
    &a^L = a^1-{\rm i}a^2
    \,.
\end{align}

\item The convention used for the covariant Lorentz generators $\mathbb{M}^{\mu\nu}$ ($\overbar{\mathbb{M}}^{\mu\nu}$) is shown in Eqs.~(\ref{eq:sl(2,C)_generators_chiral}).

\item The binomial coefficient symbol is given by 
\begin{equation}\label{eq:binomial_symbol}
  \binom{2j}{m} 
  % =  B^{2j}_{m} 
  = \frac{ (2j)! }{m! (2j-m)!}
  \;.
\end{equation} 

\item 
For tensors that are totally symmetric under exchanges of their indices we use the notation
% In order to unclutter the notation in many equations, we use the following shorthand for a set of symmetric indices of a tensor  
\begin{equation}\label{eq:symmetrized_indices_shorthand}
t^{ (\rho) }  \equiv t^{\rho_{1} \ldots \rho_{j}} \,.
\end{equation}

\item Symmetrization and antisymmetrization are defined as
\begin{align}
    [a_i,b_j]_\pm &= \left(a_ib_j\pm b_ja_i\right)\,,\nonumber\\
    [a_i,b_j]&\equiv [a_i,b_j]_- \,,\nonumber\\
     \{a_i,b_j\}&\equiv [a_i,b_j]_+
     \,.
\end{align}

\item The following notation for symmetrization over set(s) of labels or indices
\begin{subequations}\label{eq:symmetrization_indices}
\begin{align}
\underset{\left\{ (\rho) \right\} }{\mathcal{S}}  &\equiv \underset{\left\{\rho_{1} \ldots \rho_{j}\right\}}{\mathcal{S}}  \,,
\\
\underset{\left\{ (\rho) , (\sigma) \right\}}{\mathcal{S}}  &\equiv \underset{\left\{\rho_{1} \ldots \rho_{j}\right\}}{\mathcal{S}} \underset{\left\{\sigma_{1} \ldots \sigma_{j}\right\}}{\mathcal{S}} \,,
\\
& \hspace{0.25cm}\vdots  \nonumber
\end{align}
\end{subequations}
For bi-indexed tensors (see Sec.~\ref{sec:quadratic_t_products_reducible}) we need symmetrizations over pairs of indices and we introduce the notation 
\begin{subequations}
\begin{align}\label{eq:symmetrization_bi_indices}
\underset{\left\{(\mu \rho)\right\}}{\mathcal{S}} & \equiv \underset{\left\{\mu_{1}\rho_{1}, \cdots , \mu_{j}\rho_{j}\right\}}{\mathcal{S}}  \,,
\\
\underset{\left\{(\mu \rho) ,(\nu \sigma) \right\}}{\mathcal{S}}  & \equiv \underset{\left\{\mu_{1}\rho_{1}, \cdots , \mu_{j}\rho_{j}\right\}}{\mathcal{S}} \underset{\left\{\nu_{1} \sigma_{1}, \cdots, \nu_{j}\sigma_{j}\right\}}{\mathcal{S}} \,,
\\
&  \hspace{.25cm}  \vdots  \nonumber
\end{align}
\end{subequations}
where symmetrizations like ${\mathcal{S}}_{\left\{(\mu\rho)\right\}}$
are done treating the pairs of indices $\mu_{r}\rho_{r}$ as one entity.
Notice that we do not introduce any normalization coefficient as part of the symmetrization definition. All coefficients will be explicitly stated in the equations.

\item 
For products, we use the empty product convention 
\begin{equation}\label{eq:empty_prod_convention}
\prod_{l=1}^{0} (\cdots) \equiv 1
\,,
\end{equation}
meaning that the product symbol with no factors is evaluated to one.

\end{itemize}

\section{Helicity and light-front spinors} 
\label{sec:hel_lf_spinors}

Besides the canonical spinors, which are constructed using rotationless boosts, there are other spinors in use in the literature.  These are in particular the commonly used Jacob-Wick helicity spinors~\cite{Jacob:1959at} and light-front helicity spinors~\cite{Kogut:1969xa,Soper:1972xc}.  Each of these spinors corresponds to a different choice of standard boost, which relates particles in their rest frame to moving frames.  Each standard boost comes with an associated spin operator. Each common choice has different advantages that explains their use in relativistic quantum theory and their applications for multi-particle states. For detailed reviews see Refs.~\cite{Polyzou:2012ut,Keister:1991sb} and references therein.  In this appendix, we summarize the main points relevant for the constructions presented in this work and how we can accommodate the formalism outlined in this article for helicity and light-front spinors.

The standard boosts $\Lambda_\text{SB}$ are active transformations that connect rest frame states, which for massive particles are a state of a spin-$j$ SU(2) irrep with a spin quantum number $\lambda$,\footnote{This exploits the semi-direct product structure of the Poincar\'{e} group to generate induced representations from the little group irreps~\cite{Wigner:1939cj, Tung:1985na}} to a state with four momentum $p$ and the \emph{same} quantum number $\lambda$:
\begin{equation}
    |p, j\lambda \rangle \equiv U[\Lambda_\text{SB}(\bm p)] \, |\overset{\circ}{p}, j\lambda\rangle \,.
\end{equation}
These standard boosts $\Lambda_\text{SB}$ usually correspond to one of the following choices 
\begin{subequations}\label{eq:standard_boosts}
\begin{align}
    \Lambda_c(\bm p) &= e^{-{\rm i}\rho \sum\limits_{m=1}^3 \hat{p}_m\mathbb{K}_m} \,, &\text{[canonical]}\\
    \label{eq:standard_boost_hel}
    \Lambda_\text{hel}(\bm p) &= e^{-{\rm i}\phi\mathbb{J}_3}e^{-{\rm i}\theta \mathbb{J}_2}e^{+{\rm i}\phi\mathbb{J}_3} e^{-{\rm i}\rho \mathbb{K}_3} \,, &\text{[helicity]}\\
    \label{eq:standard_boost_hel'}
    \Lambda_\text{hel'}(\bm p) &= e^{-{\rm i}\phi\mathbb{J}_3}e^{-{\rm i}\theta \mathbb{J}_2} e^{-{\rm i}\rho \mathbb{K}_3} \,, &\text{[helicity$'$]}
    \\
    \Lambda\text{LF}(\bm p) &= e^{-{\rm i} \, \bm p_{\rm T} \cdot {\mathbb{G}}_{\rm T}} e^{-{\rm i}\omega \mathbb{K}_3} \,. &  \text{[light front]} 
    \label{eq:standard_boost_LF} 
\end{align}    
\end{subequations}
We remind that rapidity $\rho$ was introduced in Eq.~(\ref{eq:rho_p}), $\theta$ and $\phi$ are the polar angles of final momentum $\bm p$. In the light-front standard boost of Eq.~(\ref{eq:standard_boost_LF}) we have 
\begin{subequations}\label{eq:LF_gen}
\begin{align}
    \mathbb G_{\rm T} = (\mathbb G_1 , \mathbb G_2)  \,,\\
    \mathbb G_1 = \mathbb K_1 +\mathbb J_2 \,,\\
    \mathbb G_2 = \mathbb K_2 -\mathbb J_1 \,,\\
    e^\omega = \frac{p^+}{m} \,.
\end{align}
\end{subequations}
The canonical standard boost uses only rotationless (pure) boosts.  The helicity standard boost (due to Jacob and Wick) first has a boost along the $z$-axis to the full momentum $|\bm p|$ followed by a rotation to the final direction.  We list two possible rotations, differing by an initial rotation along the $z$-axis.  The first corresponds to the original definition by Jacob \& Wick~\cite{Jacob:1959at}, whereas the second was adopted later.  As exhibited by Eq.~(\ref{eq:ptilde_h}), the first, original, helicity form has certain advantages that allows it to be treated similarly to the light-front spinor parametrization in this formalism.  The light-front standard boost, finally, is built only using light-front boosts $\mathbb K_3, \mathbb G_1, \mathbb G_2$.  A boost along the $z$-axis (up to the final light-front momentum component $p^+$) is followed by transverse light-front boosts. 

Once a standard boost is chosen, single particle states $|p,j\lambda\rangle$ -- constructed by applying the standard boost to the rest frame states -- furnish an infinitely-dimensional unitary representation of the Poincar\'{e} group.  When boosting a moving single particle state with a Lorentz transformation, the state undergoes an additional momentum-dependent spin rotation (little group transformation), the so-called Wigner rotation $R_\text{w}$:
\begin{align} \label{eq:wigner_rot}
    &U[\Lambda]|p, j\lambda \rangle = \sum_{\lambda'} |\Lambda p, j\lambda' \rangle D^{(j)}_{\lambda' \lambda}[R_\text{w}] \,,\\
    &R_\text{w}(\Lambda,\bm p) = \Lambda_\text{SB}^{-1}(\Lambda \bm p) \Lambda \Lambda_\text{SB}(\bm p) \,,
    \label{eq:wigner_rot_expanded}
\end{align}
see also Eq.~(\ref{eq:def_Wignerrot}) for the associated transformations of the annihilation operator.
The different choices of standard boost correspond to different Wigner rotations, each of which has their specific advantages~\cite{Polyzou:2012ut}:
\begin{itemize}
    \item For the canonical spin, the Wigner rotation corresponding to a rotation $\Lambda =R$ is that rotation, $R_\text{w} = R$.  Consequently, canonical spin states are advantageous to work with when coupling spins in multi-particle states through the use of Clebsch-Gordan coefficients.
    \item For the helicity spin, the $\lambda$ quantum number corresponds to the helicity projection of the canonical spin.  Moreover, the Wigner rotation of \emph{any} Lorentz transformation is a rotation around the $z$-axis. This conserves $\lambda$ (meaning helicity is a Lorentz invariant) and the boosted state in Eq.~(\ref{eq:wigner_rot}) only acquires an extra phase.
    \item For light-front spin, the light-front boosts form a kinematical subgroup~\cite{Dirac:1949cp}.  As a result the Wigner rotation for any light-front boost is the identity $R_\text{w} = \mathbb{1}$.  There are still non-trivial Wigner rotations for light-front rotations.
\end{itemize}

As all standard boosts take the rest frame momentum $\overset{\circ}{p}$ to the same momentum $p$, different standard boost choices can be related by \emph{momentum-dependent} rest frame rotations, the so-called generalized Melosh rotations~\cite{Melosh:1974cu,Marinescu:1974yd,Polyzou:2012ut}.  The standard boosts can equivalently be written as a Melosh rotation $R_\text{M}$ followed by the canonical standard boost:
\begin{subequations}\label{eq:melosh}
\begin{align}
    &\Lambda_h(\bm p) = \Lambda_c(\bm p)R_\text{M}[\text{h}]  = \Lambda_c(\bm p) \, e^{-{\rm i}\phi\mathbb{J}_3}e^{-{\rm i}\theta \mathbb{J}_2}e^{+{\rm i}\phi\mathbb{J}_3} \,, &\text{[helicity]}\\
    &\Lambda_{h'}(\bm p) = \Lambda_c(\bm p)R_\text{M}[\text{h'}] = \Lambda_c(\bm p)\, e^{-{\rm i}\phi\mathbb{J}_3}e^{-{\rm i}\theta \mathbb{J}_2} \,, &\text{[helicity$'$]}\\
    &\Lambda_\text{LF}(\bm p) = \Lambda_c(\bm p)R_\text{M}[\text{LF}] = \Lambda_c(\bm p)\, e^{-{\rm i}\phi\mathbb{J}_3}e^{-{\rm i}\theta_\text{LF}\mathbb{J}_2} e^{+{\rm i}\phi\mathbb{J}_3} \,, & \text{[light front]}
\end{align}    
\end{subequations}
where $\theta_\text{LF} = 2\tan^{-1} \frac{p_{\rm T}}{p^++m}$, and the Euler angles of the  momentum-dependent Melosh rotation $R_\text{M}$ can be read off on the right side of the equations.  One observes that the Melosh rotation for both helicity standard boosts corresponds to the same rotations that appear after the $z$-axis boost in the standard boost  of Eqs.~(\ref{eq:standard_boost_hel}) and (\ref{eq:standard_boost_hel'}).  For both the original helicity and light-front cases, the Melosh rotations become trivial when the transverse momentum $p_T=0$ (or $\theta=0$).

For the helicity and light-front choices of standard boosts, the bispinors of Eq.~(\ref{eq:bispinor_def}) appearing in the causal fields take the form
\begin{subequations}
\label{eq:bispinors_construction}
\begin{align}
    & u_h(p,\lambda)  = \mathcal{D}^{(j)}\left[ R_z(\phi)R_y(\theta)L(|\bm p|\hat{\bm z})\right] \,\overset{\circ}{u}(\lambda) = \mathcal{D}^{(j)}[L(\bm p)] \, \overset{\circ}{u}_{h}(\lambda) \,,\\
    & u_{h'}(p,\lambda)  = \mathcal{D}^{(j)}[R_z(\phi)R_y(\theta)R_z(-\phi)L(|\bm p|\hat{\bm z})] \,\overset{\circ}{u}(\lambda) =  \mathcal{D}^{(j)}[L(\bm p)] \, \overset{\circ}{u}_{h'}(\lambda)\,,\\
    & u_{LF}(p,\lambda)  = \mathcal{D}^{(j)}\left[\exp\left(-{\rm i} \, \bm p_{\rm T} \cdot \mathbb{G}_{\rm T} \right)L(|\bm p|\hat{\bm z})\right]\, \overset{\circ}{u}(\lambda) = \mathcal{D}^{(j)}[L(\bm p)] \, \overset{\circ}{u}_{LF}(\lambda)\,, 
\end{align}    
\end{subequations}
where in the 
% second equations 
final equalities 
we exploited Eq.~(\ref{eq:melosh}) to introduce Melosh rotated rest-frame bispinors and spinors
\begin{subequations}
\label{eq:spinor_RF_meloshrotated}
\begin{align}
\overset{\circ}{u}_{i}(\lambda) & = \begin{pmatrix}
\overset{\circ}{u}_{iL,c}(\lambda) \\
\overset{\circ}{u}_{iR}^{\dot d}(\lambda)
\end{pmatrix} \,,\\
\overset{\circ}{u}_{iL,c}(\lambda) & = m^j \tensor{\left(D^{(j)}[R_M[i]]\right)}{_c^\lambda} 
% \nonumber \\
% & 
= m^j \tensor{\left( \overbar{D}^{(j)}[R_M[i]]\right)}{^{\dot d}_{\lambda}}=\overset{\circ}{u}_{iR}^{\dot d}(\lambda) \,,
\\[0.5em]
& \hspace{-.2cm} i \in \{h,h',\text{LF}\} \nonumber\,.
\end{align}    
\end{subequations}
Equivalent expressions can be written for the negative-energy spinors (or Eq.~(\ref{eq:bispinor_chargeconj}) can be applied).

The spinors have the following properties for any spin
\begin{itemize}
    \item The canonical bi-spinors transform well under parity 
[see Eqs.~(\ref{eq:beta_def}), (\ref{eq:bispinor_u_v})]
    \begin{subequations}
    \begin{align}
        \beta u(p,\lambda) = u(\bar p,\lambda) \,,\\
        \beta v(p,\lambda) = (-)^{2j} v(\bar p,\lambda) \,,
    \end{align}        
    \end{subequations}
    where $\bar{p} = (E_p,-\bm p)$. 
    As can be seen from Eqs.~(\ref{eq:general_boosts_t}) and (\ref{eq:general_boosts_tbar}), only the canonical bi-spinors transform in this simple manner, owning to the $\tilde p^\mu$ being real valued. For any other choice of standard-boost parameterization, characterized by complex parameters $\tilde p^\mu$ due to rotations, the bi-spinors are not eigenvectors of the parity operator. 
    
    \item  The helicity spinors are eigenvectors of the helicity operator
    \begin{equation}
        \frac{W^0}{|\bm p|} = \left(\begin{array}{cc}
\hat{\bm p}\cdot \bm J^{(j)} & 0 \\
0 & \hat{\bm p}\cdot \bm J^{(j)}
\end{array}\right)\,,
    \end{equation}
    where we introduced the Pauli-Lubanski vector
        \begin{align}
        W^\mu & = -\frac{1}{2}\epsilon^{\mu\nu\rho\sigma} \mathsf{M}_{\nu\rho}p_{\sigma}
        = -j\, \frac{\text{i}}{2}\epsilon^{\mu\nu\rho\sigma}\mathcal{G}_{1,\nu\rho} p_\sigma \nonumber \\
        & = j \left(\begin{array}{cc}
{\mathcal{Q}}_\text{red}^{\mu\sigma\alpha}  & 0 \\
0 & \overbar{\mathcal{Q}}_\text{red}^{\mu\sigma\alpha} 
\end{array}\right) p_{\sigma} t^{(j)}_{\alpha 0 \cdots 0}
        = \left(\begin{array}{cc}
{\mathcal{Q}}_\text{red}^{\mu\sigma\alpha}  & 0 \\
0 & \overbar{\mathcal{Q}}_\text{red}^{\mu\sigma\alpha} 
\end{array}\right) p_{\sigma} J^{(j)}_{\alpha}
        \,,
    \end{align}
where $\mathsf{M}^{\mu\nu}$ are the Lorentz generators for the bispinor representation, and we used Eq.~(\ref{eq:t_J}).    
We have
\begin{subequations}
    \begin{align}
        &\frac{W^0}{|\bm p|}  \,u_h(p,\lambda) = \lambda \,u_h(p,\lambda) \,,\\
        &\frac{W^0}{|\bm p|}  \,v_h(p,\lambda) = -\lambda \,v_h(p,\lambda) \,,
    \end{align}
\end{subequations}
and similar relations for the $u_{h'},v_{h'}$ spinors.

\item The light-front spinors are eigenvectors of the light-front helicity operator
\begin{equation}
    \frac{W^+}{p^+} = \left(\begin{array}{cc}
\frac{p^R}{p^+}J^{(j)}_L+ J^{(j)}_3 & 0 \\
0 & \frac{p^L}{p^+}J^{(j)}_R+ J^{(j)}_3
\end{array}\right)\,,
\end{equation}
and we have
\begin{subequations}
    \begin{align}
        &\frac{W^+}{p^+}  \,u_{LF}(p,\lambda) = \lambda \,u_{LF}(p,\lambda) \,,\\
        &\frac{W^+}{p^+}  \,v_{LF}(p,\lambda) = -\lambda \,v_{LF}(p,\lambda) \,.
    \end{align}
\end{subequations}
\end{itemize}

%%%%%%%%%%%%%%%%%%%%%%%%%%%%%%%%
%%%% algorithm             %%%%%
%%%%%%%%%%%%%%%%%%%%%%%%%%%%%%%%

\section{A second algorithm for \texorpdfstring{$t$}{t}-tensor construction}
\label{sec:t_algorithm}

In this appendix, we present an alternative algorithm to construct the $t$-tensors. From a viewpoint of obtaining the final results, it is less efficient than the one presented in Sec.~\ref{sec:t_tensors_gen}. It does, however, provide more of a link with the underlying Lorentz group structure. The algorithm is iterative and makes use of the exponential map of Eq.~(\ref{eq:prop_ttensor}), the traceless and symmetric nature the $t$-tensors, and the fact that  elements of the $t$-tensors with $k$ spatial indices are homogeneous polynomials of degree $k$ in the generators of rotations $\bm J^{(j)}$.  

The algorithm proceeds as follows:
\begin{enumerate}

\item As noted before, the first term in the expansion of the matrix exponential Eq.~(\ref{eq:prop_ttensor}), corresponding to the zeroth degree polynomial in $\bm J$,  is always the identity
\begin{align} \label{eq:tzero}
t^{0 \ldots 0} = \mathbb{1}
\,.
\end{align}

\item The linear polynomials are obtained from the elements with one spatial index, and are proportional to the rotation group generators
\begin{align} \label{eq:t_J}
t^{0 \ldots i \ldots 0} = \frac{2}{2j} J_i = \frac{1}{j} J_i \,,
\end{align}
where the factor $1/2j$ is needed to cancel the contributions from all $2j$ elements of the symmetric tensor with exactly one spatial index.

\item For the elements with exactly two spatial indices, we have to consider homogeneous polynomials in $\bm J$ of degree 2.  As the commutators of two $\bm J$ reduce to a linear function in $\bm J$ by making use of the $\mathfrak{su}(2)$ algebra, we only have to consider anticommutators $\{J_m, J_n\}$.  

We form the elements $t^{0 \ldots m \ldots 0 \ldots n \ldots 0}$ in such a way that they respect the traceless condition $t^{\ldots \mu \ldots \nu \ldots} g_{\mu\nu} = 0$. To fix that traceless condition, we make explicit use of the quadratic Casimir of the rotation group 
\begin{equation}\label{eq:quadratic_Casimir}
    \left\{J_k,J_l\right\}\delta_{kl} = 2j(j+1)\; \mathbb{1}\,.
\end{equation}
We can now write
\begin{align} \label{eq:t_2spatial}
t^{0 \ldots m \ldots 0 \ldots n \ldots 0}  = t^{mn 0 \ldots 0}   
% \nonumber \\
& = \left( \frac{(2j)!} {2! (2j-2)!} \right)^{-1}\left( \left\{J_m,J_n\right\} - \frac{1}{3}  \left\{J_r,J_r\right\} \delta_{mn}  \right) + \frac{1}{3} t^{0 \ldots 0}\delta_{mn} \nonumber \\ 
%& = \frac{1}{j(2j-1)} \left( \left\{J_m,J_n\right\} - \frac{1}{3} \bm J^2 \delta_{mn} \right) + \frac{1}{3} \mathbb{1} \delta_{mn}  \nonumber \\ 
& = \frac{1}{j(2j-1)} \left( \left\{J_m,J_n\right\} - \frac{2}{3}  j(j+1) \mathbb{1} \delta_{mn}\right)  + \frac{1}{3} \mathbb{1}  \delta_{mn}  \nonumber \\ 
& = \frac{1}{j(2j-1)} \left( \left\{J_{m},J_n\right\} - j  \mathbb{1} \delta_{mn}\right)   \nonumber \\ 
& = \frac{j}{(2j-1)} \left( \left\{t^{m 0 \ldots 0},t^{n 0 \ldots 0}\right\} - \frac{1}{j}  t^{0 \ldots 0} \delta_{mn}\right) \,.
\end{align}
 We divide by the binomial factor $\frac{(2j)!}{2! (2j-2)!}$ 
 to avoid overcounting permutations of equivalent indices in the symmetric tensor.
 Because the traceless condition applies to the 4-index Lorentz indices, the $t^{mn 0 \ldots 0}$ as a Cartesian rank 2 tensor is not traceless
\begin{align}\label{eq:t2_space_trace}
t^{mm 0 \ldots 0} & = \frac{1}{j(2j-1)} \left( 2 j (j+1)  - 3 j \right) \mathbb{1}  = \mathbb{1} \,.
\end{align}

\item For $t$-tensor elements with a higher number of spatial indices, symmetrizations of higher polynomials of $J$s are treated recursively in a similar fashion.  This demonstrates the link between the elements of the $t$-tensor and the Jordan algebra, built out of anticommutators of the $\bm J$.  Contrary to the commutators (Lie algebra), this Jordan algebra is representation (spin) specific.  As in Eq.~(\ref{eq:t_2spatial}), these can be related to expressions with tensor elements having fewer spatial indices.   For elements with three spatial indices, we obtain for instance
\begin{align}
 t^{mnl 0 \ldots 0} & = \left( \frac{(2j)!} {3! (2j-3)!} \right)^{-1} \frac{2}{9} \frac{1}{3!}  \underset{\left\{m,n,l\right\}}{\mathcal{S}} \left( J_{m} J_{n} J_{l} - (3j-1)  \delta_{mn} J_{l} \right)  \nonumber \\
 & \hspace{-.7cm} =  \frac{j}{(2j-2)} \frac{1}{3}  \left[ \left( t^{ m  n  0 \ldots 0} t^{ l 0 \ldots 0} + t^{ n l 0 \ldots 0} t^{ m 0 \ldots 0} + t^{ l m 0 \ldots 0} t^{ n 0 \ldots 0} \right) 
 -  \frac{1}{j} \left( \delta^{mn} t^{ l 0 \ldots 0} + \delta^{nl} t^{ m 0 \ldots 0} + \delta^{lm} t^{ l 0 \ldots 0}  \right)  \right] \,,
\end{align}
where the last three terms fix the traceless conditions.
%$t^{\ldots \lambda \ldots \mu \ldots \nu \ldots} g_{\lambda\mu} = t^{\ldots \lambda \ldots \mu \ldots \nu \ldots} g_{\lambda\nu} = t^{\ldots \lambda \ldots \mu \ldots \nu \ldots} g_{\mu\nu} =0$. 

\item
Applying the Cayley-Hamilton theorem, the matrix $J_{\hat{\bm p}} \equiv \hat{\bm p} \cdot \bm J$ satisfies its own characteristic equation. This, for the spin-$j$ case, is expressed as
\begin{equation}
(J_{\hat{\bm p}}-j)(J_{\hat{\bm p}}-j-1)...(J_{\hat{\bm p}}+j)=0 \,.
\end{equation}
As a consequence, any polynomial in $J_{\hat{\bm p}}$ of degree $>2j$ can be reduced to one of maximal degree $2j$.
Hence, the algorithm finishes after $2j$ iterations and the following set is generated
\begin{align} \label{eq:algorithm_set}
 & t^{ 0 \ldots 0} \,, \nonumber \\
 & t^{m_1 0 \ldots 0} \,, \nonumber \\
 & t^{m_1m_2 0 \ldots 0} \,, \nonumber \\
 & t^{m_1m_2m_3 0 \ldots 0} \,, \nonumber \\ 
 \vdots \nonumber\\
 & t^{m_1m_2m_3 \ldots m_{2j}} \,. 
 % \nonumber \\ 
\end{align}
  
The $n$-th element of this set, $t_{m_1 \cdots m_n 0 \cdots 0}$, is a 3-dimensional rank-$n$ symmetric tensor, which satisfies $\frac{n !}{2 !(n-2)!}$ constraints.
  It means that for element $n$ there are $\frac{(n+2) !}{2 ! n !}-\frac{n !}{2 !(n-2) !}=(2 n+1)$ independent matrices contained in $t_{m_1 \cdots m_n 0 \cdots 0}$.  The entire set has in total
\begin{equation}
    \sum_{n=0}^{2j} (2n+1) = (2j+1)^2
    \,,
\end{equation}
 independent Hermitian matrices up to $2j$-degree polynomials in $J$s.  This is a sufficient basis to generate the exponential map of Eq.~(\ref{eq:prop_ttensor}).
\end{enumerate}

With regards to the general shape of these matrices with more and more spatial indices, we have the following pattern emerging.  
\begin{enumerate}
    \item The $t^{0\ldots0}$ is diagonal (unit matrix).
    \item The $t^{m_1 0\ldots 0}$ can be written as linear combinations of $J_3$ and the raising and lowering operators $J_\pm$, so will have non-zero elements on the diagonal and the first off-diagonals.  
    \item The $t^{m_1m_2\ldots m_k 0 \ldots 0}$ with $k$ spatial indices will have non-zero matrix elements up to the $k$-th off-diagonal, or non-zero matrix elements occur only in elements ($i,i-j$) and ($i-j,i$), for $j\in [0,k]$.
\end{enumerate}
This illustrates that in order to describe matrix elements with $k$ units of spin flip, at least the $t$-tensor matrix elements with $k$ spatial indices are needed (corresponding to the $k$-th multipole, see Sec~\ref{sec:quadratic_t_vs_multipoles}).

% 
% 

%%%%%%%%%%%%%%%%%%%%%%%%%%%%%%%%%%%%
%%%    proof reduction       %%%%%%%
%%%%%%%%%%%%%%%%%%%%%%%%%%%%%%%%%%%%

\section{Proof of the cubic reduction formula}
\label{sec:proof_cubic_reduction}

In this section the $t$-tensor cubic reduction formula is proved to be given by
\begin{equation}\label{eq:tcubic_proof}
t^{\mu_{1} \cdots \mu_{2j} }  \bar{t}^{\nu_{1} \cdots \nu_{2j} } t^{\rho_{1} \cdots \rho_{2j} } = \frac{1}{[(2j)!]^2}  \underset{\left\{(\nu),(\rho)\right\}}{\mathcal{S}}   \left( \prod_{l=1}^{2j} {\mathcal{C}}^{\mu_{l} \nu_l \rho_{l} \tau_{l}} \right) t_{\tau_{1} \cdots \tau_{2j}} 
\,,
\end{equation}
with
\begin{equation}
{\mathcal{C}}^{\mu \nu \rho \tau } = g^{\mu \nu } g^{\rho \tau } - g^{\mu \rho } g^{\nu \tau } + g^{\mu \tau } g^{\nu \rho } + {\rm i} \epsilon^{\mu \nu \rho \tau} 
 \,.
\end{equation}
We start with the known spin-1/2 identity
\begin{equation}\label{eq:cubic_1/2}
    \sigma^\mu\bar{\sigma}^\nu\sigma^\rho = g^{\mu\rho} \sigma^\rho - g^{\mu\rho}\sigma^\nu+g^{\nu\rho}\sigma^{\mu}+{\rm i}\epsilon^{\mu\nu\rho\tau}\sigma_\tau = \mathcal{C}^{\mu\nu\rho\tau} \sigma_\tau
     \,.
\end{equation}
For the higher spin cases, we apply Eq.~(\ref{eq:singlet_fullreduction}) three times to reduce the cubic product $t\bar t t$ to products of all spin-1/2 $\sigma$ and $\bar \sigma$ multiplied by towers of Clebsch-Gordan coefficients:
\begin{align}\label{eq:cubic_fullreduction}
\tensor{\left(t^{(\mu)}\right)}{_{p_{2j}\dot q_{2j}}}  \tensor{\left(\bar{t}^{(\nu)}\right)}{^{\dot q_{2j} r_{2j}}} \tensor{\left(t^{(\rho)}\right)}{_{r_{2j}\dot s_{2j}}} = & \prod_{i=1}^{2j}  \tensor{(\sigma^{\mu_{i}})}{_{a_i \dot b_i}}
\tensor{(\bar \sigma^{\nu_{i}})}{^{\dot c_i d_i}}
\tensor{(\sigma^{\rho_{i}})}{_{e_i \dot f_i}}
\nonumber\\
%%%%%%%%%%%%%%%%%%%%%%%%%
& \hspace{-0.5cm} \times \tensor{\left[j-\tfrac{i-1}{2},j-\tfrac{i}{2},\tfrac{1}{2}\right]}{_{p_{2j-i+1}}^{p_{2j-i}a_i}}
\tensor{\left[j-\tfrac{i-1}{2},j-\tfrac{i}{2},\tfrac{1}{2}\right]}{_{\dot s_{2j-i+1}}^{\dot s_{2j-i}\dot f_i}}\nonumber\\
%%%%%%%%%%%%%%%%%%
& \hspace{-.5cm} \times \tensor{\delta}{^{\dot q_{2j}}_{\dot t_{2j}}}
\tensor{\left[j-\tfrac{i-1}{2},j-\tfrac{i}{2},\tfrac{1}{2}\right]}{_{\dot q_{2j-i+1}}^{\dot q_{2j-i}\dot b_i}}
\tensor{\left[j-\tfrac{i-1}{2},j-\tfrac{i}{2},\tfrac{1}{2}\right]}{^{\dot t_{2j-i+1}}_{\dot t_{2j-i}\dot c_i}}
\nonumber\\
%%%%%%%%%%%%%%%%%%%%%%%%%
& \hspace{-.5cm} \times \tensor{\delta}{_{r_{2j}}^{u_{2j}}}
\tensor{\left[j-\tfrac{i-1}{2},j-\tfrac{i}{2},\tfrac{1}{2}\right]}{^{r_{2j-i+1}}_{r_{2j-i}d_i}}
\tensor{\left[j-\tfrac{i-1}{2},j-\tfrac{i}{2},\tfrac{1}{2}\right]}{_{u_{2j-i+1}}^{u_{2j-i}e_i}}
\,.
%%%%%%%%%%%%%%%%%%%
\end{align} 
For the spinor indices, we reserve Roman indices $a,\ldots,f$ for the contractions with spin-1/2 objects and $p,\ldots,u$ Roman indices for indices that appear on the left side or are contracted between CG coefficients (and refer to higher-spin indices).  At the end of this section, we will show that the last line of the previous equation reduces to
\begin{align}\label{eq:CGtower_reduction}
\tensor{\left(\delta_\mathcal{S}\right)}{_{d_1\cdots d_{2j}}^{e_1\cdots e_{2j}}}&\equiv \tensor{\delta}{_{r_{2j}}^{u_{2j}}}\prod_{i=1}^{2j} 
\tensor{\left[j-\tfrac{i-1}{2},j-\tfrac{i}{2},\tfrac{1}{2}\right]}{^{r_{2j-i+1}}_{r_{2j-i}d_i}}
\tensor{\left[j-\tfrac{i-1}{2},j-\tfrac{i}{2},\tfrac{1}{2}\right]}{_{u_{2j-i+1}}^{u_{2j-i}e_i}}\nonumber\\ 
&= \frac{1}{(2j)!}\underset{\left\{(e) \right\}}{\mathcal{S}}\left(\tensor{\delta}{_{d_1}^{e_1}}\cdots  \tensor{\delta}{_{d_{2j}}^{e_{2j}}}\right)
 \,.
\end{align}
There is a similar relation for the Clebsch-Gordan tower in indices $b,c,q,t$ that appears on the third line in Eq.~(\ref{eq:cubic_fullreduction}):
\begin{align} \label{eq:CGtower_reduction_2}
\tensor{\left(\delta_\mathcal{S}\right)}{^{\dot b_1\cdots \dot b_{2j}}_{\dot c_1\cdots \dot c_{2j}}}&\equiv\tensor{\delta}{^{\dot q_{2j}}_{\dot t_{2j}}}\prod_{i=1}^{2j} 
\tensor{\left[j-\tfrac{i-1}{2},j-\tfrac{i}{2},\tfrac{1}{2}\right]}{_{\dot q_{2j-i+1}}^{\dot q_{2j-i}\dot b_i}}
\tensor{\left[j-\tfrac{i-1}{2},j-\tfrac{i}{2},\tfrac{1}{2}\right]}{^{\dot t_{2j-i+1}}_{\dot t_{2j-i}\dot c_i}} \nonumber\\ 
&= \frac{1}{(2j)!}\underset{\left\{(\dot b) \right\}}{\mathcal{S}}\left(\tensor{\delta}{^{\dot b_1}_{\dot c_1}}\cdots  \tensor{\delta}{^{\dot b_{2j}}_{\dot c_{2j}}}\right)
 \,.
\end{align}
 Using these identities in Eq.~(\ref{eq:cubic_fullreduction}), each term in the two symmetrizations of Eqs.~(\ref{eq:CGtower_reduction}) and (\ref{eq:CGtower_reduction_2}) generates $(2j)!$ matrix products of triples $\sigma^{\mu_i} \bar{\sigma}^{\nu_j} \sigma^{\rho_k}$. Between the different terms, the symmetrization of the \emph{spinor} indices $\{(e)\}$ and $\{(\dot b)\}$ carries over into a symmetrization in \emph{Lorentz} indices $\{(\nu)\},\{(\rho)\}$, meaning \emph{all} possible combinations $\{\mu_i,\nu_j,\rho_k\}$ of matrix products appear in the sums.  In a final step, the spin-1/2 matrix products can then be reduced using Eq.~(\ref{eq:cubic_1/2}).  These steps lead to
\begin{align}
   \tensor{\left(t^{(\mu)}\right)}{_{p_{2j}\dot q_{2j}}}  \tensor{\left(\bar{t}^{(\nu)}\right)}{^{\dot q_{2j} r_{2j}}} \tensor{\left(t^{(\rho)}\right)}{_{r_{2j}\dot s_{2j}}} & = \frac{1}{[(2j)!]^2}  \underset{\left\{(\nu),(\rho)\right\}}{\mathcal{S}}  \prod_{i=1}^{2j} 
   % \nonumber \\
   %%%%%%%%%%%%%%%%%%
   % &\qquad \times 
   \tensor{\left[j-\tfrac{i-1}{2},j-\tfrac{i}{2},\tfrac{1}{2}\right]}{_{p_{2j-i+1}}^{p_{2j-i}a_i}} 
    \nonumber\\[0.5em]
   %%%%%%%%%%%%%%%%%%
   &\hspace{3.cm} \times 
   \tensor{\left[j-\tfrac{i-1}{2},j-\tfrac{i}{2},\tfrac{1}{2}\right]}{_{\dot s_{2j-i+1}}^{\dot s_{2j-i}\dot f_i}}
   % \nonumber\\[0.5em]
   % %%%%%%%%%%%%%%%%
   % & \qquad \times 
\tensor{\left(\sigma^{\mu_{i}}\bar{\sigma}^{\nu_{i}}\sigma^{\rho_{i}} \right)}{_{a_i \dot f_i}}
   \nonumber \\[0.7em]
   %%%%%%%%%%%%%%%
   %%%%%%%%%%%%%%%
    & = \frac{1}{[(2j)!]^2}  \underset{\left\{(\nu),(\rho)\right\}}{\mathcal{S}} \prod_{i=1}^{2j}
   % \nonumber \\
   % %%%%%%%%%%%%%%%%%%
   % &\qquad \times 
   \tensor{\left[j-\tfrac{i-1}{2},j-\tfrac{i}{2},\tfrac{1}{2}\right]}{_{p_{2j-i+1}}^{p_{2j-i}a_i}} 
    \nonumber\\[0.5em]
   %%%%%%%%%%%%%%%%%%
   & \hspace{3.cm} 
   \times \tensor{\left[j-\tfrac{i-1}{2},j-\tfrac{i}{2},\tfrac{1}{2}\right]}{_{\dot s_{2j-i+1}}^{\dot s_{2j-i}\dot f_i}}
   % \nonumber\\[0.5em]
   % %%%%%%%%%%%%%%%%
   % & \qquad \times 
   \mathcal{C}^{\mu_{i} \nu_{i} \rho_{i} \tau_{i}} \tensor{\left(\sigma_{\tau_{i}}\right)}{_{a_i \dot f_i}}
    \,.
\end{align}
Finally, using Eq.~(\ref{eq:singlet_fullreduction}) to rebuild the $t$-tensor, we arrive at the desired Eq.~(\ref{eq:tcubic_proof}).

What remains is to prove Eq.~(\ref{eq:CGtower_reduction}).  From Eq.~(\ref{eq:cubic_fullreduction}) and the transformation properties of the $t$-tensors [Eqs.~(\ref{eq:t_cov})] and CG coefficients [Eqs.~(\ref{eq:CG_transf})], we can infer that $\delta_\mathcal{S}$ is an invariant tensor in its spin-1/2 spinor indices. The symmetry in the Lorentz indices of the $t$-tensors, combined with the fact that spinor indices of $\delta_\mathcal{S}$ are contracted with identical $\sigma$ or $\bar\sigma$ means that any permutation within spinor indices $\{d_i\}$ and $\{e_i\}$ can be compensated by a permutation of Lorentz indices and $\delta_\mathcal{S}$ has to be completely symmetric in both sets of spin-1/2 spinor indices.  As an antisymmetric invariant tensor, the Levi-Civita $\epsilon^{ij}$ cannot be used to construct $\delta_\mathcal{S}$ and the only available invariant SL(2,$\mathbb{C}$) tensor is $\tensor{\delta}{_{d}^{e}}$.  This fixes the overall form to the symmetric one of Eq.~(\ref{eq:CGtower_reduction}).  To show that the normalization constant is $(2j)!$, we can contract Eq.~(\ref{eq:CGtower_reduction}) in its indices: 
\begin{align}\label{eq:CGtower_reduction_proof}
    \tensor{\left(\delta_\mathcal{S}\right)}{_{e_1\cdots e_{2j}}^{e_1\cdots e_{2j}}}&\equiv \tensor{\delta}{_{r_{2j}}^{u_{2j}}} \prod_{i=1}^{2j}  \tensor{\left[j-\tfrac{i-1}{2},j-\tfrac{i}{2},\tfrac{1}{2}\right]}{^{r_{2j-i+1}}_{r_{2j-i} \, e_i}}
\tensor{\left[j-\tfrac{i-1}{2},j-\tfrac{i}{2},\tfrac{1}{2}\right]}{_{u_{2j-i+1}}^{u_{2j-i} \, e_i}}\nonumber\\ 
    &= \frac{1}{(2j)!}\left(\tensor{\delta}{_{e_1}^{d_1}}\cdots  \tensor{\delta}{_{e_{2j}}^{d_{2j}}}\right)\underset{\left\{e_1 \cdots e_{2j} \right\}}{\mathcal{S}}\left(\tensor{\delta}{_{d_1}^{e_1}}\cdots  \tensor{\delta}{_{d_{2j}}^{e_{2j}}}\right)
    \,.
\end{align}
In the towers of Clebsch-Gordan coefficients of the first equality, one can use the orthogonality properties to reduce this expression to $\tensor{\delta}{_{r_{2j}}^{r_{2j}}}=2j+1$.  To simplify the expression on the second line, the cycle structure for the permutations in the symmetrization play a role.  For each term in the symmetrization, the contractions evaluate to $2^k$ with $k$ the number of fixed points (disjoint cycles) in that particular permutation.  In the contraction of the Kronecker deltas, each fixed point contracts to a trace of one Kronecker delta, which yields a factor of 2 as the $d_i,e_i$ indices are associated with spin-1/2. 
With these steps, we obtain for Eq.~(\ref{eq:CGtower_reduction_proof}) that
\begin{equation} \label{eq:stirling}
    2j+1 = \frac{1}{(2j)!}\sum_k c(2j,k) 2^k \,,
\end{equation}
where $c(n,k)$ is the number of permutations of $n$ objects with $k$ cycles, which correspond to the unsigned Stirling numbers of the first kind~\cite{wiki:Stirling_numbers_of_the_first_kind}, and where all permutations sum to $\sum_{k=1}^{n}c(n,k)=n!$.   These unsigned Stirling numbers obey the generating equation
\begin{equation}
    x^{\bar n} \equiv x(x+1)\cdots(x+n-1) = \sum_{k=0}^n c(n,k)x^k \,.
\end{equation}
Evaluating this generating equation for $x=2, n=2j$ recovers Eq.~(\ref{eq:stirling}).  This completes the proof.

%%%%%%%%%%%%%%%%%%%%%%%%%%%%%%%%
%%%% algebra of  Covariant Tensors            %%%%%
%%%%%%%%%%%%%%%%%%%%%%%%%%%%%%%%

\section{Useful relations among the  reduction tensors}
% {Useful relations among the covariant tensors}
\label{sec:coeff_relations}

Here we summarize a number of important algebraic relations satisfied by the invariants rank-4 coefficient tensors defined in  Eqs.~(\ref{eq:coefficients_reduction_cubic}), (\ref{eq:coefficients_reduction_cubic_bar}), (\ref{eq:cubic_coefficients_Cred}) and by the covariant rank-3 coefficient tensors defined in  Eqs.~(\ref{eq:coefficients_reduction_quadratic}) and (\ref{eq:quadratic_coefficients_Qred}).
Their defining equations are repeated here for convenience, 
\begin{subequations}\label{eq:set_C_tensors}
\begin{align}
%\label{eq:coefficients_reduction_cubic}
{\mathcal{C}}^{\mu \rho \sigma \alpha } & = g^{\mu \rho } g^{\sigma \alpha } - g^{\mu \sigma } g^{\rho \alpha } + g^{\mu \alpha } g^{\rho \sigma } + {\rm i} \epsilon^{\mu \rho \sigma \alpha } \,, 
\\
%\label{eq:coefficients_reduction_cubic_bar}
\overbar{\mathcal{C}}^{\mu \rho \sigma \alpha } & = g^{\mu \rho } g^{\sigma \alpha } - g^{\mu \sigma } g^{\rho \alpha } + g^{\mu \alpha } g^{\rho \sigma } - {\rm i} \epsilon^{\mu \rho \sigma \alpha } \,,
\\
%\label{eq:coefficients_reduction_cubic}
{\mathcal{C}}^{\mu \rho \sigma \alpha }_\text{red} & = - g^{\mu \sigma } g^{\rho \alpha } + g^{\mu \alpha } g^{\rho \sigma } + {\rm i} \epsilon^{\mu \rho \sigma \alpha } \,,
\\
\overbar{\mathcal{C}}^{\mu \rho \sigma \alpha }_\text{red} & = - g^{\mu \sigma } g^{\rho \alpha } + g^{\mu \alpha } g^{\rho \sigma } - {\rm i} \epsilon^{\mu \rho \sigma \alpha } 
\,,
\end{align}
\end{subequations}
and
\begin{subequations}\label{eq:set_Q_tensors}
%\label{eq:coefficients_reduction_quadratic}
\begin{align}
{\mathcal{Q}}^{\mu\rho\alpha} = {\mathcal{C}}^{\mu\rho\sigma\alpha} \eta_{\sigma} & =  g^{\mu \rho } \eta ^{\alpha } - g^{\rho \alpha } \eta^{\mu } + g^{\mu \alpha } \eta^{\rho } + {\rm i} \epsilon^{\mu \rho \sigma \alpha } \eta_{\sigma } \,,
\\
\overbar{\mathcal{Q}}^{\mu\rho\alpha} = \overbar{\mathcal{C}}^{\mu\rho\sigma\alpha} \eta_{\sigma} & = g^{\mu \rho } \eta ^{\alpha } - g^{\rho \alpha } \eta^{\mu } + g^{\mu \alpha } \eta^{\rho } - {\rm i} \epsilon^{\mu \rho \sigma \alpha } \eta_{\sigma } \,,
\\
{\mathcal{Q}}_\text{red}^{\mu\rho\alpha} = {\mathcal{C}}_\text{red}^{\mu\rho\sigma\alpha} \eta_{\sigma} & =  - g^{\rho \alpha } \eta^{\mu } + g^{\mu \alpha } \eta^{\rho } + {\rm i} \epsilon^{\mu \rho \sigma \alpha } \eta_{\sigma } \,,
\\
\overbar{\mathcal{Q}}_\text{red}^{\mu\rho\alpha} = \overbar{\mathcal{C}}_\text{red}^{\mu\rho\sigma\alpha} \eta_{\sigma} & = - g^{\rho \alpha } \eta^{\mu } + g^{\mu \alpha } \eta^{\rho } - {\rm i} \epsilon^{\mu \rho \sigma \alpha } \eta_{\sigma } 
\,,
\end{align}    
\end{subequations}
where, $\eta^{\sigma} \equiv (1,0,0,0)$, see Eq.~(\ref{eq:def_eta}).

Many calculations involve the invariant $\mathcal{C}$-coefficients under permutation of the first three indices.  For these permutations it is advantageous to relate them back to the ordering used in Eq.~(\ref{eq:coefficients_reduction_cubic}).  
% In particular, use is made of the following identities, 
% 
The coefficient tensors satisfy the following indices permutations 
\begin{subequations}
\begin{align}
\label{eq:covariant_coefficients_C_permutations}
{\mathcal{C}}^{\sigma \rho \mu \alpha } & =  \overbar{\mathcal{C}}^{\mu \rho \sigma \alpha } = \left({\mathcal{C}}^{\mu \rho \sigma \alpha }\right)^* \,,
\\
\label{eq:covariant_coefficients_Cred_permutations}
{\mathcal{C}}_{\text{red}}^{\rho \mu \sigma \alpha } & =  {\mathcal{C}}_{\text{red}}^{\mu \rho \alpha \sigma} = - {\mathcal{C}}_{\text{red}}^{\mu \rho \sigma \alpha } \,,
\\
\label{eq:covariant_coefficients_Q_permutations}
{\mathcal{Q}}^{\mu \alpha \rho} & =  \overbar{\mathcal{Q}}^{\mu \rho \alpha } = \left({\mathcal{Q}}^{\mu \rho \alpha }\right)^* \,,
\\
\label{eq:covariant_coefficients_Qred_permutations}
{\mathcal{Q}}_{\text{red}}^{\rho \mu \alpha } & = - {\mathcal{Q}}_{\text{red}}^{\mu \rho \alpha } \,.
\end{align}
\end{subequations}

The analogous identities for $\overbar{\mathcal{C}},\overbar{\mathcal{C}}_{\text{red}},\overbar{\mathcal{Q}}_{\text{red}}$-coefficients are obtained by barring all previous equations, which corresponds, in this case, to taking the complex conjugate. 
Note that Eqs.~(\ref{eq:covariant_coefficients_Cred_permutations}) and (\ref{eq:covariant_coefficients_Qred_permutations}) shows that both ${\mathcal{C}}_{\text{red}}^{\rho \mu \sigma \alpha }$ and ${\mathcal{Q}}_{\text{red}}^{\mu \rho \alpha}$ are antisymmetric in their first two indices.  This plays a role in the permutation symmetry of the Lorentz indices in the quadratic products of $t$-tensors Eq.~(\ref{eq:quadratic_products}), where they generate the bi-index structure of Eqs.~(\ref{eq:bi_index}).

The ${\mathcal{Q}}_{\text{red}}$ and $\overbar{\mathcal{Q}}_{\text{red}}$ also satisfy the following useful relations:
\begin{subequations}\label{eq:contraction_Qred}
\begin{align}\label{eq:contraction_Qred_eta}
\eta_\alpha {\mathcal{Q}}_{\text{red}}^{\mu \rho \alpha} & = \eta_\alpha \overbar{\mathcal{Q}}_{\text{red}}^{\mu \rho \alpha} = 0 \,, 
\\[0.5em]
\label{eq:contraction_Qred_g}
g_{\mu \rho} {\mathcal{Q}}_{\text{red}}^{\mu \rho \alpha} & = g_{\mu \rho} \overbar{\mathcal{Q}}_{\text{red}}^{\mu \rho \alpha} = 0 
\,,
\end{align}
\end{subequations}
and are (anti-)self-dual with regard to the first two indices
\begin{subequations}\label{eq:Qred_antiselfdual}
    \begin{align}
    &\frac{\rm i}{2}\epsilon^{\mu\nu\rho\sigma}\tensor{(\mathcal{Q}_\text{red})}{_{\rho\sigma}^\alpha} = -\mathcal{Q}_\text{red}^{\mu\nu\alpha}\,,\\
    &\frac{\rm i}{2}\epsilon^{\mu\nu\rho\sigma}\tensor{(\overbar{\mathcal{Q}}_\text{red})}{_{\rho\sigma}^\alpha} = +\overbar{\mathcal{Q}}_\text{red}^{\mu\nu\alpha}\,.
    \end{align}
\end{subequations}
From Eq.~(\ref{eq:contraction_Qred_eta}) we find that only spatial indices contribute in contractions in the last index of ${\mathcal{Q}}_{\text{red}}^{\mu \rho \alpha}$, which is an important property used in Sec.~\ref{sec:quadratic_t_vs_multipoles} when contracting with $t$-tensors.  Their (anti-)self-dual nature helps in making the connections with the SL(2,$\mathbb{C}$) irrep content of certain objects encountered in the constructions in this work.

The algebra among $t$-tensors, {i.e.} the reduction formulas for products of them, relies on contractions between the covariant tensors in Eqs.~(\ref{eq:set_C_tensors}) and (\ref{eq:set_Q_tensors}). 
% ${\mathcal{Q}}$, $\overbar{\mathcal{Q}}$, ${\mathcal{C}}$, and $\overbar{\mathcal{C}}$. 
Some of these contractions can be simplified by making use of the identity
\begin{equation}
% \label{eq:identities}
\label{eq:one_Levi-Civita}
\epsilon^{\mu \nu \rho \beta} = \eta^{\mu} \epsilon^{0 \nu \rho \beta} + \eta^{\nu} \epsilon^{\mu 0 \rho \beta} + \eta^{\rho} \epsilon^{\mu \nu 0 \beta} +  \eta^{\beta} \epsilon^{\mu \nu \rho 0}
\,.
\end{equation}

Below we state relations involving unbarred tensors. 
As stated above, the analogous equations with barred tensors are obtained by complex conjugation.
The most important contraction between $\mathcal{Q}$-tensors is
\begin{align}\label{eq:QQ_to_C}
\mathcal{Q}^{\mu_1 \rho_1 \alpha} {\mathcal{Q}}^{\mu_2 \rho_2}{}_{\alpha} & = \mathcal{C}^{\mu_1 \rho_1 \rho_2 \mu_2} \,.
\end{align}

The $\mathcal{Q}_{\text{red}}$-tensors further satisfy
\begin{subequations}\label{eq:algebra_quadratic_coeff}
\begin{align} \label{eq:app_algebra_quadratic_coeff_a}  
\mathcal{Q}_\text{red}^{\mu_1 \rho_1 \alpha}  \mathcal{Q}_\text{red}^{\mu_2 \rho_2}{}_{\alpha} & = - \mathcal{C}_\text{red}^{\mu_1 \rho_1 \mu_2 \rho_2} \,,
\\[0.5em]
\label{eq:algebra_quadratic_coeff_b}  
\mathcal{Q}_\text{red}^{\mu_1 \rho_1}{}_{\alpha_1} \mathcal{Q}_\text{red}^{\mu_2 \rho_2}{}_{\alpha_2} \mathcal{Q}_\text{red}^{\alpha_1\alpha_2\xi} & = \mathcal{Q}_\text{red}^{\mu_1 \rho_1}{}_{\alpha_1} \mathcal{Q}_\text{red}^{\mu_2 \rho_2}{}_{\alpha_2} \  {\rm i} \eta_{\alpha_3} \epsilon^{\alpha_1 \alpha_2 \alpha_3 \xi} 
\nonumber \\
& 
= g^{\mu_1 \mu_2} \mathcal{Q}_\text{red}^{\rho_1 \rho_2 \xi}-g^{\mu_1 \rho_2} \mathcal{Q}_\text{ref}^{\rho_1 \mu_2 \xi}+g^{\rho_1 \rho_2} \mathcal{Q}_\text{red}^{\mu_1 \mu_2 \xi}-g^{\rho_1 \mu_2} \mathcal{Q}_\text{red}^{\mu_1 \rho_2 \xi}
\,.
\end{align}
\end{subequations}
Note that with Eqs.~(\ref{eq:algebra_quadratic_coeff_b}) and (\ref{eq:sl(2,C)_generators_chiral}), the Lie algebra commutation relations for the spin $j$ representation can be written as
\begin{align}\label{eq:Lie_algebra_reps_Lorentz}
\mathcal{Q}_\text{red}^{\mu_1 \rho_1}{}_{\alpha_1} \mathcal{Q}_\text{red}^{\mu_2 \rho_2}{}_{\alpha_2} \mathcal{Q}_\text{red}^{\alpha_1 \alpha_2 \xi}  t_{\xi 0 \cdots 0} 
& = - \frac{\rm i}{j} \left( g^{\mu_1 \mu_2} \mathbb{M}^{\rho_1 \rho_2} - g^{\mu_1 \rho_2} \mathbb{M}^{\rho_1 \mu_2} + g^{\rho_1 \rho_2} \mathbb{M}^{\mu_1 \mu_2} - g^{\rho_1 \mu_2} \mathbb{M}^{\mu_1 \rho_2} \right)
\nonumber \\
& = \frac{1}{j} \left[ \mathbb{M}^{\mu_1 \rho_1} , \mathbb{M}^{\mu_2 \rho_2} \right] 
\,.
\end{align}
Introducing the following definition
\begin{equation}\label{eq:g_tilde}
\tilde{g}^{\beta_1 \beta_2} = g^{\beta_1 \beta_2} - \eta^{\beta_1} \eta^{\beta_2}
\,,
\end{equation}
the other kind of contractions among $\mathcal{Q}_{\text{red}}$-tensors can be written as
\begin{subequations}\label{eq:algebra_quadratic_coeff_2}
\begin{align} \label{eq:app_algebra_quadratic_coeff_2_a}
\mathcal{Q}_{\text{red}}^{\mu \rho \beta } \left(\mathcal{Q}^{\text{red}}_{\mu \rho }{}\right)^{\alpha } = & 4 \tilde{g}^{\alpha \beta } \,,
\\[0.5em]
\mathcal{Q}_{\text{red}}^{\mu_1 \rho_1 \alpha_1} \left(\mathcal{Q}^{\text{red}}_{\mu_1}{}\right)^{\rho_2 \alpha_2} = & \tilde{g}^{\rho_1 \rho_2} \tilde{g}^{\alpha_1 \alpha_2} - \tilde{g}^{\rho_1\alpha_2} \tilde{g}^{\rho_2\alpha_1} + \tilde{g}^{\rho_1\alpha_1} \tilde{g}^{\rho_2\alpha_2} 
\\
& + \eta ^{\rho_1} \eta ^{\rho_2} \tilde{g}^{\alpha_1\alpha_2} + {\rm i} \eta^{\rho_1} \eta_{\sigma} \epsilon^{\sigma \alpha_1 \rho_2 \alpha_2} + {\rm i} \eta^{\rho_2} \eta_{\sigma} \epsilon^{\sigma \alpha_2 \rho_1 \alpha_1}
\,.
\end{align}
\end{subequations}

Below we include contractions between  ${\mathcal{C}}_{\text{red}}^{\rho \mu \sigma \alpha }$ and 2 to 4 copies of ${\mathcal{Q}}_{\text{red}}^{\mu \rho \alpha}$.  
As the last index of the ${\mathcal{Q}}_{\text{red}}^{\mu \rho \alpha}$ is contracted with a $t$-tensor, we can use symmetrization in this last index to produce simpler results than the more general contractions.  
These results are
\begin{subequations}\label{eq:contraction_Cred_Qred_Qred}
\begin{align}
\label{eq:contraction_Cred_Qred_Qred_b}
\frac{1}{2} \underset{\left\{(\beta)\right\}}{\mathcal{S}} {\mathcal{C}}^{\text {red }}_{\mu_1 \rho_1 \mu_2 \rho_2} {\mathcal{Q}}_{\text{red}}^{\mu_1 \rho_1  \beta_1} {\mathcal{Q}}_{\text{red}}^{\mu_2 \rho_2  \beta_2} = & \; 4^2 (\eta^{\beta_1} \eta^{\beta_2} - g^{\beta_1 \beta_2}) = -4^2 \ \tilde{g}^{\beta_1 \beta_2} \,,
\\
\label{eq:contraction_Cred_Qred_Qred_c}
\frac{1}{3!}\underset{\left\{(\beta)\right\}}{\mathcal{S}} {\mathcal{C}}^{\text {red }}_{\mu_1 \rho_1 \mu_2 \mu_3} {\mathcal{Q}}_{\text{red}}^{\mu_1 \rho_1  \beta_1} {\mathcal{Q}}_{\text{red}}^{\mu_2 \rho_2  \beta_2} {\mathcal{Q}}_{\text{red}}^{\mu_3 \rho_3  \beta_3} 
= & \; \frac{1}{3!}\underset{\left\{(\beta)\right\}}{\mathcal{S}} \left( - 4 \tilde{g}^{\beta_1 \beta_2} \mathcal{Q}_{\text{red}}^{\rho_2 \rho_3 \beta_3} \right) \,, 
\\
\label{eq:contraction_Cred_Qred_Qred_d}
\frac{1}{4!}\underset{\left\{(\beta)\right\}}{\mathcal{S}} 
{\mathcal{C}}^{\text {red }}_{\mu_1 \mu_2 \mu_3 \mu_4} {\mathcal{Q}}_{\text{red}}^{\mu_1 \rho_1 \beta_1} {\mathcal{Q}}_{\text{red}}^{\mu_2 \rho_2 \beta_2} {\mathcal{Q}}_{\text{red}}^{\mu_3 \rho_3 \beta_3} {\mathcal{Q}}_{\text{red}}^{\mu_4 \rho_4 \beta_4} 
& = {\mathcal{C}}_{\text {red }}^{\rho_1 \rho_2 \rho_3 \rho_4} \frac{1}{4!}\underset{\left\{(\beta)\right\}}{\mathcal{S}} 
\left( \tilde{g}^{\beta_1 \beta_2} \tilde{g}^{\beta_3 \beta_4} \right) 
\,. 
\end{align}
\end{subequations}
Including the contractions with the $t$-tensors, we then have
\begin{subequations}\label{eq:contraction_Cred_Qred_Qred_t}
\begin{align}
\label{eq:contraction_Cred_Qred_Qred_t_a}
{\mathcal{C}}^{\text {red }}_{\mu_1 \rho_1 \mu_2 \rho_2} {\mathcal{Q}}_{\text{red}}^{\mu_1 \rho_1  \beta_1} {\mathcal{Q}}_{\text{red}}^{\mu_2 \rho_2  \beta_2} t_{\beta_1 \beta_2 \cdots} & = 16 \  t_{0 0 \cdots } \,,
\\[0.5em]
\label{eq:contraction_Cred_Qred_Qred_t_b}
{\mathcal{C}}^{\text {red }}_{\mu_1 \rho_1 \mu_2 \mu_3} {\mathcal{Q}}_{\text{red}}^{\mu_1 \rho_1  \beta_1} {\mathcal{Q}}_{\text{red}}^{\mu_2 \rho_2  \beta_2} {\mathcal{Q}}_{\text{red}}^{\mu_3 \rho_3  \beta_3} t_{\beta_1 \beta_2 \beta_3 \cdots}  & = 4 \ \mathcal{Q}_{\text{red}}^{\rho_2 \rho_3 \beta} t_{\beta 0 0 \cdots } \,,
\\[0.5em]
\label{eq:contraction_Cred_Qred_Qred_t_c}
{\mathcal{C}}^{\text {red }}_{\mu_1 \mu_2 \mu_3 \mu_4} {\mathcal{Q}}_{\text{red}}^{\mu_1 \rho_1 \beta_1} {\mathcal{Q}}_{\text{red}}^{\mu_2 \rho_2 \beta_2} {\mathcal{Q}}_{\text{red}}^{\mu_3 \rho_3 \beta_3} {\mathcal{Q}}_{\text{red}}^{\mu_4 \rho_4 \beta_4} t_{\beta_1 \beta_2 \beta_3 \beta_4 \cdots} & = {\mathcal{C}}_{\text {red }}^{\rho_1 \rho_2 \rho_3 \rho_4} t_{0 0 0 0 \cdots }
\,.
\end{align}
\end{subequations}

Other contractions that frequently appear are the full contractions
\begin{subequations}\label{eq:contractions_C_C}
\begin{align}
\label{eq:full_contraction_C_C}
{\mathcal{C}}_{\mu_1 \mu_2  \mu_3 \mu_4}  {\mathcal{C}}^{\mu_1 \mu_2  \mu_3 \mu_4} & = 64  \,,
\\[0.5em]
\label{eq:full_contraction_C_Cred}
{\mathcal{C}}_{\mu_1 \mu_2  \mu_3 \mu_4}  {\mathcal{C}}_{\text {red }}^{\mu_1 \mu_2  \mu_3 \mu_4} & =  48  \,,
\\[0.5em]
\label{eq:full_contraction_Cred_Cred}
{\mathcal{C}}^{\text {red }}_{\mu_1 \mu_2  \mu_3 \mu_4}  {\mathcal{C}}_{\text {red }}^{\mu_1 \mu_2  \mu_3 \mu_4} & = 48 
\,,
\end{align}
\end{subequations} 
and the partial contractions
\begin{subequations}\label{eq:partial_contractions_C_C}
\begin{align}
\label{eq:2_indices_contraction_Cred_Cred_1}
\mathcal{C}^{\text{red}}_{\mu_1 \rho_1}{}^{\mu_2 \rho_2} \mathcal{C}_{\text{red}}^{\mu_1 \rho_1 \mu_4 \rho_4} = & -4 \mathcal{C}_{\text{red}}^{\mu_2 \rho_2 \mu_4 \rho_4}
\,,
\\[0.5em]
\label{eq:2_indices_contraction_Cred_Cred_2}
\mathcal{C}^{\text{red}}_{\mu_1}{}^{\rho_1}{}_{\mu_2}{}^{\rho_2} \mathcal{C}_{\text{red}}^{\mu_1 \mu_2 \mu_4 \rho_4} = & \mathcal{C}_{\text{red}}^{\rho_1 \rho_2 \mu_4 \rho_4}
\,,
\\[0.5em]
\label{eq:3_indices_contraction_Cred_Cred_1}
\mathcal{C}^{\text{red}}_{\mu_1 \rho_1 \mu_2}{}^{\rho_2} \mathcal{C}_{\text{red}}^{\mu_1 \mu_2 \mu_3 \rho_1} = & 12 g^{\rho_2 \mu_3}
\,,
\\[0.5em]
\label{eq:3_indices_contraction_Cred_Cred_2}
\mathcal{C}^{\text{red}}_{\mu_1}{}^{\rho_1}{}_{\mu_2 \mu_3} \mathcal{C}_{\text{red}}^{\mu_1 \mu_2 \mu_3 \mu_4} = & \overbar{\mathcal{C}}^{\text{red}}_{\mu_1 \mu_2 \mu_3}{}^{\rho_1} \mathcal{C}_{\text{red}}^{\mu_1 \mu_2 \mu_3 \mu_4} = 3 g^{\rho_1 \mu_4}
\,.
\end{align}
\end{subequations}

% \subsection{The \texorpdfstring{${\mathcal{R}}^{\mu\sigma\alpha}$}{R tensor} tensor }
% \label{sec:R_tensors}

The ${\mathcal{R}}^{\mu\sigma\alpha}$ tensors were introduced in Sec.~\ref{sec:quadratic_coeff} in the reduction of quadratic products $tt$ and $\bar t\bar t$.  They are defined in Eq.~(\ref{eq:R_coefficients_reduction_quadratic}) as
\begin{subequations}
%\label{eq:R_coefficients_reduction_quadratic}
\begin{align}
{\mathcal{R}}^{\mu\sigma\alpha}={\mathcal{C}}^{\mu\rho\sigma\alpha}\eta_\rho & =  \eta^{\mu} g^{\sigma \alpha} - \eta^{\alpha} g^{\mu \sigma } + \eta^{\sigma} g^{\mu \alpha } + {\rm i} \epsilon^{\mu \rho \sigma \alpha } \eta_{\rho} \,,
\\
\overbar{\mathcal{R}}^{\mu\sigma\alpha}=\overbar{\mathcal{C}}^{\mu\rho\sigma\alpha}\eta_\rho & = \eta^{\mu} g^{\sigma \alpha} - \eta^{\alpha} g^{\mu \sigma } + \eta^{\sigma} g^{\mu \alpha } - {\rm i} \epsilon^{\mu \rho \sigma \alpha } \eta_{\rho}  = \left( {\mathcal{R}}^{\mu\sigma\alpha} \right)^* \,.
\end{align}    
\end{subequations}
All their properties can be derived from the relations
\begin{subequations}
\label{eq:R_vs_Q}
\begin{align}
{\mathcal{R}}^{\mu\sigma\alpha}= {\mathcal{Q}}^{\alpha\sigma\mu} \,,
\\
\overbar{\mathcal{R}}^{\mu\sigma\alpha}=\overbar{\mathcal{Q}}^{\alpha\sigma\mu} \,.
\end{align}    
\end{subequations}
Here we only stress the most relevant for our purposes
\begin{equation}
\label{eq:R_symmetries}
{\mathcal{R}}^{\sigma\mu\alpha}= \left( {\mathcal{R}}^{\mu\sigma\alpha} \right)^* = \overbar{\mathcal{R}}^{\mu\sigma\alpha} \,,
%\\
%\overbar{\mathcal{R}}^{\mu\sigma\alpha}=\overbar{\mathcal{Q}}^{\alpha\sigma\mu} \,.  
\end{equation}
which means that under an interchange of the first two indices the real part of ${\mathcal{R}}$ is symmetric, and its imaginary part is anti-symmetric.
In addition we have,
\begin{subequations}
\label{eq:R_contractions_with_eta}
\begin{align}
\eta_{\mu} {\mathcal{R}}^{\mu\sigma\alpha}= g^{\sigma\alpha} \,,
\\
\eta_{\sigma} {\mathcal{R}}^{\mu\sigma\alpha}=g^{\mu\alpha} \,.
\end{align}    
\end{subequations}

%%%%%%%%%%%%%%%%%%%%%%%%%%%%%%%%
%%%%   Generalized quadratic reductions and CR  %%%%%
%%%%%%%%%%%%%%%%%%%%%%%%%%%%%%%%

\section{Generalized quadratic reductions}
% {Generalized quadratic reductions and commutation relations}
\label{sec:generalized_quadratic_reductions_and_CR}

In this section, we provide useful results related to products and (anti)commutation relations among particular cases of $t$-tensors. By explicit calculation, we find
\begin{align}\label{eq:generalized_quadratic_reduction_for_tt}
t_{\alpha_{1} \cdots \alpha_{m} 0 \cdots 0} t_{\beta_{1} 0 \cdots 0} = & \left( \prod_{l=m+1}^{2j} \eta^{\alpha_l} \right) \left( \prod_{r=2}^{2j} \eta^{\beta_r} \right) t_{(\alpha)} {t}_{(\beta)}
\nonumber\\
= & \frac{1}{2 j}\left[\left(2 j-m\right) t_{\alpha_{1} \cdots \alpha_{m} \beta_{1} 0 \cdots 0}+\sum_{n=1}^{m} {\mathcal{R}}_{\alpha_{n} \beta_{1} \xi} t^{\xi}_{ \alpha_{1} \cdots \alpha_{n-1} \alpha_{n+1} \cdots \alpha_{m} 0 \cdots 0}\right]
\,.
\end{align}
Here, $1\leq m \leq 2j$, and we use Eq.~(\ref{eq:def_tt_reduction_quadratic}) to reduce the quadratic product $t_{(\alpha)} {t}_{(\beta)}$, as well as the properties listed in Appendix~\ref{sec:coeff_relations}.

For the extremal cases $m=1$ and $m=2j$, 
Eq.~(\ref{eq:generalized_quadratic_reduction_for_tt}) reduces to
\begin{subequations}
\begin{align}
\label{eq:generalized_quadratic_reduction_for_tt_m_1}
t_{\alpha_1 0 \cdots 0} t_{\beta_1 0 \cdots 0} = & \frac{1}{2 j}\left[\left(2 j-1\right) t_{\alpha_{1} \beta_{1} 0 \cdots 0} + {\mathcal{R}}_{\alpha_{1} \beta_{1} \xi} t^{\xi}_{0 \cdots 0}\right]
\,,
\\
\label{eq:generalized_quadratic_reduction_for_tt_m_2j}
t_{\alpha_1 \cdots \alpha_{2j}} t_{\beta_1 0 \cdots 0} = & \frac{1}{2 j} \sum_{n=1}^{2j} {\mathcal{R}}_{\alpha_{n} \beta_{1} \xi} t^{\xi}_{ \alpha_{1} \cdots \alpha_{n-1} \alpha_{n+1} \cdots \alpha_{2j} }
\,.
\end{align}
\end{subequations}

% \subsection{Commutation relations}
% \label{sec:CR_generalized_quadratic_reductions}

The following commutation relations are used in Sec.~\ref{sec:covariantly_independent_terms_in_quadratic_t_products} to prove that every term in the sum of Eq.~(\ref{eq:quadratic_products}) transform covariantly independent from the others,
\begin{align}\label{eq:CR_for_tt}
{\left[t_{\beta, 0 \cdots 0}, t_{\alpha_1 \cdots \alpha_m 0 \cdots 0}\right] } & =\frac{1}{j} {\rm i} \operatorname{Im}\left[\sum_{n=1}^m R_{\beta \alpha_n \xi} t^{\xi}_{\alpha_1 \cdots \alpha_{n-1} \alpha_{n+1} \cdots \alpha_m 0 \cdots 0} \right] \nonumber\\
& = \frac{1}{j} \sum_{n=1}^m  {\rm i} \epsilon_{\beta \rho \alpha_n \xi} \, \eta^{\rho} \, t^{\xi}_{\alpha_1 \cdots \alpha_{n-1} \alpha_{n+1} \cdots \alpha_m 0 \cdots 0}
\,,
\end{align}
where Eqs.~(\ref{eq:generalized_quadratic_reduction_for_tt}) and (\ref{eq:R_symmetries}) are used.

% \subsection{Covariant multipoles}
% \label{sec:covariant_multipoles_generalized_quadratic_reductions}

The $\mathfrak{sl}(2,\mathbb{C})$ multipole of order $m$ is defined by Eq.~(\ref{eq:sl2C_multipoles})
\begin{equation}
\label{eq:App_sl2C_multipoles}
{\mathcal{M}}_{m}^{\mu_1 \rho_1 , \cdots , \mu_{m} \rho_{m}} 
= \frac{1}{m!} \underset{\left\{\left(\mu \rho\right)\right\}}{\mathcal{S}} \prod_{r=1}^m \mathbb{M}^{\mu_{r} \rho_{r}}  - \left({\text{Traces}}\right)
\,,
%\nonumber
\end{equation}
for which the symmetrized products of spin-$j$ representation of Lorentz generators  ${\mathbb{M}}^{\mu \rho}$ [Eq.~(\ref{eq:sl(2,C)_generators_chiral})] are needed.
Using the relation between $t$-tensors and Lorentz generators ${\mathbb{M}}^{\mu \rho}$ provided by Eq.~(\ref{eq:sl(2,C)_generators_chiral}), together with repeated use of Eq.~(\ref{eq:generalized_quadratic_reduction_for_tt}), one can find a relation between the aforementioned symmetrized products and $t$-tensors. 
For example, the first term $m=2$
%, which also corresponds to the first $m$-even case 
is given by
\begin{align}\label{eq:symm_prod_Lorentz_Gen_as_t_tensors_m2}
\frac{1}{2!} \underset{\left\{\left(\mu \rho\right)\right\}}{\mathcal{S}} \prod_{r=1}^2 \mathbb{M}^{\mu_{r} \rho_{r}} & = \frac{1}{2!} \underset{\left\{\left(\mu \rho\right)\right\}}{\mathcal{S}} \prod_{r=1}^2 ({\rm i} j) {\mathcal{Q}}_{\text{red}}^{\mu_r \rho_r \alpha_r} t_{\alpha_r 0 \cdots 0}
= \prod_{r=1}^2 ({\rm i} j) {\mathcal{Q}}_{\text{red}}^{\mu_r \rho_r \alpha_r} \frac{1}{2!} \underset{\left\{\left(\mu \rho\right)\right\}}{\mathcal{S}}  \prod_{l=1}^2 t_{\alpha_l 0 \cdots 0}
 \nonumber\\
& = \left(\frac{{\rm i}}{2}\right)^2 \left[ 2! \binom{2j}{2} \left[ \prod_{r=1}^2 {\mathcal{Q}}_{\text{red}}^{\mu_r \rho_r \alpha_r} \right] t_{\alpha_1 \alpha_2 0 \cdots 0} 
+ {\mathcal{C}}_{\text{red}}^{\mu_1 \rho_1 \mu_2 \rho_2} t_{0 \cdots 0}
 \right]
\,,
\end{align}
where we have used Eqs.~(\ref{eq:generalized_quadratic_reduction_for_tt_m_1}), (\ref{eq:R_vs_Q}), (\ref{eq:contraction_Qred_eta}) and (\ref{eq:app_algebra_quadratic_coeff_a}).
The next case $m=3$ corresponds to
\begin{align}\label{eq:symm_prod_Lorentz_Gen_as_t_tensors_m3}
\frac{1}{3!} \underset{\left\{\left(\mu \rho\right)\right\}}{\mathcal{S}} \prod_{r=1}^3 \mathbb{M}^{\mu_{r} \rho_{r}} = \; & \frac{1}{3!} \underset{\left\{\left(\mu \rho\right)\right\}}{\mathcal{S}} \prod_{r=1}^3 ({\rm i} j) {\mathcal{Q}}_{\text{red}}^{\mu_r \rho_r \alpha_r} t_{\alpha_r 0 \cdots 0} \nonumber\\
& \hspace{-.5cm} = \left(\frac{{\rm i}}{2}\right)^3 
3! \binom{2j}{3} \left[ \prod_{r=1}^3 {\mathcal{Q}}_{\text{red}}^{\mu_r \rho_r \alpha_r} \right] t_{\alpha_1 \alpha_2 \alpha_3 0 \cdots 0} - \left(\frac{{\rm i}}{2}\right)^3 \frac{(3j-1)2j}{3} \underset{\left\{\left(\mu \rho\right)\right\}}{\mathcal{S}}  {\mathcal{C}}_{\text{red}}^{\mu_1 \rho_1 \mu_2 \rho_2} {\mathcal{Q}}_{\text{red}}^{\mu_3 \rho_3 \alpha_3} t_{\alpha_3 0 \cdots 0}
\,.
\end{align}
The pattern continues for higher order terms, which take the general form
\begin{align}\label{eq:symm_prod_Lorentz_Gen_as_t_tensors_m}
\frac{1}{m!} \underset{\left\{\left(\mu \rho\right)\right\}}{\mathcal{S}} \prod_{r=1}^m \mathbb{M}^{\mu_{r} \rho_{r}} = & \left(\frac{{\rm i}}{2}\right)^m  m! \binom{2j}{m} \ProdQt{}_{m}^{(\mu\rho)} + \sum_{n=1}^{\lfloor \frac{m}{2} \rfloor} d_{m,n} \underset{\left\{\left(\mu \rho\right)\right\}}{\mathcal{S}}  \ProdQt{}_{m-2n}^{(\mu\rho)}  \prod_{l = m-2n+1, m-2n+3, \cdots}^{m-1} {\mathcal{C}}_{\text{red}}^{\mu_{l} \rho_{l} \mu_{l+1} \rho_{l+1}}
\,,
\end{align}
where we use the notation introduced in Eq.~(\ref{eq:cov_independent_substructures_T}).
We did not find a simple closed form for the coefficients $d_{m,n}$. However, that does not concern us here as we still need to subtract the traces to get to the desired result of Eq.~(\ref{eq:App_sl2C_multipoles}). The important point is that all the terms of the sum in the right-hand side of Eq.~(\ref{eq:symm_prod_Lorentz_Gen_as_t_tensors_m}) will not survive these trace subtractions and will be absorbed (partially) in the $\mathcal{T}_m^{(\mu\nu)}$. 
These terms have the same form as the ${\mathcal{C}}_{\text{red}}$-{\it traces} we found in Sec.~\ref{sec:orthogonal_basis_quadratic_t}. 
This shows that the $\mathfrak{sl}(2,\mathbb{C})$ multipoles must be ${\mathcal{C}}_{\text{red}}$-{\it traceless} in the sense of Eq.~(\ref{eq:Cred_Tau_m}), and they are related to our orthogonal basis $\mathcal{T}_{m}^{(\mu\rho)}$ [see Eq.~(\ref{eq:Tau_m})] only by a multiplicative factor. 
Moreover, the multiplicative factor is that of the highest-order term $\ProdQt{}_{m}^{(\mu\rho)}$ in Eq.~(\ref{eq:symm_prod_Lorentz_Gen_as_t_tensors_m}), which is the one containing the non-trace part of the symmetrized product of Lorentz generators.  
We conclude that the $m$th multipoles are given by
\begin{equation}\label{eq:App_sl(2,C)_multipole_tTensors}
% \label{eq:App_sl(2,C)_multipole_tTensors_a}
{\mathcal{M}}_{m}^{\mu_1 \rho_1 ,\cdots, \mu_m \rho_m}  = \frac{{\rm i}^m}{2^m} m! \binom{2j}{m}  {\mathcal{T}}_m^{\mu_1 \rho_1 ,\cdots, \mu_m \rho_m} 
\,, \qquad 0 \leq m \leq 2j
\,.
\end{equation}
It is worth emphasizing that the subtracted traces appearing in the definition of ${\mathcal{T}}_m^{(\mu\rho)}$ (right-hand side of Eq.~(\ref{eq:App_sl(2,C)_multipole_tTensors})) are not related by an overall multiplicative factor with the traces in the definition of the covariant multipoles Eq.~(\ref{eq:App_sl2C_multipoles}). The reason is that the symmetrized product of Lorentz generators already contains trace-like terms with multiplicative factors that are term dependent, see Eq.~(\ref{eq:symm_prod_Lorentz_Gen_as_t_tensors_m}).

As usual, all the corresponding expressions for the right-handed chiral representation are obtained by {\it barring} those given for the left-handed one.

\acknowledgments

We are thankful to C\'{e}dric Lorc\'{e} for useful discussions during the preparation of this work. 
We thank Jeroen Demeyer for pointing us towards the Stirling numbers to complete the proof in Appendix~\ref{sec:proof_cubic_reduction}.   
We also want to thank Christian Weiss and Adam Freese for encouraging and invigorating discussions about many of the topics presented here. This work is partially supported by NSF grants PHY-2111442 and PHY-2239274.  W.C. acknowledges support from an ORAU Ralph E. Powe award during the early stages of this work.  F.V. acknowledges support from an MPS-ASCEND fellowship NSF grant PHY-2316701, and a JLab EIC Center postdoc fellowship during early stages of this work.
Work supported in part by the U.S. Department of Energy, Office of Science, Office of Nuclear Physics under contract DE-AC05-06OR23177.

%%%%%%%%%%%%%%%%%%%%%%%%%%%%%%%%
%%%% bibliography            %%%%%
%%%%%%%%%%%%%%%%%%%%%%%%%%%%%%%%
%\bibliographystyle{apsrev4-2}
\bibliography{anyspin_amplitudes}

\end{document}